\documentclass[twocolumn,notitlepage,prx,superscriptaddress,longtable, longbibliography]{revtex4-1}
\usepackage{mhchem}
\usepackage{amsthm}
\usepackage{amsmath}
\usepackage{amssymb}
\usepackage{mathdots}
\usepackage{graphicx}
\usepackage{mathrsfs}
\usepackage{longtable}
\usepackage{multirow}
\makeatletter

\@ifundefined{textcolor}{}
{%
 \definecolor{BLACK}{gray}{0}
 \definecolor{WHITE}{gray}{1}
 \definecolor{RED}{rgb}{1,0,0}
 \definecolor{GREEN}{rgb}{0,1,0}
 \definecolor{BLUE}{rgb}{0,0,1}
 \definecolor{CYAN}{cmyk}{1,0,0,0}
 \definecolor{MAGENTA}{cmyk}{0,1,0,0}
 \definecolor{YELLOW}{cmyk}{0,0,1,0}
}

\usepackage{braket}
\usepackage[backref=true,
bookmarksnumbered=true,
bookmarks=true,
bookmarksopen=true,
colorlinks=true,
citecolor=blue,
linkcolor=blue,
anchorcolor=green,
urlcolor=blue,unicode=false]{hyperref}

\let\baraccent=\= 
\renewcommand{\=}[1]{\stackrel{#1}{=}} 

\begin{document}
\title{Floquet higher-order topological insulators and superconductors with space-time symmetries}
\author{Yang Peng}
\email{yang.peng@csun.edu}
\affiliation{Department of Physics and Astronomy, California State University, Northridge, Northridge, California 91330, USA}
\affiliation{Institute of Quantum Information and Matter and Department of Physics,California Institute of Technology,
Pasadena, California 91125, USA}
\affiliation{Walter Burke Institute for Theoretical Physics, California Institute of Technology, Pasadena, California 91125, USA}

\begin{abstract}
Floquet higher-order topological insulators and superconductors (HOTI/SCs) with an order-two space-time symmetry or antisymmetry are classified. 
This is achieved by considering unitary loops, whose nontrivial topology leads to the anomalous Floquet topological phases,  
subject to a space-time symmetry/antisymmetry.  By mapping these unitary loops to static Hamiltonians with an order-two crystalline symmetry/antisymmetry, 
one is able to obtain the $K$ groups for the unitary loops and thus complete the classification of Floquet HOTI/SCs.
Interestingly, we found that for every order-two nontrivial space-time symmetry/antisymmetry involving a half-period time translation, 
there exists a unique order-two static crystalline symmetry/antisymmetry, such that the two symmetries/antisymmetries 
give rise to the same topological classification. Moreover, by exploiting the frequency-domain formulation of the Floquet problem,
a general recipe that constructs model Hamiltonians for Floquet HOTI/SCs is provided, which can be used to 
understand the classification of Floquet HOTI/SCs from an intuitive and complimentary perspective.
\end{abstract}

\maketitle
\section{Introduction}
The interplay between symmetry and topology leads to 
various of topological phases. 
For a translationally invariant noninteracting gapped system,
the topological phase is characterized by the band structure topology, 
as well as the symmetries the system respects. 
Along with these thoughts, a classification was obtained 
for topological insulators and superconductors (TI/SC) \cite{Hasan2010,Qi2011,Bernevig2013book}
in the ten Altland-Zirnbauer (AZ) symmetry classes
\cite{Schnyder2008,Kitaev2009,Ryu2010,Teo2010,Chiu2016},
which is determined by the presence or absence of three types of nonspatial
symmetries, i.e. the time-reversal, particle-hole and chiral symmetries. 

One nice feature of these tenfold-way phases is the bulk-boundary correspondence, 
namely, a topologically nontrival bulk band structure implies the existence
of codimension-one gapless boundary modes on the surface, irrespective the surface orientation. 
(The codimension is defined as the difference between the bulk dimension and 
the dimension of the boundary where the gapless mode propagates).

When considering more symmetries beyond the nonspatial ones, the topological classification is enriched. 
Topological crystalline insulators \cite{Fu2011, Chiu2013, Shiozaki2014, Ando2015, Kruthoff2017} are 
such systems protected by crystalline symmetries.
They are able to host codimension-one gapless boundary modes only when the boundary is invariant under the crystalline symmetry operation. 
For example, topological crystalline insulators protected by reflection symmetry \cite{Chiu2013} 
can support gapless modes only on the reflection invariant boundary.
On the other hand, inversion symmetric topological crystalline insulators do not 
necessarily give rise to codimension-one gapless boundary modes \cite{Turner2010, Hughes2011}, 
because no boundary is invariant under inversion.

Remarkably, it was recently demonstrated that a crystal with a crystalline-symmetry compatible bulk topology
may manifest itself through protected boundary modes of codimension greater than one \cite{Slager2015, 
Benalcazar2017, Peng2017, Langbehn2017, Benalcazar2017s, Song2017, Schindler2018, Geier2018, Khalaf2018, Khalaf2018prx, Trifunovic2019}.
Such insulating and superconducting phases are called higher-order topological insulators and
superconductors (HOTI/SCs). Particularly, an $n$th order TI/SC can support codimension-$n$ boundary modes. 
(The strong TI/SCs in the tenfold-way phases with protected boundary modes at codimension one
can be called as first-order TI/SCs according to this definition.)
A higher-order bulk-boundary correspondence between the bulk topology and gapless boundary modes at different codimensions 
was derived in Ref.~\cite{Trifunovic2019} based on $K$-theory. 

Beyond equilibrium or static conditions,
it is known that topological phases also exist, and one of the famous examples 
is the Floquet topological insulator,
which is proposed to be brought from a static band insulator
by applying a periodic drive, such as a circularly polarized radiation or
an alternating Zeeman field \cite{Oka2009,Inoue2010,Kitagawa2011,Lindner2011,Lindner2013}.
A complete classification of the Floquet topological insulators (as well as superconductors) in the AZ symmetry
classes has been obtained in Ref.~\cite{Roy2017, Yao2017}, which can be regarded as a generalization
of the classfication for static tenfold-way TI/SCs.

In a periodically driven, or Floquet, system, the nontrivialty can arise from the nontrivial topology of
the unitary time-evolution operator $U(t)$ (with period $T$), which can be decomposed into two parts as $U(t) = e^{-iH_F
t}P(t)$. 
Here, the first part describes the stroboscopic evolution at time of multiples of $T$ in terms of
a static effective Hamiltonian $H_F$, and the second part is known as the micromotion operator $P(t) = P(t+T)$
describing the evolution within a given time period \cite{Shirley1965}. (We will make this decomposition 
more explicit later).
Thus, the nontrivial topology can separately arise from $H_F$ as in a static topological phase, or 
from the nontrivial winding of $P(t)$ over one period. 
Whereas Floquet topological phase in former situation is very similar to a static topological phase as it has a static
limit, the latter is purely dynamical and cannot exist if the time-periodic term in the Hamiltonian vanishes. Therefore,
systems belong to the latter case are more interesting and are known as the anomalous Floquet topological phases.

\begin{figure}[t]
  \centering
  \includegraphics[width=0.4\textwidth]{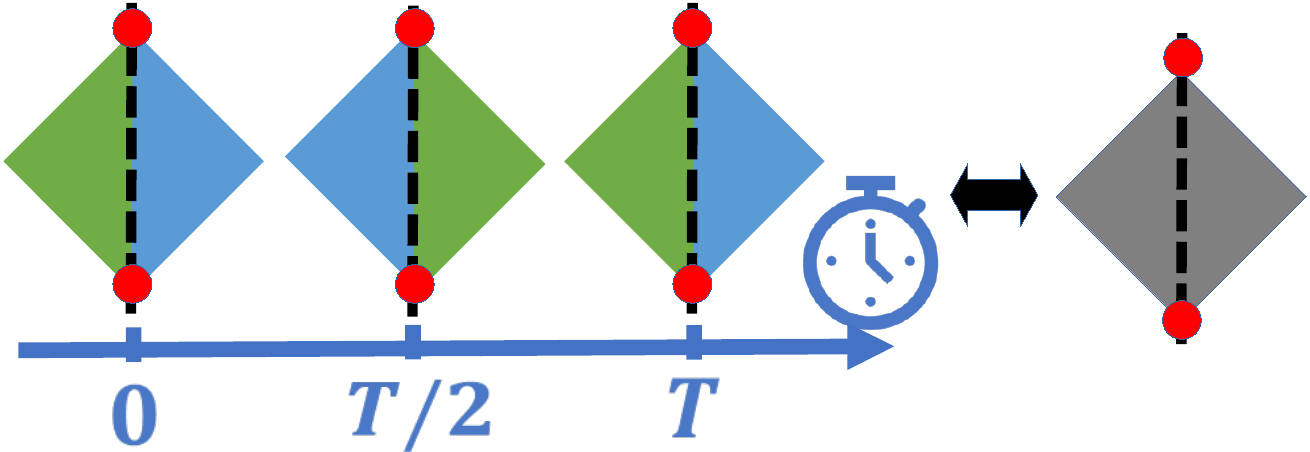}
  \caption{Floquet second-order TI/SCs protected by time-glide symmetry/antisymmetry can be mapped to 
    static second-order TI/SCs protected by reflection symmetry/antisymmetry.
    The dashed line indicates the reflection (time-glide) plane. }
  \label{fig:FHOTI}
\end{figure}

In a Floquet system, energy is not conserved because of the excplicit time-depdendence of the Hamiltonian. 
However, one can define quasienergies as eigenvalues of $H_F = \frac{i}{T}\ln U(T)$, 
which are only defined modulo the periodic driving
frequency $\omega = 2\pi/T$. This can be intuitively understood due to the existence of energy quanta $\omega$
that can be absorbed and emitted. 
Similar to static topological phases, the quasienergy spectrum can be different with different boundary conditions. 
Particularly, inside a bulk quasienergy gap (when periodic boundary condition is applied), 
there may exist topologically protected boundary modes.

In Floquet topological phases protected by nonspatial symmetries (tenfold-way phases),  the bulk-boundary correspondence 
is also expected to hold \cite{Roy2017}, namely the number of boundary modes inside a particular bulk gap
can be fully obtained from the topology of the evolution operator $U(t)$, when periodic boundary condition is applied. 
Interestingly, when there exists a symmetry relating states at quasienergies $\epsilon$ and $-\epsilon$, then the 
topological protected boundary modes will appear inside the quasienergy gap at $0$ and $\omega/2$, since these 
are quasienergies that are invariant under the above symmetry operation. 

In particular, a bulk micromotion operator with nontrivial topology 
is able to produce gapless Floquet codimension-one boundary modes at quasienergy $\omega/2$ (which will be made clear later).
The natural following question to ask is that how can we create Floquet higher-order topological phases, 
with protected gapless modes at arbitrary codimensions. In particular, we want to have the topological nontriviality 
arise from the micromotion operator, otherwise we just need to have $H_F$ as a Hamiltonian for a static higher-order topological phase. 

Similar to the static situation, when only nonspatial symmetries are involved, the tenfold-way Floquet
topological phases are all first-order phases which can only support codimension-one boundary modes. 
Higher-order phases are yet possible when symmetries relating different spatial points of the system 
are involved. These symmetries can be static crystalline symmetries, as well as
space-time symmetries which relates systems at different times. 

Recently, the authors in Refs.~\cite{Huang2018, Bomantara2019,Rodriguez2019, Peng2019, Seshadri2019, Nag2019} constructed
Floquet second-order TI/Scs. Particularly, the authors of Ref.~\cite{Peng2019} were able to
construct Floquet corner modes by exploiting the time-glide symmetry~\cite{Morimoto2017}, 
which combines a half-period time translation and a spatial reflection, as illustrated in the left part of
Fig.~\ref{fig:FHOTI}.

It turns out that the roles played by such space-time symmetries in Floquet systems
cannot be trivially replaced by spatial symmetries. 
As pointed out in Ref.~\cite{Peng2019}, in protecting anomalous Floquet boundary modes, 
the space-time symmetries generally have different commutation relations with the 
nonspatial symmetries, compared to what the corresponding spatial symmetries do. 

Since the use of space-time symmetries opens up new possibilities in 
engineering Floquet topological phases, especially the Floquet HOTI/SCs,
it is important to have a thorough topological classification, 
as well as a general recipe of model construction for such systems.

In this work, we completely classify Floquet HOTI/SCs with an order-two space-time symmetry/antisymmetry 
realized by an operator $\hat{\mathcal{O}}$, which can be either unitary or antiunitary. 
By order-two, it means that the symmetry/antisymmetry operator
twice trivially acts on the time-periodic Hamiltonian $H(t)$, namely
\begin{equation}
  [\hat{\mathcal{O}}^{2},H(t)]=0,\quad\hat{\mathcal{O}}=\hat{\mathcal{U}},\hat{\mathcal{A}}
\label{eq:order_two_symmetry}
\end{equation}
where $\hat{\mathcal{O}}$ can be either unitary $\hat{\mathcal{U}}$
or antiunitary $\hat{\mathcal{A}}$.

We further provide a general recipe of constructing tight-binding Hamiltonians for such Floquet HOTI/SCs in 
different symmetry classes. 
Note that the order-two static crystalline
symmetries/antisymmetries considered in Ref.~\cite{Shiozaki2014} will be a subset of the symmetries/antisymmetries considered in this work. 

Our classification and model construction of Floquet HOTI/SCs involve two complementary approaches.
The first approach is based on the classification of gapped unitaries \cite{Roy2017,Morimoto2017},
namely the time-evolution operator $U(t)$ at time
$t \in [0,T)$, with $U(T)$ gapped in its eigenvalues' phases.
It turns out that the gapped unitaries can be (up to homotopy equivalence) decomposed 
as a unitary loop (which is actually the micromotion operator) and a unitary evolution under
the static Floquet Hamiltonian $H_F$. 
Thus, a general gapped unitary is classified by separately considering the unitary loop
and the static Hamiltonian $H_F$, where the latter is well known for systems
in AZ classes as well as systems with additional crystaline symmetries. 
The classification of unitary loops on the other hand is less trivial
since it is responsible for the existence of anomalous Floquet phases \cite{Rudner2013},
especially when we are considering space-time symmetries. 

We focus on the classification of Floquet unitary loops in this work.
In particular, a hermitian map between unitary loops and hermitian matrices is introduced, 
which is inspired by the dimensional reduction map used in the classification of TI/SCs with
scattering matrices \cite{Fulga2012}.  
The key observation is that the symmetry constraints on the unitary loops 
share the same features as the ones on scattering matrices.
This hermitian map has advantages over the one used in earlier works \cite{Roy2017, Morimoto2017}, 
because it simply maps a unitary loop with a given order-two space-time symmetry/antisymmetry
to a static Hamiltonian of a topological crystaline insulator with an order-two crystalline symmetry/antisymmetry.
This enable us to exploit the full machinary of $K$ theory,
to define $K$ groups, as well as the $K$ subgroup series introduced in Ref.~\cite{Trifunovic2019}, 
for the unitary loops subject to space-time symmetries/antisymmetries.


\begin{table}
\caption{\label{tab:summary} 
  Nontrivial space-time symmetry/antisymmetry with subscript $T/2$ vs. static spatial symmetry/antisymmetry with
  subscript $0$, sharing the same $K$ groups at the same dimension. $\hat{\mathcal{U}}$, $\hat{\mathcal{A}}$, 
  $\overline{\mathcal{U}}$ and $\overline{\mathcal{A}}$ denote unitary symmetry, antiunitary symmetry, unitary
  antisymmetry and antiunitary antisymmetry, respectively.
  The commutation (anticommutation) relations with
  coexisting nonspatial symmetries are denoted as additional subscripts $+$ ($-$), while the superscript indicates the square
  of the operator. In the case of classes BDI, DIII, CII, CII, the first and second $\pm$ correspond to
   time-reversal, and particle-hole symmetries, respectively. 
 }
\begin{ruledtabular}
\centering
\begin{tabular}{ccc}
AZ class& Space-time & Static \\
\hline
\multirow{2}{*}{A} &$\hat{\mathcal{U}}_{T/2}^+$ &$\hat{\mathcal{U}}_{0}^+$ \\
&$\overline{\mathcal{A}}_{T/2}^{\pm}$ &$\overline{\mathcal{A}}_{0}^{\mp}$ \\
\hline
\multirow{2}{*}{AIII} &$\hat{\mathcal{U}}_{T/2,\pm}^{+}$ & $\hat{\mathcal{U}}_{0,\mp}^{+}$\\
&$\hat{\mathcal{A}}_{T/2,\pm}^{\pm}$  & $\hat{\mathcal{A}}_{0,\mp}^{\pm}$\\
\hline
\multirow{2}{*}{AI, AII} &$\hat{\mathcal{U}}_{T/2,\pm}^+$ &$\hat{\mathcal{U}}_{0,\pm}^+$ \\
&$\overline{\mathcal{U}}_{T/2,\pm}^+$ &$\overline{\mathcal{U}}_{0,\mp}^+$  \\
\hline
C, D &$\hat{\mathcal{U}}_{T/2,\pm}^+$ &$\hat{\mathcal{U}}_{0,\mp}^+$ \\
\hline
BDI, DIII, CII, CI &$\hat{\mathcal{U}}_{T/2,\pm \pm}$ &$\hat{\mathcal{U}}_{0,\pm \mp}$
\end{tabular}
\end{ruledtabular}
\end{table}

Based on this approach, we obtain the first important result of the this work,
namely, for every order-two nontrivial space-time (anti)unitary symmetry/antisymmetry, which involves
a half-period time translation, there always exists a unique order-two
static spatial (anti)unitary symmetry/antisymmetry, such that the two symmetries/antisymmetries corresopond
to the same $K$ group and thus the same classification. 
This result is illustrated in Fig.~\ref{fig:FHOTI} for the case of time-glide vs. reflection symmetries.
The explicit relations are summarized in Table~\ref{tab:summary}.
Because of these relations, all results for the classification \cite{Benalcazar2017, Peng2017, Langbehn2017,
Benalcazar2017s, Song2017, Schindler2018, Geier2018, Khalaf2018, Khalaf2018prx} as well as the higher-order bulk-boundary
correspondence \cite{Trifunovic2019} of static HOTI/SCs can be applied directly to the anomalous Floquet HOTI/SCs. 

In the second approach, by exploiting the frequency-domain formulation,
we obtain the second important result of this work, which is
a general recipe of constructing harmonically driven Floquet HOTI/SCs from static HOTI/SCs. 
This recipe realizes the $K$ group isomorphism of systems with a space-time symmetry and 
systems with a static crystalline symmetry at the microscopic level of Hamiltonians,
and therefore provide a very intuitive way of understanding the classification table obtained from the formal $K$ theory.

The rest of the paper is organized as follows. We first introduce the symmetries,
both nonspatial symmetries and the order-two space-time symmetries,
for Floquet system in Sec.~\ref{sec:symmetries}. 
Then, in Sec.~\ref{sec:hermitian_map}, we introduce a hermitian map which 
enables us to map the classification of unitary loops
to the classification of static Hamiltonians. 
In Sec.~\ref{sec:classification_order_two}, by using the hermitian map,
we explicitly map the classification of unitary loops in all possible symmetry classes
supporting an order-two symmetry, to the classification 
of static Hamiltonians with an order-two crystalline symmetry. 
In Sec.~\ref{sec:Kgroup}, we derive the corresponding $K$ groups for unitary loops 
in all possible symmetry classes and dimensions. 
In Sec.~\ref{sec:FHOTI_extension}, we introduce the
$K$ subgroup series for unitary loops, which enables us to completely
classify Floquet HOTI/SCs. 
In Sec.~\ref{sec:FHOTI_frequency_domain}, the frequency-domain formulation is introduced,
which provides a complimentary perspective on the topological classification
of Floquet HOTI/SCs.
In Sec.~\ref{sec:models}, we introduce a general recipe of constructing
harmonically driven Floquet HOTI/SCs, and provide examples in different situations. 
Finally, we conclude our work in Sec.~\ref{sec:conclusion}.

Note that it is possible to skip the $K$-theory classification sections from \ref{sec:hermitian_map} to 
\ref{sec:FHOTI_extension}, and understand the main results in terms of the frequency-domain formulation.   

\section{Floquet basics \label{sec:floquet_basics}}
In a Floquet system, the Hamiltonian 
\begin{equation}
H(t+T)=H(t)
\end{equation}
is periodic in time with period $T=2\pi/\omega$, where $\omega$
is the angular frequency. In a $d$-dimensional system with translational symmetry
and periodic boundary condition, we have well defined Bloch wave vector $\boldsymbol{k}$
in the $d$ dimensional Brillouin zone $T^d$ (torus).
The system can thus be characterized by a time-periodic Bloch Hamiltonian
$H(\boldsymbol{k},t)$. 

In the presence of a $d_{\mathrm{def}}-$dimensional topological defect, the
wave vector $\boldsymbol{k}$ is no longer a good quantum number due
to the broken translational symmetry. However, the topological
properties of the defect can be obtained by considering a large
$D=(d-d_{\mathrm{def}}-1)$-dimensional surface,
on which the translational symmetry is asymptotically restored so
that $\boldsymbol{k}$ can be defined, surrounding the defect.
We will denote $\boldsymbol{r}$ as the real space coordinate
on this surrounding surface, or a $D$-sphere $S^{D}$,
which will determine the topological classification.
Thus, we have a time-periodic ($t\in S^{1}$) Bloch Hamiltonian $H(\boldsymbol{k},\boldsymbol{r},t)$
defined on $T^{d}\times S^{D+1}$. In the following, we will denote the dimension
of such a system with a topological defect as a pair $(d,D)$. 

The topological properties for a given Hamiltonian $H(\boldsymbol{k},\boldsymbol{r},t)$, can
be derived from its time-evolution operator 
\begin{equation}
  U(\boldsymbol{k},\boldsymbol{r},t_0+t,t_0)=\hat{\mathscr{T}}\exp\left[-i\int_{t_0}^{t_0+t}dt'\,H(\boldsymbol{k},\boldsymbol{r},t')\right],
\end{equation}
where $\hat{\mathscr{T}}$ denotes the time-ordering operator.
The Floquet effective Hamiltonian $H_F(\boldsymbol{k},\boldsymbol{r})$ is defined as 
\begin{equation}
U(\boldsymbol{k},\boldsymbol{r},T+t_0,t_0) = \exp(-iH_F(\boldsymbol{k},\boldsymbol{r})T).
\end{equation}
Note that different $H_F$s defined at different $t_0$s are related by unitary transformations, 
and thus the eigenvalues of the Floquet effective Hamiltonian are uniquely defined independent of $t_0$.
which is independent of $t_0$. 
We also introduce $\epsilon_{n}(\boldsymbol{k},\boldsymbol{r})\in [-\pi/T,\pi/T]$ to denote the $n$th eigenvalue of
$H_F(\boldsymbol{k},\boldsymbol{r})$, and call it the $n$th quasienergy band.
Although $H_F$ captures the stroboscopic evolution of the system, it does not
produce a complete topological classification of the Floquet phases. 
It is known that one can have the so called anomalous Floquet phases even when  
$H_F$ is a trivial Hamiltonian.

To fully classify the Floquet phases, we need information of the evolution operator
at each $t$ within the period. 
In order to have a well defined phase, we will only consider gapped unitary
evolution operators, whose quasienergy bands are gapped at a particular
quasienergy $\epsilon_{\rm gap}$. 
Thus, given a set of symmetries the system respects, one needs to classify these gapped unitaries defined from each gapped 
quasienergies $\epsilon_{\rm gap}$.  
The most common considered gapped energies in a system with particle-hole
or chiral symmetry are $0$ and $\omega/2$, since such energies respect the symmetry.
Note that the $\epsilon_{\rm gap}=\omega/2$ case is more interesting since they correspond to anomalous Floquet
phases \cite{Rudner2013}, which has no static analog. 
When neither of the two above mentioned symmetries exists, the gapped energy can take any value, but one
can always deform the Hamiltonian such that the gapped energy appears at $\omega/2$ without changing the topological
classification. Hence, in the following we will fix $\epsilon_{\rm gap}=\omega/2$.

It is evident that the initial time $t_0$ in the evolution operator does not affect the classification, since
it corresponds to different ways of defining the origin of time. Thus, from now on, we will set $t_0=0$ and
denote
\begin{equation}
U(\boldsymbol{k},\boldsymbol{r},t) = U(\boldsymbol{k},\boldsymbol{r},t,0).
\end{equation}

A less obvious fact, is that one can define
the symmetrized time-evolution operator \cite{Roy2017}
centered around time $\tau$ as
\begin{equation}
U_{\tau}(\boldsymbol{k},\boldsymbol{r},t)=\mathscr{T}\exp\left[-i\int_{\tau - \frac{t}{2}}^{\tau +
\frac{t}{2}}dt'\,H(\boldsymbol{k},\boldsymbol{r},t')\right]\label{eq:symm-evolution}, 
\end{equation}
which will also give rise to the same topological classification. 
This statement is proved in Appendix~\ref{app:proof_symmetric_evolution}.
In fact, $U_{\tau}(\boldsymbol{k},\boldsymbol{r},T)$ leads to the same quasienergy band structure
independent of the choice of $\tau$. This is because (the explicit $\boldsymbol{k},\boldsymbol{r}$ 
dependence is omitted) 
\begin{equation}
U_\tau(T)  = W U_0(T) W^\dagger 
\end{equation}
with unitary matrix $W = U(\tau+T/2)U^\dagger(T/2)$.
Thus, $U_\tau(T)$s at different $\tau$s are related by unitary transformation,
and we will in the following use $U_{\tau}(\boldsymbol{k},\boldsymbol{r},t)$ 
to classify Floquet topological phases. 

For classification purpose, we need to setup the notion of homotopy equivalence between
unitary evolutions. Let us consider evolution operators gapped at a given quasienergy.
Following the definition in Ref.~\cite{Roy2017}, we say two evolution operators $U_{1}$ and $U_{2}$ are homotopic, 
denoted as $U_{1}\approx U_{2}$,
if and only if there exists a continuous unitary-matrix-valued function $f(s)$, 
with $s\in[0,1]$, such that 
\begin{equation}
	f(0)=U_1,\quad f(1)=U_2,
\end{equation}
where $f(s)$ is a gapped evolution operator for all intermediat $s$. 
It is worth mentioning that when dealing with symmetrized evolution operators instead of ordinary
evolution operators, the definition of homotopy equivalence is similar except one needs to impose
that the interpolation function $f(s)$ for all $s$ is also a gapped symmetrized evolution operator. 
When comparing evolution operators with different number of bands, the equivalence relation
of stable homotopy can be further introduced. Such a equivalence relation is denoted as $U_{1}\sim U_{2}$
if there exist two trivial unitaries $U_{n_1}^0$ and $U_{n_2}^0$, with $n_1$ and $n_2$ bands respectively,
such that
\begin{equation}
U_1 \oplus U_{n_1}^0 \approx  U_2 \oplus U_{n_2}^0,
\end{equation}
where $\oplus$ denotes the direct sum of matrices. 

We will now define how to make compositions between two symmetrized evolution operators.
Using the notation in Ref.~\cite{Roy2017}, we write the evolution due to $U_{\tau,1}$ followed by 
$U_{\tau,2}$ as $U_{\tau,1} * U_{\tau,2}$, which is given by the symmetrized evolution 
under Hamiltonian $H(t)$ given by
\begin{equation}
H(t)=\begin{cases}
H_{2}(2t+\frac{T}{2}-\tau) & \tau-\frac{T}{2}\leq t\leq\tau-\frac{T}{4}\\
H_{1}(2t-\tau) & \tau-\frac{T}{4}\leq t\leq\tau+\frac{T}{4}\\
H_{2}(2t-\frac{T}{2}-\tau) & \tau+\frac{T}{4}\leq t\leq\tau+\frac{T}{2},
\end{cases}	
\end{equation}
where $H_1(t)$ and $H_2(t)$ are the corresponding Hamiltonians used for the evolution operators $U_{\tau,1}$ 
and $U_{\tau,2}$, respectively. 

As proved in Ref.~\cite{Roy2017}, with such definitions of homotopy and compositions of evolution operators, 
one can obtain the following two important theorems. First, every gapped symmetrized evolution operator $U_{\tau}$
is homotopic to a composition of a unitary loop $L_{\tau}$, followed by a constant Hamiltonian evolution $C_{\tau}$, 
unique up to homotopy. Here the unitary loop is a special time evolution operator such that it becomes an identity
operator after a full period evolution. Second, $L_{\tau,1}*C_{\tau,1} \approx L_{\tau,2} * C_{\tau,2}$ if and only if
$L_{\tau,1}\approx L_{\tau,2}$ and $C_{\tau,1}\approx C_{\tau,2}$, $L_{\tau,1}$, $L_{\tau,2}$ are unitary loops, 
and $C_{\tau,1}$, $C_{\tau,2}$ are constant Hamiltonian evolutions.  For completeness, we put the proof of the two
theorems in Appendix~\ref{app:two_theorems}.

Because of these two theorems, classifying generic time-evolution operators
reduces to classifying separately the unitary loops and the constant Hamiltonian evolutions.
Since the latter is exactly the same as classifying static Hamiltonians, we will in this work only focus
on the classification of unitary loops.
In the following, all the following time-evolution operators are unitary loops, which additionally satisfy
$U_{\tau}(\boldsymbol{k},\boldsymbol{r},t) = U_{\tau}(\boldsymbol{k},\boldsymbol{r},t+T)$.

\section{Symmetries in Floquet systems \label{sec:symmetries}}
In this section, we will summarize the transformation properties of the time evolution opeerator
under various of symmetry operators. 
\subsection{Nonspatial symmetries}
Let us first look at the nonspatial symmetries and consider systems belong to
one of the ten AZ classes (see Table
\ref{tab:AZ-symmetry-classes}), 
determmined by the presence or absence of time-reversal, particle-hole
and chiral symmetries, which are defined by the operators $\hat{\mathcal{T}}=\mathcal{U}_{T}\mathcal{\hat{K}}$,
$\hat{\mathcal{C}}=\mathcal{U}_{C}\hat{\mathcal{K}}$ and
$\hat{\mathcal{S}}=\mathcal{U}_{S}=\hat{\mathcal{T}}\hat{\mathcal{C}}$
respectively, such that
\begin{gather}
\hat{\mathcal{T}}H(\boldsymbol{k},\boldsymbol{r},t)\hat{\mathcal{T}}^{-1}=H(-\boldsymbol{k},\boldsymbol{r},-t) \nonumber\\
\hat{\mathcal{C}}H(\boldsymbol{k},\boldsymbol{r},t)\hat{\mathcal{C}}^{-1}=-H(-\boldsymbol{k},\boldsymbol{r},t) \nonumber\\
\hat{\mathcal{S}}H(\boldsymbol{k},\boldsymbol{r},t)\hat{\mathcal{S}}^{-1}=-H(\boldsymbol{k},\boldsymbol{r},-t)
\label{eq:AZsym_H}.
\end{gather}
where $\hat{\mathcal{T}}=\mathcal{U}_{T}\mathcal{\hat{K}}$, $\hat{\mathcal{C}}=\mathcal{U}_{C}\hat{\mathcal{K}}$
are antiunitary operators with unitary matrices $\mathcal{U}_{T},\mathcal{U_{C}}$
and complex conjugation operator $\hat{\mathcal{K}}$. 
Here $\boldsymbol{r}$ is invariant in the above equations, because of the nonspatial
nature of the symmetries.

For a Floquet system, the action of symmetry operations
$\hat{\mathcal{T}}$, $\hat{\mathcal{C}}$, and $\hat{\mathcal{S}}$ 
on the symmetrized unitary loops $U_{\tau}(\boldsymbol{k},\boldsymbol{r},t)$
can be summarized as
\begin{gather}
\hat{\mathcal{T}}U_\tau(\boldsymbol{k},\boldsymbol{r},t)\hat{\mathcal{T}}^{-1}=U_{-\tau}^{\dagger}(-\boldsymbol{k},\boldsymbol{r},t)
\label{eq:time-reversal-U} \\
\hat{\mathcal{C}}U_\tau(\boldsymbol{k},\boldsymbol{r},t)\hat{\mathcal{C}}^{-1}=U_\tau(-\boldsymbol{k},\boldsymbol{r},t)
\label{eq:particle-hole-U} \\
\hat{\mathcal{S}}U_\tau(\boldsymbol{k},\boldsymbol{r},t)\hat{\mathcal{S}}^{-1}=U_{-\tau}^{\dagger}(\boldsymbol{k},\boldsymbol{r},t)
\label{eq:chiral-U}
\end{gather}
which follow directly from Eqs.~(\ref{eq:AZsym_H}).

For later convenience, we further introduce notations $\text{\ensuremath{\epsilon_{T}=}}\mathcal{U}_{T}\mathcal{U}_{T}^{*}=\hat{\mathcal{T}}^{2}=\pm1$,
$\epsilon_{C}=\mathcal{U}_{C}\mathcal{U}_{C}^{*}=\hat{\mathcal{C}}^{2}=\pm1$,
and $\epsilon_{S}=\mathcal{U}_{S}^{2}=\hat{\mathcal{S}}^{2}=1$ respectively.

\subsection{Order-two space-time symmetry \label{sec:order-two}}
In addition to the nonspatial symmetries, let us assume
the system supports an order-two space-time symmetry realized by $\hat{\mathcal{O}}$, 
as defined in Eq.~(\ref{eq:order_two_symmetry}). 
Moreover, we assume $\hat{\mathcal{O}}$ commute or anticommute with the 
operators for the nonspatial symmetries of the system. 

Under the order-two space-time
symmetry operation $\hat{\mathcal{O}}$,
the momentum $\boldsymbol{k}$ transforms as \cite{Shiozaki2014}
\begin{equation}
\boldsymbol{k}\to\begin{cases}
\hat{\mathcal{O}}\boldsymbol{k}=(-\boldsymbol{k}_{\parallel},\boldsymbol{k}_{\perp}) & {\rm for}\
\hat{\mathcal{O}}=\hat{\mathcal{U}}  \\
-\hat{\mathcal{O}}\boldsymbol{k}=(\boldsymbol{k}_{\perp},-\boldsymbol{k}_{\parallel}) & {\rm for}\ \hat{\mathcal{O}}=\hat{\mathcal{A}},
\end{cases}
\end{equation}
where the second equality assumes we are in the diagonal basis of $\hat{\mathcal{O}}$, 
$\boldsymbol{k}_\parallel = (k_1,k_2,\dots,k_{d_{\parallel}})$, and $\boldsymbol{k}_\perp = (k_{d_{\parallel}+1}, k_{d_{\parallel}+2},\dots, k_{d})$.

While the nonspatial symmetries leave the spatial coordinate $\boldsymbol{r}$ invariant, 
the order-two space-time symmetry transforms $\boldsymbol{r}$ nontrivially.
To determine the transformation law, we follow Ref.~\cite{Shiozaki2014} and consider a
$D$-dimensional sphere $S^{D}$ surrounding the topological defect, whose coordinates
in Euclidean space are determined by
\begin{equation}
\boldsymbol{n}^2 = a^2, \quad \boldsymbol{n}=(n_1,n_2,\dots,n_{D+1}),
\end{equation}
with radius $a>0$. Since $\hat{\mathcal{O}}$ maps $S^D$ into itself, 
we have 
\begin{equation}
\boldsymbol{n} \to (-\boldsymbol{n}_{\parallel},\boldsymbol{n}_{\perp}),
\end{equation}
with $n_{\parallel}=(n_1,n_2,\dots,n_{D_{\parallel}})$, and $n_{\perp} =
(n_{D_{\parallel}+1},n_{D_\parallel+2},\dots,n_{D+1})$ in a diagonal basis
of $\hat{\mathcal{O}}$. When $D_{\parallel}\leq D$,
we can introduce the coordinate $\boldsymbol{r}\in S^{D}$
by 
\begin{equation}
r_{i} = \frac{n_i}{a-n_{D+1}}, \quad (i=1,\dots,D),
\end{equation}
which leads to 
\begin{equation}
	\boldsymbol{r}\to(-\boldsymbol{r}_{\parallel},\boldsymbol{r}_{\perp}).
\end{equation}
Here, $\boldsymbol{r}_{\parallel}= (r_1,r_2,\dots,r_{D_{\parallel}})$ and $\boldsymbol{r}_{\perp} = (r_{D_{\parallel+1}},
r_{D_{\parallel}+2},\dots,r_{D})$.

Thus, we need to introduce $(d,d_{\parallel},D,D_{\parallel})$ to characterize
the dimension of the system according to the transformation properties
of the coordinates, where $d$ and D are defined the same as defined
previously, while $d_{\parallel}$ and $D_{\parallel}$ denote the
dimensions of the flipping momenta and the defect surrounding coordinates,
respectively. For example, a unitary symmetry with $(d,d_{\parallel},D,D_{\parallel})=(2,1,1,1)$ correspond to the
reflection in 2D with a point defect on the reflection line, 
while a unitary symmetry with $(d,d_{\parallel},D,D_\parallel) = (3,2,2,2)$ is
a two-fold rotation in 3D with a point defect on the rotation axis.

Next, let us consider the action of the order-two space-time symmetry on the time arguement.
For unitary symmetries, an action on $t$ can generically have the form $t \to t+s$.
Due to the periodicity in $t$ and the order-two nature of the symmetry, $s$ can either be $0$ or $T/2$.

For antiunitary symmetries, we have $t \to -t+s$. When the system does not support
time-reversal or chiral symmetry,  as in classes A, C, and D, the constraints due
to time-periodicity and the order-two nature do not restrict the value $s$ takes. 
Hence, $s$ is an arbitrary real number in this situation. 

However, when the system has at least one of the time-reversal and chiral symmetries, 
denoted as $\hat{\mathcal{P}}$, $s$ will be restricted to take 
a few values as shown in the following. 
The composite operation $\hat{\mathcal{P}}\hat{\mathcal{O}}$ 
shift the time as $t \to -s + t$. On the other hand, since $\hat{\mathcal{P}}\hat{\mathcal{O}}$ 
is another order-two symmetry, $s$ can be either $0$ or $T/2$ (note that $s$ is defined modulo
$T$).

To summarize, for a Hamiltonian $H(\boldsymbol{k},\boldsymbol{r},t)$
living in dimension $(d,d_{\parallel},D,D_{\parallel})$,
under the action of $\hat{\mathcal{O}}$,
it transforms as
\begin{gather}
\hat{\mathcal{U}}_{s}H(\boldsymbol{k},\boldsymbol{r},t)\hat{\mathcal{U}}_{s}^{-1}=H(-\boldsymbol{k}_{\parallel},\boldsymbol{k}_{\perp},-\boldsymbol{r}_{\parallel},\boldsymbol{r}_{\perp},t+s)
\label{eq:unitary_symmetry}
\\
\hat{\mathcal{A}_{s}}H(\boldsymbol{k},\boldsymbol{r},t)\hat{\mathcal{A}}_{s}^{-1}=H(\boldsymbol{k}_{\parallel},-\boldsymbol{k}_{\perp},-\boldsymbol{r}_{\parallel},\boldsymbol{r}_{\perp},-t+s)
\end{gather}
in the diagonal basis of $\hat{\mathcal{O}}$.
for unitary and antiunitary symmetries.

Let us suppose $\hat{\mathcal{O}}^{2}=\epsilon_{O}=\pm1$,
and $\hat{\mathcal{O}}$ commutes or anticommutes with coexisting
nonspatial symmetries according to
\begin{equation}
\hat{\mathcal{O}}\hat{T}=\eta_{T}\hat{T}\hat{\mathcal{O}},\quad\hat{\mathcal{O}}\hat{\mathcal{C}}=\eta_{C}\hat{\mathcal{C}}\hat{\mathcal{O}},\quad\hat{\mathcal{O}}\hat{\mathcal{S}}=\eta_{S}\hat{\mathcal{S}}\hat{\mathcal{O}},
\end{equation}
where $\eta_{T}=\pm1$, $\eta_{C}=\pm1$, and $\eta_{S}=\pm1$. Note
that when $\mathcal{\hat{O}}=\hat{\mathcal{U}}$, we can always set $\epsilon_{O}=1$
with the help of multiplying $\hat{\mathcal{O}}$ by imaginary unit $i$, but this
changes the (anti)commutation relation with $\hat{\mathcal{T}}$ and/or
$\hat{\mathcal{C}}$ at the same time.

One can also consider an order-two antisymmetry $\overline{\mathcal{O}}$
defined by
\begin{gather}
\overline{\mathcal{U}}_{s}H(\boldsymbol{k},\boldsymbol{r},t)\overline{\mathcal{U}}_{s}^{-1}=-H(-\boldsymbol{k}_{\parallel},\boldsymbol{k}_{\perp},-\boldsymbol{r}_{\parallel},\boldsymbol{r}_{\perp},-t+s)\nonumber \\
\overline{\mathcal{A}}_{s}H(\boldsymbol{k},\boldsymbol{r},t)\overline{\mathcal{A}}_{s}^{-1}=-H(\boldsymbol{k}_{\parallel},-\boldsymbol{k}_{\perp},-\boldsymbol{r}_{\parallel},\boldsymbol{r}_{\perp},t+s),
\end{gather}
where $\overline{\mathcal{O}}$ can be either unitary $\overline{\mathcal{U}}$
or antiunitary $\overline{\mathcal{A}}$. Such an antisymmetry can
be realized by combining any of order-two symmetries with chiral or
particle-hole symmetry. Similar to $\hat{\mathcal{O}}$, we define
$\overline{\mathcal{O}}^{2}=\epsilon_{\overline{O}}$, 
$\overline{\mathcal{O}}\hat{\mathcal{T}}=\overline{\eta}_{T}\hat{\mathcal{T}}\overline{\mathcal{O}}$,
$\overline{\mathcal{O}}\hat{\mathcal{C}}=\overline{\eta}_{C}\hat{\mathcal{C}}\overline{\mathcal{O}}$,
and $\overline{\mathcal{O}}\hat{\mathcal{S}}=\overline{\eta}_{S}\hat{\mathcal{S}}\overline{\mathcal{O}}$. 
The values that the time shift $s$ takes are similar to the ones in the case of symmetries. 
We have $s=0,T/2$ for $\overline{\mathcal{U}}_s$. For $\overline{\mathcal{A}}_s$, $s$ is arbitrary
in classes A, C and D, whereas $s=0,T/2$ the rest of classes.

The actions of symmetry/antisymmetry operators $\hat{\mathcal{O}}$ and $\overline{\mathcal{O}}$, either unitary or antiunitary,
on the unitary loops can be summarized as follows
\begin{gather}
\hat{\mathcal{U}}_{s}U_{\tau}(\boldsymbol{k},\boldsymbol{r},t)\hat{\mathcal{U}}_{s}^{-1}
=U_{\tau+s}(-\boldsymbol{k}_{\parallel},\boldsymbol{k}_{\perp},-\boldsymbol{r}_{\parallel},\boldsymbol{r}_{\perp},t), \\
\hat{\mathcal{A}}_{s}U_{\tau}(\boldsymbol{k},\boldsymbol{r},t)\hat{\mathcal{A}}_{s}^{-1}
=U_{s-\tau}^{\dagger}(\boldsymbol{k}_{\parallel},-\boldsymbol{k}_{\perp},-\boldsymbol{r}_{\parallel},\boldsymbol{r}_{\perp},t),
\\
\overline{\mathcal{U}}_{s}U_{\tau}(\boldsymbol{k},\boldsymbol{r},t)\overline{\mathcal{U}}_{s}^{-1}=U_{s-\tau}^{\dagger}(-\boldsymbol{k}_{\parallel},\boldsymbol{k}_{\perp},-\boldsymbol{r}_{\parallel},\boldsymbol{r}_{\perp},t), \\
\overline{\mathcal{A}}_{s}U_{\tau}(\boldsymbol{k},\boldsymbol{r},t)\overline{\mathcal{A}}_{s}^{-1}=U_{s+\tau}(\boldsymbol{k}_{\parallel},-\boldsymbol{k}_{\perp},-\boldsymbol{r}_{\parallel},\boldsymbol{r}_{\perp},t).
\end{gather}

In the following, we will discuss each symmetry/antisymmetry operator separately, and choose a particular value of $\tau$ for each
case, since we know the classification would not depend on what the value $\tau$ takes. 

For $\hat{\mathcal{U}}_s$ and $\overline{\mathcal{A}}_s$ $s=0, T/2$, and we take $\tau={T/2}$.
By using
\begin{equation}
 U_{\tau+T/2}(\boldsymbol{k},\boldsymbol{r},t) = U_{\tau}^{\dagger}(\boldsymbol{k},\boldsymbol{r},T-t),
\end{equation}
and omitting the subscript $\tau$ from $U_{\tau}(\boldsymbol{k},\boldsymbol{r},t)$ from now on for simplicity,
we get
\begin{gather}
\hat{\mathcal{U}}_{0}U(\boldsymbol{k},\boldsymbol{r},t)\hat{\mathcal{U}}_{0}^{-1}=U(-\boldsymbol{k}_{\parallel},\boldsymbol{k}_{\perp},-\boldsymbol{r}_{\parallel},\boldsymbol{r}_{\perp},t)
\nonumber \\
\hat{\mathcal{U}}_{T/2}U(\boldsymbol{k},\boldsymbol{r},t)\hat{\mathcal{U}}_{T/2}^{-1}=U^{\dagger}(-\boldsymbol{k}_{\parallel},\boldsymbol{k}_{\perp},-\boldsymbol{r}_{\parallel},\boldsymbol{r}_{\perp},T-t)
\nonumber
\\
\overline{\mathcal{A}}_{0}U(\boldsymbol{k},\boldsymbol{r},t)\overline{\mathcal{A}}_{0}^{-1}=U(\boldsymbol{k}_{\parallel},-\boldsymbol{k}_{\perp},-\boldsymbol{r}_{\parallel},\boldsymbol{r}_{\perp},t)
\nonumber \\
\overline{\mathcal{A}}_{T/2}U(\boldsymbol{k},\boldsymbol{r},t)\overline{\mathcal{A}}_{T/2}^{-1}=U^{\dagger}(\boldsymbol{k}_{\parallel},-\boldsymbol{k}_{\perp},-\boldsymbol{r}_{\parallel},\boldsymbol{r}_{\perp},T-t).
\end{gather}

When considering $\overline{\mathcal{U}}_s$ and $\hat{\mathcal{A}}_s$ in classes A, C and D, 
we can choose $\tau = s/2$,  which gives
\begin{gather}
\hat{\mathcal{A}}_{s}U(\boldsymbol{k},\boldsymbol{r},t)\hat{\mathcal{A}}_{s}^{-1}
=U^{\dagger}(\boldsymbol{k}_{\parallel},-\boldsymbol{k}_{\perp},-\boldsymbol{r}_{\parallel},\boldsymbol{r}_{\perp},t)
\nonumber
\\
\overline{\mathcal{U}}_{s}U(\boldsymbol{k},\boldsymbol{r},t)\overline{\mathcal{U}}_{s}^{-1}=U^{\dagger}(-\boldsymbol{k}_{\parallel},\boldsymbol{k}_{\perp},-\boldsymbol{r}_{\parallel},\boldsymbol{r}_{\perp},t)
.
\end{gather}
This implies that the value $s$ here actually does not play a role in determining topological classification. 

In the remaining classes, we have $s=0, T/2$, and we will choose $\tau = T/2$. This leads to
\begin{gather}
\hat{\mathcal{A}}_{0}U(\boldsymbol{k},\boldsymbol{r},t)\hat{\mathcal{A}}_{0}^{-1}=U^{\dagger}(\boldsymbol{k}_{\parallel},-\boldsymbol{k}_{\perp},-\boldsymbol{r}_{\parallel},\boldsymbol{r}_{\perp},t)\nonumber
\\
\hat{\mathcal{A}}_{T/2}U(\boldsymbol{k},\boldsymbol{r},t)\hat{\mathcal{A}}_{T/2}^{-1}=U(\boldsymbol{k}_{\parallel},-\boldsymbol{k}_{\perp},-\boldsymbol{r}_{\parallel},\boldsymbol{r}_{\perp},T-t)
\nonumber
\\
\overline{\mathcal{U}}_{0}U(\boldsymbol{k},\boldsymbol{r},t)\overline{\mathcal{U}}_{0}^{-1}=U^{\dagger}(-\boldsymbol{k}_{\parallel},\boldsymbol{k}_{\perp},-\boldsymbol{r}_{\parallel},\boldsymbol{r}_{\perp},t)
\nonumber \\
\overline{\mathcal{U}}_{T/2}U(\boldsymbol{k},\boldsymbol{r},t)\overline{\mathcal{U}}_{T/2}^{-1}=U(-\boldsymbol{k}_{\parallel},\boldsymbol{k}_{\perp},-\boldsymbol{r}_{\parallel},\boldsymbol{r}_{\perp},T-t).
\end{gather}

\section{Hermitian map \label{sec:hermitian_map}}
One observation that can be made from 
Eqs.~(\ref{eq:time-reversal-U}--\ref{eq:chiral-U})
is that at fixed $\boldsymbol{r}$ and $t$,
the transformation properties for the unitary loops $U(\boldsymbol{k},\boldsymbol{r},t)$ 
under the actions of $\hat{\mathcal{T}}$, $\hat{\mathcal{C}}$, and $\hat{\mathcal{S}}$
are exactly the same as the ones for unitary boundary 
reflection matrices introduced in, for example, Refs.~\cite{Fulga2012, Peng2017}.
In these works, an effective hermitian matrix can be constructed
from a given reflection matrix, which maps the classification of reflection matrices
into the classification of hermitian matrices.

Here, we can borrow the same hermitian mapping defined as
\begin{equation}
\mathcal{H}(\boldsymbol{k},\boldsymbol{r},t)=\mathcal{U}_{S}U(\boldsymbol{k},\boldsymbol{r},t)\label{eq:effective_hamiltonian_chiral}
\end{equation}
if $U(\boldsymbol{k},\boldsymbol{r},t)$ has a chiral symmetry, and
\begin{equation}
\mathcal{H}(\boldsymbol{k},\boldsymbol{r},t)=\left(\begin{array}{cc}
0 & U(\boldsymbol{k},\boldsymbol{r},t)\\
U^{\dagger}(\boldsymbol{k},\boldsymbol{r},t) & 0
\end{array}\right)\label{eq:effective_hamiltonian_nonchiral}
\end{equation}
if $U(\boldsymbol{k},\boldsymbol{r},t)$ does not have a chiral symmetry.
In the latter case, $\mathcal{H}(\boldsymbol{k},\boldsymbol{r},t)$ aquires
a new chiral symmetry
\begin{equation}
\mathcal{U}_{S}'\mathcal{H}(\boldsymbol{k},\boldsymbol{r},t)=-\mathcal{H}(\boldsymbol{k},\boldsymbol{r},t)\mathcal{U}_{S}',
\end{equation}
with $\mathcal{U}_{S}=\rho_{z}\otimes\mathbb{I}$, where we have introduced
a set of Pauli matrices $\rho_{x,y,z}$ in the enlarged space.

Note that when the unitary loop $\mathcal{U}(\boldsymbol{k},\boldsymbol{r},t)$ does not have a chiral symmetry, 
our hermitian map is the same as the one used in Refs.~\cite{Roy2017,Morimoto2017}.
When the unitary loop does have a chiral symmetry, however, we chose a new map which maps the 
unitary loop into a hermitian matrix without unitary symmetry. 

The advantage of the hermitian map defined here over the one in the previous works
will become clear soon. Note that the hermitian matrix $\mathcal{H}(\boldsymbol{k},\boldsymbol{r},t)$
can be regarded as a static spatially modulated Hamiltonian in
$(d,D+1)$ dimension, because the time arguement transforms like a spatial coordinate similar to $\boldsymbol{r}$. 
The classification of unitary loops in $(d,D)$ dimension in a given symmetry class,
is then the same as the classification of static Hamiltonians in $(d,D+1)$
dimension in the symmetry class shifted upward by one ($s\to s-1$) (mod 2 or 8 depending for complex or real symmetry classes),
where $s$ is used to order the symmetry classes according to Table~\ref{tab:AZ-symmetry-classes}.
Thus, one can directly apply the classification scheme of the static Hamiltonians $H(\boldsymbol{k},\boldsymbol{r})$
using $K$ theory, as was done in Ref.~\cite{Shiozaki2014}.
This is provided by a homotopy classification of maps from the base space $(\boldsymbol{k},\boldsymbol{r})\in S^{d+D}$
to the classifying space of Hamiltonians $H(\boldsymbol{k},\boldsymbol{r})$ subject to the given symmetries,
which we denoted as $\mathcal{C}_{s}$ or $\mathcal{R}_{s}$ as shown in the table.

\begin{table}
\caption{\label{tab:AZ-symmetry-classes}AZ symmetry classes and their classifying
spaces. The top two rows ($s=0,1\mod2$) are complex AZ classes, while
the rest eight rows $(s=0,\dots,7\mod8)$ are real AZ classes. The
third to fifth columns denote the absence (0) or presence ($\epsilon_{T},\epsilon_{C}=$$\pm1$
or $\eta_{S}=1$) of time-reversal ($\hat{\mathcal{T}}$), particle-hole
($\hat{\mathcal{C}}$) and chiral symmetries ($\hat{\mathcal{S}}$).
$\mathcal{C}_{s}$ ($\mathcal{R}_{s}$) denotes the classifying space
of $s$ complex (real) AZ class.}
\begin{ruledtabular}
\centering
\begin{tabular}{ccccccc}
$s$ & AZ class & $\hat{\mathcal{T}}$ & $\hat{\mathcal{C}}$ & $\hat{S}$ & $\mathcal{C}_{s}$ or $\mathcal{R}_{s}$ & $\pi_{0}(\mathcal{C}_{s})$ or $\pi_{0}(\mathcal{R}_{s})$\\
\hline 
$0$ & A & 0 & 0 & 0 & $\mathcal{C}_{0}$ & $\mathbb{Z}$\\
$1$ & AIII & 0 & 0 & 1 & $\mathcal{C}_{1}$ & 0\\
\hline 
$0$ & AI & $+1$ & 0 & 0 & $\mathcal{R}_{0}$ & $\mathbb{Z}$\\
$1$ & BDI & $+1$ & +1 & 1 & $\mathcal{R}_{1}$ & $\mathbb{\mathbb{Z}}_{2}$\\
$2$ & D & 0 & $+1$ & 0 & $\mathcal{R}_{2}$ & $\mathbb{\mathbb{Z}}_{2}$\\
$3$ & DIII & $-1$ & +1 & 1 & $\mathcal{R}_{3}$ & 0\\
$4$ & AII & $-1$ & 0 & 0 & $\mathcal{R}_{4}$ & $2\mathbb{Z}$\\
$5$ & CII & $-1$ & $-1$ & 1 & $\mathcal{R}_{5}$ & 0\\
$6$ & C & 0 & $-1$ & 0 & $\mathcal{R}_{6}$ & 0\\
$7$ & CI & +1 & $-1$ & 1 & $\mathcal{R}_{7}$ & 0\\
\end{tabular}
\end{ruledtabular}
\end{table}

Because of the Bott periodicity in the periodic table of
static TI/SCs \cite{Schnyder2008,Kitaev2009,Ryu2010,Teo2010,Chiu2016},
the classification is unchanged when simultaneously shifting the dimension $D \to D+1$ and the symmetry class upward by
one $s\to s-1$ (mod 2 or 8 for complex or real symmetry classes).
It turns out that the classification of unitary loops is the same as the classification of
the static Hamiltonian in the same symmetry class and with the same dimension $(d,D)$.
In the following, we will explicitly derive the action of the hermitian map
on each symmetry classes.

\subsection{Classes A and AIII}
We first consider the two complex classes. 
Under the hermitian map defined above, classifying unitary
loops in $(d,D)$ dimension in class A is the same as classifying hermitian
matrices in $(d,D+1)$ dimension in class
AIII. On the other hand, classifying unitary loops in
$(d,D)$ dimension in class AIII is the same as classifying hermitian
matrices in $(d,D+1)$ dimension in class
A.

\subsection{Classes AI and AII}
Now we turn to real symmetry classes.
Since classes AI and AII have only time-reversal symmetry, we need
to apply the hermitian map defined in Eq.(\ref{eq:effective_hamiltonian_nonchiral}).
By using Eq.~(\ref{eq:time-reversal-U}) with $\tau=T/2$, or
\begin{equation}
\mathcal{U}_{T}U^{T}(\boldsymbol{k},\boldsymbol{r},t)=U(-\boldsymbol{k},\boldsymbol{r},t)\mathcal{U}_{T},
\end{equation}
we have effective time-reversal symmetry
\begin{equation}
\mathcal{U}_{T}'\mathcal{H}^{*}(\boldsymbol{k},\boldsymbol{r},t)=\mathcal{H}(-\boldsymbol{k},\boldsymbol{r},t)\mathcal{U}_{T}'
\end{equation}
with $\mathcal{U}_{T}'=\rho_{x}\otimes\mathcal{U}_{T}$, and effective
particle-hole symmetry
\begin{equation}
\mathcal{U}_{C}'\mathcal{H}^{*}(\boldsymbol{k},\boldsymbol{r},t)=-\mathcal{H}^{*}(-\boldsymbol{k},\boldsymbol{r},t)\mathcal{U}_{C}'
\end{equation}
with $\mathcal{U}_{C}'=i\rho_{y}\otimes\mathcal{U}_{T}$.

Note that the effective time-reversal and particle-hole symmetries combines
into the chiral symmetry as expected. The types of the effective time-reversal
and particle-hole symmetries of $\mathcal{H}(\boldsymbol{k},t)$ are
determined from
\begin{gather}
\mathcal{U}_{T}'\mathcal{U}_{T}'^{*}=\rho_{0}\otimes(\mathcal{U}_{T}\mathcal{U}_{T}^{*})\\
\mathcal{U}_{C}'\mathcal{U}_{C}'^{*}=-\rho_{0}\otimes(\mathcal{U}_{T}\mathcal{U}_{T}^{*}),
\end{gather}
where $\rho_{0}$ is the two-by-two identity matrix in the extended
space. Under the hermitian map, classifying unitary loops in
($d,D)$ dimension in classes AI and AII, are the same as classifying
hermitian matrices in $(d,D+1)$ dimension
in classes CI and DIII.

\subsection{Classes C and D}
Let us consider classes C and D with only particle-hole symmetry. 
We need to apply the hermitian map defined in Eq.(\ref{eq:effective_hamiltonian_nonchiral}).
By using Eq.(\ref{eq:particle-hole-U}), one can define effective
time-reversal symmetry with $\mathcal{U}_{T}'=\rho_{0}\otimes\mathcal{U}_{C}$,
and particle-hole symmetry with $\mathcal{U}_{C}'=\rho_{z}\otimes\mathcal{U}_{C}$,
such that
\begin{gather}
\mathcal{U}_{T}'\mathcal{H}^{*}(\boldsymbol{k},\boldsymbol{r},t)=\mathcal{H}(-\boldsymbol{k},\boldsymbol{r},t)\mathcal{U}_{T}'\\
\mathcal{U}_{C}'\mathcal{H^{*}}(\boldsymbol{k},\boldsymbol{r},t)=-\mathcal{H}(-\boldsymbol{k},\boldsymbol{r},t)\mathcal{U}_{C}'.
\end{gather}

Note that $\mathcal{U}_{T}'$ and $\mathcal{U}_{C}'$ combines into
the chiral symmetry as expected. The types of these effective symmetries
are determined by
\begin{gather}
\mathcal{U}_{T}'\mathcal{U}_{T}'^{*}=\rho_{0}\otimes(\mathcal{U}_{C}\mathcal{U}_{C}^{*})\\
\mathcal{U}_{C}'\mathcal{U}_{C}'^{*}=\rho_{0}\otimes(\mathcal{U}_{C}\mathcal{U}_{C}^{*}).
\end{gather}
Under the hermitian map, classifying unitary loops in $(d,D)$
dimension in classes C and D, are the same as classifying hermitian
matrices in $(d,D+1)$ dimension in
classes CII and BDI.

\subsection{classes CI, CII, DIII, and BDI}
Here we consider symmetry classes where time-reversal, particle-hole,
and chiral symmetries are all present In this case, $\mathcal{U}_{S}=\mathcal{U}_{T}\mathcal{U}_{C}^{*}$.
By $\mathcal{U}_{S}^{2}=1,$ we have $\mathcal{U}_{T}^{*}\mathcal{U}_{C}\mathcal{U}_{S}^{*}=1$.
This can be used to show that
\begin{equation}
\mathcal{U}_{S}\mathcal{U}_{C}=\mathcal{U}_{C}\mathcal{U}_{S}^{*}(\mathcal{U}_{C}\mathcal{U}_{C}^{*})(\mathcal{U}_{T}\mathcal{U}_{T}^{*}).
\end{equation}
Notice that $\mathcal{U}_{C}\mathcal{U}_{C}^{*}=\pm1$ and $\mathcal{U}_{T}\mathcal{U}_{T}^{*}=\pm1$
are just numbers.

The effective Hamiltonian $\mathcal{H}(\boldsymbol{k},t)$ defined
in Eq.(\ref{eq:effective_hamiltonian_chiral}) has the property
\begin{equation}
\mathcal{H}(\boldsymbol{k},\boldsymbol{r,}t)\mathcal{U}_{C}=(\mathcal{U}_{C}\mathcal{U}_{C}^{*})(\mathcal{U}_{T}\mathcal{U}_{T}^{*})\mathcal{U}_{C}\mathcal{H}(-\boldsymbol{k},\boldsymbol{r},t)^{*}.
\end{equation}
This gives rise to time-reversal or particle-hole symmetry depending
on $(\mathcal{U}_{C}\mathcal{U}_{C}^{*})(\mathcal{U}_{T}\mathcal{U}_{T}^{*})=1$
or $-1$, respectively. Therefore, under the hermitian map,
the unitary loops in $(d,D)$ dimension in classes CI, CII, DIII,
and BDI, map to hermitian matrices
in $(d,D+1)$ dimension in classes C, AII, D, and AI, respectively. 

\section{Classfication with additional order-two space-time symmetry \label{sec:classification_order_two}}
After introducing the hermitian map which reduces the classification of 
unitary loops to the classification of static hermitian matrices, or Hamiltonians, 
in the AZ symmetry classes, let us now assume the system
supports an additional order-two space-time symmetry/antisymmetry, which
is either unitary or antiunitary, as defined in Sec.~\ref{sec:order-two}.
In the following, we will focus on each class separately. 

\subsection{Complex symmetry classes}
The complex classes A and AIII are characterized by the absence of
time-reversal and particle-hole symmetries.

\subsubsection{Class A}
Let us start with Class A, with additional symmetry 
realized by $\hat{\mathcal{O}}$ or $\overline{\mathcal{O}}$, 
whose properties are summarized as $({\rm A},\hat{\mathcal{O}}^{\epsilon_{O}})$ or $({\rm A,\overline{\mathcal{O}}^{\epsilon_{\overline{O}}}})$.
For unitary symmetry realized by $\hat{\mathcal{U}}$
and $\overline{\mathcal{U}}$, one can fix $\epsilon_{U}=1$ or $\epsilon_{\overline{U}}=1$. 

\paragraph{$\hat{\mathcal{O}}=\hat{\mathcal{U}}_{0}$}
We have
\begin{equation}
\hat{\mathcal{U}}_{0}'\mathcal{H}(\boldsymbol{k},\boldsymbol{r},t)\hat{\mathcal{U}}_{0}'^{-1}=\mathcal{H}(-\boldsymbol{k}_{\parallel},\boldsymbol{k}_{\perp},-\boldsymbol{r}_{\parallel},\boldsymbol{r}_{\perp},t),
\end{equation}
where $\hat{\mathcal{U}}_{0}'=$ $\rho_{0}\otimes\hat{\mathcal{U}}_{0}$
behaves as an order-two crystalline symmetry if one regards $t\in S^{1}$
as an additional defect surrounding parameter. Recall that $\mathcal{H}(\boldsymbol{k},\boldsymbol{r},t)$
has chiral symmetry realized by operator $\hat{\mathcal{S}}'=\mathcal{U}_{S}'=\rho_{z}\otimes\mathbb{I}$,
we have
\begin{equation}
[\hat{\mathcal{U}}_{0}',\hat{\mathcal{S}}']=0.
\end{equation}
This means under the hermitian map, unitary loops with symmetry $({\rm A},\hat{\mathcal{U}}_{0}^{+})$
in dimension $(d,d_{\parallel},D,D_{\parallel})$ are mapped to static
Hamiltonians with symmetry $({\rm AIII},\hat{\mathcal{U}}_{+}^{+})$
in dimension $(d,d_{\parallel},D+1,D_{\parallel}).$ Here,  we use
the notation $({\rm AIII},\hat{\mathcal{O}}_{\eta_{\Gamma}}^{\epsilon_{O}})$
to denote class AIII with an additional symmetry realized by $\hat{\mathcal{O}}$,
which squares to $\epsilon_{O}$ and commutes ($\eta_{S}=1$) or anticommutes
$(\eta_{S}=-1)$ with the chiral symmetry operator $\hat{\mathcal{S}}'$.
One can also replace $\hat{\mathcal{O}}$ by $\overline{\mathcal{O}}$
to define class AIII with an additional antisymmetry in the similar way.

\paragraph{$\hat{\mathcal{O}}=\hat{\mathcal{U}}_{T/2}$}
We have
\begin{equation}
\hat{\mathcal{U}}_{T/2}'\mathcal{H}(\boldsymbol{k},\boldsymbol{r},t)\hat{\mathcal{U}}_{T/2}'^{-1}=\mathcal{H}(-\boldsymbol{k}_{\parallel},\boldsymbol{k}_{\perp},-\boldsymbol{r}_{\parallel},\boldsymbol{r}_{\perp},T-t)
\end{equation}
where $\hat{\mathcal{U}}_{T/2}'=\rho_{x}\otimes\hat{\mathcal{U}}_{T/2}$,
which satisfies $\{\hat{\mathcal{U}}_{T/2}',\hat{\mathcal{S}}'\}=0$
and $\hat{\mathcal{U}}_{T/2}'^{2}=1$. Since $t\in S^{1}$, if we
shift the origin by defining $t=\frac{T}{2}+t'$, and use $t'\in S^{1}$
instead of $t$, then the map $t\to T-t$ becomes $t'\to-t'$. Now
$t'$ can be regards as an additional defect surrounding coordinate
which flips under the order-two symmetry. Under the hermitian
map, unitary loops with symmetry $({\rm A},\hat{\mathcal{U}}_{T/2}^{+})$
in dimension $(d,d_{\parallel},D,D_{\parallel})$ are mapped to static
Hamiltonians with symmetry $({\rm AIII},\hat{\mathcal{U}}_{-}^{+})$
in dimension $(d,d_{\parallel},D+1,D_{\parallel}+1).$ 

\paragraph{$\overline{\mathcal{O}}=\overline{\mathcal{U}}_{s}$}
The unitary antisymmetry $\overline{\mathcal{U}}_{s}$
leads to an order-two symmetry on $\mathcal{H}(\boldsymbol{k},\boldsymbol{r},t)$
with
\begin{equation}
\overline{\mathcal{U}}_{s}'\mathcal{H}(\boldsymbol{k},\boldsymbol{r},t)\overline{\mathcal{U}}_{s}'^{-1}=\mathcal{H}(-\boldsymbol{k}_{\parallel},\boldsymbol{k}_{\perp},-\boldsymbol{r}_{\parallel},\boldsymbol{r}_{\perp},t),
\end{equation}
where $\overline{\mathcal{U}}_{s}'=\rho_{x}\otimes\overline{\mathcal{U}}_{s}$.
Moreover, we have $\overline{\mathcal{U}}_{s}'^{2}=1$ and $\{\overline{\mathcal{U}}_{s}',\hat{\mathcal{S}'}\}=0$.
Under the hermitian map, unitary loops with symmetry $({\rm A},\overline{\mathcal{U}}_{s}^{+})$
in dimension $(d,d_{\parallel},D,D_{\parallel})$ are mapped to static
Hamiltonians with symmetry $({\rm AIII},\hat{\mathcal{U}}_{-}^{+})$
in dimension $(d,d_{\parallel},D+1,D_{\parallel}).$ 

\paragraph{$\hat{\mathcal{O}}=\hat{\mathcal{A}}_{s}$}
We have
\begin{equation}
\hat{\mathcal{A}}_{s}'\mathcal{H}(\boldsymbol{k},\boldsymbol{r},t)\hat{\mathcal{A}}_{s}'^{-1}=\mathcal{H}(\boldsymbol{k}_{\parallel},-\boldsymbol{k}_{\perp},-\boldsymbol{r}_{\parallel},\boldsymbol{r}_{\perp},t)
\end{equation}
with $\hat{\mathcal{A}}_{s}'=\rho_{x}\otimes\hat{\mathcal{A}}_{s}$.
Moreover, we have $\{\hat{\mathcal{A}}_{s}',\hat{\mathcal{S}'}\}=0$
and $\hat{\mathcal{A}}_{s}'^{2}=\hat{\mathcal{A}}_{s}^{2}$. Thus,
under the hermitian map, unitary loops with symmetry $({\rm A},\hat{\mathcal{A}}_{s}^{\pm})$
in dimension $(d,d_{\parallel},D,D_{\parallel})$ are mapped to static
Hamiltonians with symmetry $({\rm AIII},\hat{\mathcal{A}}_{-}^{\pm})$
in dimension $(d,d_{\parallel},D+1,D_{\parallel})$. 

\paragraph{$\overline{\mathcal{O}}=\overline{\mathcal{A}}_{0}$}
We have
\begin{equation}
\overline{\mathcal{A}}_{0}'\mathcal{H}(\boldsymbol{k},\boldsymbol{r},t)\overline{\mathcal{A}}_{0}'^{-1}=\mathcal{H}(\boldsymbol{k}_{\parallel},-\boldsymbol{k}_{\perp},-\boldsymbol{r}_{\parallel},\boldsymbol{r}_{\perp},t),
\end{equation}
with $\overline{\mathcal{A}}_{0}'=\rho_{0}\otimes\overline{\mathcal{A}}_{0}$,
which satisfies $\overline{\mathcal{A}}_{0}'^{2}=\overline{\mathcal{A}}_{0}^{2}$
and $\text{[\ensuremath{\overline{\mathcal{A}}_{0}',\hat{\mathcal{S}}'}]=0}$.
Under the hermitian map, unitary loops with symmetry $({\rm A},\overline{\mathcal{A}}_{0}^{\pm})$
in dimension $(d,d_{\parallel},D,D_{\parallel})$ are mapped to static
Hamiltonians with symmetry $({\rm AIII},\hat{\mathcal{A}}_{+}^{\pm})$
in dimension $(d,d_{\parallel},D+1,D_{\parallel})$.

\paragraph{$\overline{\mathcal{O}}=\overline{\mathcal{A}}_{T/2}$}
We have
\begin{equation}
\overline{\mathcal{A}}_{T/2}'\mathcal{H}(\boldsymbol{k},\boldsymbol{r},t)\overline{\mathcal{A}}_{T/2}'^{-1}=\mathcal{H}(\boldsymbol{k}_{\parallel},-\boldsymbol{k}_{\perp},-\boldsymbol{r}_{\parallel},\boldsymbol{r}_{\perp},T-t)
\end{equation}
with $\overline{\mathcal{A}}_{T/2}'=\rho_{x}\otimes\overline{\mathcal{A}}_{T/2}$,
which satisfies $\overline{\mathcal{A}}_{T/2}'^{2}=\overline{\mathcal{A}}_{T/2}^{2}$,
$\{\overline{\mathcal{A}}_{T/2}',\hat{\mathcal{S}}'\}=0$. Under
the hermitian map, unitary loops with symmetry $({\rm A},\overline{\mathcal{A}}_{T/2}^{\pm})$
in dimension $(d,d_{\parallel},D,D_{\parallel})$ are mapped to static
Hamiltonians with symmetry $({\rm AIII},\hat{\mathcal{A}}_{-}^{\pm})$
in dimension $(d,d_{\parallel},D+1,D_{\parallel}+1)$.

\subsubsection{Class AIII}
In class AIII, we have a chiral symmetry realized by $\hat{\mathcal{S}}$.
We assume an additional order-two symmetry $\hat{\mathcal{U}}_{\eta_{S}}^{\epsilon_{U}}$
or antisymmetry $\overline{\mathcal{U}}_{\overline{\eta}_{S}}^{\epsilon_{\overline{U}}}$.
Moreover, we can fix $\epsilon_{U}=1$ and $\epsilon_{\overline{U}}=1$
for unitary symmetries and antisymmetries realized by $\hat{\mathcal{U}}$
and $\overline{\mathcal{U}}$ respectively. For unitary (anti)symmetries,
note that $\overline{\mathcal{U}}_{\eta_{S}}$ in class AIII is essentially
the same as $\hat{\mathcal{U}}_{\eta_{S}}$, because they can be converted
to each other by $\overline{\mathcal{U}}_{\eta_{s}}=\hat{\mathcal{S}}\hat{\mathcal{U}}_{\eta_{S}}$.
Similarly, for antiunitary (anti)symmetries, $\hat{\mathcal{A}}_{\eta_{S}}^{\epsilon_{A}}$
and $\overline{\mathcal{A}}_{\eta_{S}}^{\epsilon_{A}\eta_{S}}$ are
equivalent since $\hat{\mathcal{A}}_{\eta_{S}}^{\epsilon_{A}}=\hat{\mathcal{S}}\overline{\mathcal{A}}_{\eta_{S}}^{\epsilon_{A}\eta_{S}}.$
Hence in the following, we only discuss unitary and antiunitary symmetries.

\paragraph{$\hat{\mathcal{O}}=\hat{\mathcal{U}}_{0}$}
We have
\begin{equation}
\hat{\mathcal{U}}_{0}\mathcal{H}(\boldsymbol{k},\boldsymbol{r},t)\hat{\mathcal{U}}_{0}^{-1}=\eta_{S}\mathcal{H}(-\boldsymbol{k}_{\parallel},\boldsymbol{k}_{\perp},-\boldsymbol{r}_{\parallel},\boldsymbol{r}_{\perp},t).
\end{equation}
Under the hermitian map, unitary loops with symmetry $({\rm AIII},\hat{\mathcal{U}}_{0,+}^{+})$
and $({\rm AIII},\hat{\mathcal{U}}_{0,-}^{+})$ in dimension $(d,d_{\parallel},D,D_{\parallel})$
are mapped to static Hamiltonians with symmetry $({\rm A},\hat{\mathcal{U}}^{+})$
and $({\rm A},\overline{\mathcal{U}}^{+})$ in dimension $(d,d_{\parallel},D+1,D_{\parallel})$,
respectively. 

\paragraph{$\hat{\mathcal{O}}=\hat{\mathcal{U}}_{T/2}$}
We have
\begin{equation}
\hat{\mathcal{S}}\hat{\mathcal{U}}_{T/2}\mathcal{H}(\boldsymbol{k},\boldsymbol{r},t)(\hat{\mathcal{S}}\hat{\mathcal{U}}_{T/2})^{-1}=\eta_{S}\mathcal{H}(-\boldsymbol{k}_{\parallel},\boldsymbol{k}_{\perp},-\boldsymbol{r}_{\parallel},\boldsymbol{r}_{\perp},T-t).
\end{equation}
Under the hermitian map, unitary loops with symmetry $({\rm AIII},\hat{\mathcal{U}}_{T/2,+}^{+})$
and $({\rm AIII},\hat{\mathcal{U}}_{T/2,-}^{+})$ in dimension $(d,d_{\parallel},D,D_{\parallel})$
are mapped to static Hamiltonians with symmetry $({\rm A},\hat{\mathcal{U}}^{+})$
and $({\rm A},\overline{\mathcal{U}}^{+})$ in dimension $(d,d_{\parallel},D+1,D_{\parallel}+1)$,
respectively.

\paragraph{$\hat{\mathcal{O}}=\hat{\mathcal{A}}_{0}$}

We have
\begin{equation}
\hat{\mathcal{S}}\hat{\mathcal{A}}_{0}\mathcal{H}(\boldsymbol{k},\boldsymbol{r},t)(\hat{\mathcal{S}}\hat{\mathcal{A}}_{0})^{-1}=\eta_{S}\mathcal{H}(\boldsymbol{k}_{\parallel},-\boldsymbol{k}_{\perp},-\boldsymbol{r}_{\parallel},\boldsymbol{r}_{\perp},t).
\end{equation}
Under the hermitian map, unitary loops with symmetry $({\rm AIII},\hat{\mathcal{A}}_{0,+}^{\pm})$
and $({\rm AIII},\hat{\mathcal{A}}_{0,-}^{\pm})$ in dimension $(d,d_{\parallel},D,D_{\parallel})$
are mapped to static Hamiltonians with symmetry $({\rm A},\hat{\mathcal{A}}^{\pm})$
and $({\rm A},\overline{\mathcal{A}}^{\mp})$ in dimension $(d,d_{\parallel},D+1,D_{\parallel})$,
respectively.

\paragraph{$\hat{\mathcal{O}}=\hat{\mathcal{A}}_{T/2}$}
We have
\begin{equation}
\hat{\mathcal{A}}_{T/2}\mathcal{H}(\boldsymbol{k},\boldsymbol{r},t)\hat{\mathcal{A}}_{T/2}^{-1}=\eta_{S}\mathcal{H}(\boldsymbol{k}_{\parallel},-\boldsymbol{k}_{\perp},-\boldsymbol{r}_{\parallel},\boldsymbol{r}_{\perp},t).
\end{equation}
Under the hermitian map, unitary loops with symmetry $({\rm AIII},\hat{\mathcal{A}}_{T/2,+}^{\pm})$
and $({\rm AIII},\hat{\mathcal{A}}_{T/2,-}^{\pm})$ in dimension $(d,d_{\parallel},D,D_{\parallel})$
are mapped to static Hamiltonians with symmetry $({\rm A},\hat{\mathcal{A}}^{\pm})$
and $({\rm A},\overline{\mathcal{A}}^{\pm})$ in dimension $(d,d_{\parallel},D+1,D_{\parallel}+1)$,
respectively.

\subsection{Real symmetry classes}
Now let us consider real symmetry classes, where at least
one antiunitary symmetry is present.  

In classes AI and AII, only time reversal symmetry is present. We have
the following equivalence relations between the additional order-two
symmetries/antisymmetries
\begin{gather}
\hat{\mathcal{U}}_{\eta_{T}}^{\epsilon_{U}}=i\hat{\mathcal{U}}_{-\eta_{T}}^{-\epsilon_{U}}=\hat{\mathcal{T}}\hat{\mathcal{A}}_{\eta_{T}}^{\eta_{T}\epsilon_{T}\epsilon_{U}}=i\hat{\mathcal{T}}\hat{\mathcal{A}}_{\eta_{T}}^{\eta_{T}\epsilon_{T}\epsilon_{U}}\\
\overline{\mathcal{U}}_{\overline{\eta}_{T}}^{\overline{\epsilon}_{U}}=i\overline{\mathcal{U}}_{-\overline{\eta}_{T}}^{-\overline{\epsilon}_{U}}=\hat{\mathcal{T}}\overline{\mathcal{A}}_{\overline{\eta}_{T}}^{\overline{\eta}_{T}\epsilon_{T}\overline{\epsilon}_{U}}=i\hat{\mathcal{T}}\overline{\mathcal{A}}_{-\overline{\eta}_{T}}^{-\overline{\eta}_{T}\epsilon_{T}\overline{\epsilon}_{U}},
\end{gather}
where $\epsilon_{U}=\hat{\mathcal{U}}^{2}$, $\epsilon_{\overline{U}}=\overline{\mathcal{U}}^{2}$,
$\epsilon_{T}=\hat{\mathcal{T}}^{2}$. We only need to consier four
cases $\mathcal{\hat{U}}_{+}^{+}$, $\hat{\mathcal{U}}_{-}^{+}$,
$\overline{\mathcal{U}}_{+}^{+}$, and $\overline{\mathcal{U}}_{-}^{+}$. 

In classes C and D, the particle-hole symmetry leads to the following
equivalence relations between the additional order-two symmetries/antisymmetries
\begin{gather}
\hat{\mathcal{U}}_{\eta_{C}}^{\epsilon_{U}}=i\hat{\mathcal{U}}_{-\eta_{C}}^{-\epsilon_{U}}=\hat{\mathcal{C}}\overline{\mathcal{A}}_{\eta_{C}}^{\eta_{C}\epsilon_{C}\epsilon_{U}}=i\hat{\mathcal{C}}\overline{\mathcal{A}}_{-\eta_{C}}^{\eta_{C}\epsilon_{C}\epsilon_{U}}\\
\overline{\mathcal{U}}_{\overline{\eta}_{C}}^{\overline{\epsilon}_{U}}=i\overline{\mathcal{U}}_{-\overline{\eta}_{T}}^{-\overline{\epsilon}_{U}}=\hat{\mathcal{C}}\hat{\mathcal{A}}_{\overline{\eta}_{C}}^{\overline{\eta}_{C}\epsilon_{C}\overline{\epsilon}_{U}}=i\hat{\mathcal{C}}\hat{\mathcal{A}}_{-\overline{\eta}_{C}}^{\overline{\eta}_{C}\epsilon_{C}\overline{\epsilon}_{U}},
\end{gather}
where $\epsilon_{C}=\hat{\mathcal{C}}^{2}$. We just need to consider
four cases $\mathcal{\hat{U}}_{+}^{+}$, $\hat{\mathcal{U}}_{-}^{+}$,
$\overline{\mathcal{U}}_{+}^{+}$, and $\overline{\mathcal{U}}_{-}^{+}$.

Finally, in classes BDI, DIII, CII and CI, with time-reversal, particle-hole
and chiral symmetries all together, we have
\begin{align}
& \hat{\mathcal{U}}_{\eta_{T},\eta_{C}}^{\epsilon_{U}}=i\hat{\mathcal{U}}_{-\eta_{T},-\eta_{C}}^{-\epsilon_{U}}=\hat{\mathcal{T}}\hat{\mathcal{A}}_{\eta_{T},\eta_{C}}^{\eta_{T}\epsilon_{T}\epsilon_{U}}=i\hat{\mathcal{T}}\hat{\mathcal{A}}_{-\eta_{T},\eta_{C}}^{\eta_{T}\epsilon_{T}\epsilon_{U}}
\nonumber \\
&=\hat{\mathcal{C}}\overline{\mathcal{A}}_{\eta_{T},\eta_{C}}^{\eta_{C}\epsilon_{C}\epsilon_{U}}=i\hat{\mathcal{C}}\overline{\mathcal{A}}_{-\eta_{T},-\eta_{C}}^{\eta_{C}\epsilon_{C}\epsilon_{U}}
\end{align}
\begin{align}
&\overline{\mathcal{U}}_{\overline{\eta}_{T},\overline{\eta}_{C}}^{\overline{\epsilon}_{U}}=i\overline{\mathcal{U}}_{-\overline{\eta}_{T},-\overline{\eta}_{C}}^{-\overline{\epsilon}_{U}}=\hat{\mathcal{T}}\overline{\mathcal{A}}_{\overline{\eta}_{T},\overline{\eta}_{C}}^{\overline{\eta}_{T}\epsilon_{T}\overline{\epsilon}_{U}}=i\hat{\mathcal{T}}\overline{\mathcal{A}}_{-\overline{\eta}_{T},-\overline{\eta}_{C}}^{\overline{\eta}_{T}\epsilon_{T}\overline{\epsilon}_{U}}\nonumber
\\
&=\hat{\mathcal{C}}\hat{\mathcal{A}}_{\overline{\eta}_{T},\overline{\eta}_{C}}^{\overline{\eta}_{C}\epsilon_{C}\overline{\epsilon}_{U}}=i\hat{\mathcal{C}}\hat{\mathcal{A}}_{-\overline{\eta}_{T},-\overline{\eta}_{C}}^{\overline{\eta}_{C}\epsilon_{C}\overline{\epsilon}_{U}}.
\end{align}
Hence, only four cases $\hat{\mathcal{U}}_{+,+}^{+}$,
$\hat{\mathcal{U}}_{+,-}^{+}$, $\hat{\mathcal{U}}_{-,-}^{+}$ and
$\hat{\mathcal{U}}_{-,+}^{+}$ need to be considered. 

\subsubsection{Classes AI and AII}
\paragraph{$\hat{\mathcal{O}}=\hat{\mathcal{U}}_{0}$ }
The new hermitian matrix $\mathcal{H}(\boldsymbol{k},\boldsymbol{r},t)$
under the hermitian map defined by Eq.(\ref{eq:effective_hamiltonian_nonchiral})
aquires new time-reversal and particle-hole symmetries, realized by
$\hat{\mathcal{T}}'=\rho_{x}\otimes\hat{\mathcal{T}}$ and $\hat{\mathcal{C}'}=i\rho_{y}\otimes\hat{\mathcal{T}}$,
respectively. Due to the order-two symmetry realized by $\hat{\mathcal{U}}_{0}$,
we have
\begin{equation}
\hat{\mathcal{U}}_{0}'\mathcal{H}(\boldsymbol{k},\boldsymbol{r},t)\hat{\mathcal{U}}_{0}'^{-1}=\mathcal{H}(-\boldsymbol{k}_{\parallel},\boldsymbol{k}_{\perp},-\boldsymbol{r}_{\parallel},\boldsymbol{r}_{\perp},t),
\end{equation}
with $\hat{\mathcal{U}}_{0}'=\rho_{0}\otimes\hat{\mathcal{U}}_{0}$.
Moreover, we have 
\begin{gather}
\hat{\mathcal{U}}_{0}'\hat{\mathcal{T}'}=\eta_{T}\hat{\mathcal{T}}'\hat{\mathcal{U}}_{0}'\\
\hat{\mathcal{U}}_{0}'\hat{\mathcal{C}}'=\eta_{T}\hat{\mathcal{C}}'\hat{\mathcal{U}}_{0}'.
\end{gather}
and $\hat{\mathcal{U}}_{0}'^{2}=\epsilon_{U}$. Under the hermitian
map, unitary loops with symmetry $({\rm AI},\text{\ensuremath{\hat{\mathcal{U}}_{0,\eta_{T}}^{\epsilon_{U}}}})$
and $({\rm AII},\hat{\mathcal{U}}_{0,\eta_{T}}^{\epsilon_{U}})$ in
dimension $(d,d_{\parallel},D,D_{\parallel})$ are mapped to static
Hamiltonians with symmetry $({\rm CI},\hat{\mathcal{U}}_{\eta_{T},\eta_{T}}^{\epsilon_{U}})$
and $({\rm DIII},\hat{\mathcal{U}}_{\eta_{T},\eta_{T}}^{\epsilon_{U}})$
in dimension $(d,d_{\parallel},D+1,D_{\parallel})$, respectively.

\paragraph{$\hat{\mathcal{O}}=\hat{\mathcal{U}}_{T/2}$}
Due to the order-two symmetry realized by $\hat{\mathcal{U}}_{T/2}$,
we have
\begin{equation}
\hat{\mathcal{U}}_{T/2}'\mathcal{H}(\boldsymbol{k},\boldsymbol{r},t)\hat{\mathcal{U}}_{T/2}'^{-1}=\mathcal{H}(-\boldsymbol{k}_{\parallel},\boldsymbol{k}_{\perp},-\boldsymbol{r}_{\parallel},\boldsymbol{r}_{\perp},T-t),
\end{equation}
with $\hat{\mathcal{U}}_{T/2}'=\rho_{x}\otimes\mathcal{\hat{\mathcal{U}}}_{T/2}$,
which satisfies
\begin{gather}
\hat{\mathcal{U}}_{T/2}'\hat{\mathcal{T}'}=\eta_{T}\hat{\mathcal{T}}'\hat{\mathcal{U}}_{T/2}'\\
\hat{\mathcal{U}}_{T/2}'\hat{\mathcal{C}}'=-\eta_{T}\hat{\mathcal{C}}'\hat{\mathcal{U}}_{T/2}'.
\end{gather}
and $\hat{\mathcal{U}}_{T/2}'^{2}=\epsilon_{U}$. Under the
hermitian map, unitary loops with symmetry $({\rm AI},\text{\ensuremath{\hat{\mathcal{U}}_{T/2,\eta_{T}}^{\epsilon_{U}}}})$
and $({\rm AII},\hat{\mathcal{U}}_{T/2,\eta_{T}}^{\epsilon_{U}})$
in dimension $(d,d_{\parallel},D,D_{\parallel})$ are mapped to static
Hamiltonians with symmetry $({\rm CI},\hat{\mathcal{U}}_{\eta_{T},-\eta_{T}}^{\epsilon_{U}})$
and $({\rm DIII},\hat{\mathcal{U}}_{\eta_{T},-\eta_{T}}^{\epsilon_{U}})$
in dimension $(d,d_{\parallel},D+1,D_{\parallel}+1)$, respectively.

\paragraph{$\overline{\mathcal{O}}=\text{\ensuremath{\overline{\mathcal{U}}_{0}}}$}
Due to the order-two antisymmetry realized by $\overline{\mathcal{U}}_{0}$,
we have
\begin{equation}
\overline{\mathcal{U}}_{0}'\mathcal{H}(\boldsymbol{k},\boldsymbol{r},t)\overline{\mathcal{U}}_{0}'^{-1}=\mathcal{H}(-\boldsymbol{k}_{\parallel},\boldsymbol{k}_{\perp},-\boldsymbol{r}_{\parallel},\boldsymbol{r}_{\perp},t),
\end{equation}
with $\overline{\mathcal{U}}_{0}'=\rho_{x}\otimes\overline{\mathcal{U}}_{0}$,
which satisfies
\begin{gather}
\hat{\mathcal{U}}_{0}'\hat{\mathcal{T}'}=\overline{\eta}_{T}\hat{\mathcal{T}}'\hat{\mathcal{U}}_{0}'\\
\hat{\mathcal{U}}_{0}'\hat{\mathcal{C}}'=-\overline{\eta}_{T}\hat{\mathcal{C}}'\hat{\mathcal{U}}_{0}',
\end{gather}
and $\overline{\mathcal{U}}_{0}'^{2}=\overline{\epsilon}_{U}$. 
Under the hermitian map, unitary loops with symmetry $({\rm AI},\text{\ensuremath{\overline{\mathcal{U}}_{0,\overline{\eta}_{T}}^{\overline{\epsilon}_{U}}}})$
and $({\rm AII},\text{\ensuremath{\overline{\mathcal{U}}_{0,\overline{\eta}_{T}}^{\overline{\epsilon}_{U}}}})$
in dimension $(d,d_{\parallel},D,D_{\parallel})$ are mapped to static
Hamiltonians with symmetry $({\rm CI},\hat{\mathcal{U}}_{\overline{\eta}_{T},-\overline{\eta}_{T}}^{\overline{\epsilon}_{U}})$
and $({\rm DIII},\hat{\mathcal{U}}_{\overline{\eta}_{T},-\overline{\eta}_{T}}^{\overline{\epsilon}_{U}})$
in dimension $(d,d_{\parallel},D+1,D_{\parallel})$, respectively.

\paragraph{$\overline{\mathcal{O}}=\overline{\mathcal{U}}_{T/2}$}
Due to the order-two antisymmetry realized by $\overline{\mathcal{U}}_{T/2}$,
we haves
\begin{equation}
\overline{\mathcal{U}}_{T/2}'\mathcal{H}(\boldsymbol{k},\boldsymbol{r},t)\overline{\mathcal{U}}_{T/2}'^{-1}=\mathcal{H}(-\boldsymbol{k}_{\parallel},\boldsymbol{k}_{\perp},-\boldsymbol{r}_{\parallel},\boldsymbol{r}_{\perp},T-t),
\end{equation}
with $\overline{\mathcal{U}}_{T/2}'=\rho_{0}\otimes\overline{\mathcal{U}}_{T/2}$,
which satisfies
\begin{gather}
\hat{\mathcal{U}}_{0}'\hat{\mathcal{T}'}=\overline{\eta}_{T}\hat{\mathcal{T}}'\hat{\mathcal{U}}_{0}'\\
\hat{\mathcal{U}}_{0}'\hat{\mathcal{C}}'=\overline{\eta}_{T}\hat{\mathcal{C}}'\hat{\mathcal{U}}_{0}',
\end{gather}
and $\overline{\mathcal{U}}_{T/2}'^{2}=\overline{\epsilon}_{U}$.
Under the hermitian map, unitary loops with symmetry $({\rm AI},\text{\ensuremath{\overline{\mathcal{U}}_{T/2,\overline{\eta}_{T}}^{\overline{\epsilon}_{U}}}})$
and $({\rm AII},\text{\ensuremath{\overline{\mathcal{U}}_{T/2,\overline{\eta}_{T}}^{\overline{\epsilon}_{U}}}})$
in dimension $(d,d_{\parallel},D,D_{\parallel})$ are mapped to static
Hamiltonians with symmetry $({\rm CI},\hat{\mathcal{U}}_{\overline{\eta}_{T},\overline{\eta}_{T}}^{\overline{\epsilon}_{U}})$
and $({\rm DIII},\hat{\mathcal{U}}_{\overline{\eta}_{T},\overline{\eta}_{T}}^{\overline{\epsilon}_{U}})$
in dimension $(d,d_{\parallel},D+1,D_{\parallel}+1)$, respectively.

\subsubsection{Classes C and D}
\paragraph{$\hat{\mathcal{O}}=\hat{\mathcal{U}}_{0}$}
The new hermitian matrix $\mathcal{H}(\boldsymbol{k},\boldsymbol{r},t)$
under the hermitian map defined by Eq.(\ref{eq:effective_hamiltonian_nonchiral})
aquires new time-reversal and particle-hole symmetries, realized by
$\hat{\mathcal{T}}'=\rho_{0}\otimes\hat{\mathcal{C}}$ and $\hat{\mathcal{C}'}=\rho_{z}\otimes\hat{\mathcal{T}}$,
respectively. Due to the order-two symmetry realized by $\hat{\mathcal{U}}_{0}$,
we have
\begin{equation}
\hat{\mathcal{U}}_{0}'\mathcal{H}(\boldsymbol{k},\boldsymbol{r},t)\hat{\mathcal{U}}_{0}'^{-1}=\mathcal{H}(-\boldsymbol{k}_{\parallel},\boldsymbol{k}_{\perp},-\boldsymbol{r}_{\parallel},\boldsymbol{r}_{\perp},t),
\end{equation}
with $\hat{\mathcal{U}}_{0}'=\rho_{0}\otimes\hat{\mathcal{U}}_{0}$.
which satisfies
\begin{gather}
\hat{\mathcal{U}}_{0}'\hat{\mathcal{T}'}=\eta_{C}\hat{\mathcal{T}}'\hat{\mathcal{U}}_{0}'\\
\hat{\mathcal{U}}_{0}'\hat{\mathcal{C}}'=\eta_{C}\hat{\mathcal{C}}'\hat{\mathcal{U}}_{0}',
\end{gather}
and $\hat{\mathcal{U}}_{0}'^{2}=\epsilon_{U}$. Under the hermitian
map, unitary loops with symmetry $({\rm C},\text{\ensuremath{\hat{\mathcal{U}}_{0,\eta_{C}}^{\epsilon_{U}}}})$
and $({\rm D},\hat{\mathcal{U}}_{0,\eta_{C}}^{\epsilon_{U}})$ in
dimension $(d,d_{\parallel},D,D_{\parallel})$ are mapped to static
Hamiltonians with symmetry $({\rm CII},\hat{\mathcal{U}}_{\eta_{C},\eta_{C}}^{\epsilon_{U}})$
and $({\rm BDI},\hat{\mathcal{U}}_{\eta_{C},\eta_{C}}^{\epsilon_{U}})$
in dimension $(d,d_{\parallel},D+1,D_{\parallel})$, respectively.

\paragraph{$\hat{\mathcal{O}}=\hat{\mathcal{U}}_{T/2}$}
Due to the order-two symmetry realized by $\hat{\mathcal{U}}_{T/2}$,
, we have
\begin{equation}
\hat{\mathcal{U}}_{T/2}'\mathcal{H}(\boldsymbol{k},\boldsymbol{r},t)\hat{\mathcal{U}}_{T/2}'^{-1}=\mathcal{H}(-\boldsymbol{k}_{\parallel},\boldsymbol{k}_{\perp},-\boldsymbol{r}_{\parallel},\boldsymbol{r}_{\perp},T-t),
\end{equation}
with $\hat{\mathcal{U}}_{T/2}'=\rho_{x}\otimes\mathcal{\hat{\mathcal{U}}}_{T/2}$,
which satisfies
\begin{gather}
\hat{\mathcal{U}}_{T/2}'\hat{\mathcal{T}'}=\eta_{C}\hat{\mathcal{T}}'\hat{\mathcal{U}}_{T/2}'\\
\hat{\mathcal{U}}_{T/2}'\hat{\mathcal{C}}'=-\eta_{C}\hat{\mathcal{C}}'\hat{\mathcal{U}}_{T/2}'
\end{gather}
and $\hat{\mathcal{U}}_{T/2}'^{2}=\epsilon_{U}$. Under the
hermitian map, unitary loops with symmetry $({\rm C},\text{\ensuremath{\hat{\mathcal{U}}_{T/2,\eta_{C}}^{\epsilon_{U}}}})$
and $({\rm D},\hat{\mathcal{U}}_{T/2,\eta_{C}}^{\epsilon_{U}})$ in
dimension $(d,d_{\parallel},D,D_{\parallel})$ are mapped to static
Hamiltonians with symmetry $({\rm CII},\hat{\mathcal{U}}_{\eta_{C},-\eta_{C}}^{\epsilon_{U}})$
and $({\rm BDI},\hat{\mathcal{U}}_{\eta_{C},-\eta_{C}}^{\epsilon_{U}})$
in dimension $(d,d_{\parallel},D+1,D_{\parallel}+1)$, respectively.

\paragraph{$\overline{\mathcal{O}}=\text{\ensuremath{\overline{\mathcal{U}}_{s}}}$}
Due to the order-two antisymmetry realized by $\overline{\mathcal{U}}_{s}$,
we have
\begin{equation}
\overline{\mathcal{U}}_{s}'\mathcal{H}(\boldsymbol{k},\boldsymbol{r},t)\overline{\mathcal{U}}_{s}'^{-1}=\mathcal{H}(-\boldsymbol{k}_{\parallel},\boldsymbol{k}_{\perp},-\boldsymbol{r}_{\parallel},\boldsymbol{r}_{\perp},t),
\end{equation}
with $\overline{\mathcal{U}}_{s}'=\rho_{x}\otimes\overline{\mathcal{U}}_{s}$,
which satisfies
\begin{gather}
\hat{\mathcal{U}}_{s}'\hat{\mathcal{T}'}=\overline{\eta}_{C}\hat{\mathcal{T}}'\hat{\mathcal{U}}_{s}'\\
\hat{\mathcal{U}}_{s}'\hat{\mathcal{C}}'=-\overline{\eta}_{C}\hat{\mathcal{C}}'\hat{\mathcal{U}}_{s}',
\end{gather}
and $\overline{\mathcal{U}}_{s}'^{2}=\overline{\epsilon}_{U}$. Hence,
under the hermitian map, unitary loops with symmetry $({\rm C},\text{\ensuremath{\overline{\mathcal{U}}_{s,\overline{\eta}_{C}}^{\overline{\epsilon}_{U}}}})$
and $({\rm D},\text{\ensuremath{\overline{\mathcal{U}}_{s,\overline{\eta}_{C}}^{\overline{\epsilon}_{U}}}})$
in dimension $(d,d_{\parallel},D,D_{\parallel})$ are mapped to static
Hamiltonians with symmetry $({\rm CII},\hat{\mathcal{U}}_{\overline{\eta}_{C},-\overline{\eta}_{C}}^{\overline{\epsilon}_{U}})$
and $({\rm BDI},\hat{\mathcal{U}}_{\overline{\eta}_{C},-\overline{\eta}_{C}}^{\overline{\epsilon}_{U}})$
in dimension $(d,d_{\parallel},D+1,D_{\parallel})$, respectively.

\subsubsection{Classes CI, CII, DIII, and BDI}
In these classes, the time-reversal, particle-hole and chiral symmetries
are all present. Without loss of generality, we assume $\hat{S}=\hat{\mathcal{T}}\hat{\mathcal{C}}$
and $\hat{\mathcal{S}}^{2}=1$. The hermitian matrix $\mathcal{H}(\boldsymbol{k},\boldsymbol{r},t)$
defined according to Eq.(\ref{eq:effective_hamiltonian_chiral}) has
either time-reversal or particle-hole symmetry realized by 
\begin{equation}
(\epsilon_{C}\epsilon_{T})\hat{\mathcal{C}}\mathcal{H}(-\boldsymbol{k},\boldsymbol{r},t)=\mathcal{H}(\boldsymbol{k},\boldsymbol{r},t)\hat{\mathcal{C}},
\end{equation}
depending on whether $\epsilon_{C}\epsilon_{T}$ is $1$ or $-1$. 

\paragraph{$\hat{\mathcal{O}}=\hat{\mathcal{U}}_{0}$}
Due to the order-two symmetry realized by $\hat{\mathcal{U}}_{0}$,
we have
\begin{equation}
\hat{\mathcal{U}}_{0}\mathcal{H}(\boldsymbol{k},\boldsymbol{r},t)\hat{\mathcal{U}}_{0}^{-1}=\eta_{T}\eta_{C}\mathcal{H}(-\boldsymbol{k}_{\parallel},\boldsymbol{k}_{\perp},-\boldsymbol{r}_{\parallel},\boldsymbol{r}_{\perp},t),
\end{equation}
with $\hat{\mathcal{U}}_{0}\hat{\mathcal{C}}=\eta_{C}\hat{\mathcal{C}}\hat{\mathcal{U}}_{0}$
and $\hat{\mathcal{U}}_{0}^{2}=\epsilon_{U}$. Under the hermitian
map, unitary loops in dimension $(d,d_{\parallel},D,D_{\parallel})$
with a given symmetry are mapped to static Hamiltonians in dimension
$(d,d_{\parallel},D+1,D_{\parallel})$ with another symmetry according
to 
\begin{equation}
({\rm X},\hat{\mathcal{U}}_{0,\eta_{T},\eta_{C}}^{\epsilon_{U}})\to\begin{cases}
({\rm Y},\hat{\mathcal{U}}_{\eta_{C}}^{\epsilon_{U}}) & \eta_{T}\eta_{C}=1\\
({\rm Y},\overline{\mathcal{U}}_{\eta_{C}}^{\epsilon_{U}}) & \eta_{T}\eta_{C}=-1
\end{cases},
\end{equation}
with ${\rm X}={\rm {\rm CI,CII,DIII,BDI}}$, and ${\rm Y}={\rm {\rm C,AII,D,AI}}$
respectively.

\paragraph{$\hat{\mathcal{O}}=\hat{\mathcal{U}}_{T/2}$}
Due to the order-two symmetry realized by $\hat{\mathcal{U}}_{T?2}$,
we have
\begin{align}
& (\hat{\mathcal{S}}\hat{\mathcal{U}}_{T/2})\mathcal{H}(\boldsymbol{k},\boldsymbol{r},t)(\hat{\mathcal{S}}\hat{\mathcal{U}}_{T/2})^{-1} \nonumber \\
&=\eta_{T}\eta_{C}\mathcal{H}(-\boldsymbol{k}_{\parallel},\boldsymbol{k}_{\perp},-\boldsymbol{r}_{\parallel},\boldsymbol{r}_{\perp},T-t).
\end{align}
Moreover, we have $(\hat{\mathcal{S}}\hat{\mathcal{U}}_{T/2})\hat{\mathcal{C}}=\eta_{C}\epsilon_{C}\epsilon_{T}\hat{\mathcal{C}}(\hat{\mathcal{S}}\hat{\mathcal{U}}_{T/2})$,
$(\hat{\mathcal{S}}\hat{\mathcal{U}}_{T/2})^{2}=\eta_{T}\eta_{C}\epsilon_{U}$.
Under the hermitian map, unitary loops in dimension $(d,d_{\parallel},D,D_{\parallel})$
with a given symmetry are mapped to static Hamiltonians in dimension
$(d,d_{\parallel},D+1,D_{\parallel}+1)$ with another symmetries according
to 
\begin{align}
&({\rm X},\hat{\mathcal{U}}_{T/2,\eta_{T},\eta_{C}}^{\epsilon_{U}})\to \nonumber \\
&\begin{cases}
({\rm Y},\hat{\mathcal{U}}_{\eta_{C}\epsilon_{C}\epsilon_{T}}^{\epsilon_{U}}) & \eta_{T}\eta_{C}=1\\
({\rm Y},\overline{\mathcal{U}}_{\eta_{C}\epsilon_{C}\epsilon_{T}}^{-\epsilon_{U}})=({\rm Y},\overline{\mathcal{U}}_{-\eta_{C}\epsilon_{C}\epsilon_{T}}^{\epsilon_{U}}) & \eta_{T}\eta_{C}=-1
\end{cases}.
\end{align}
with ${\rm X}={\rm {\rm CI,CII,DIII,BDI}}$, and ${\rm Y}={\rm {\rm C,AII,D,AI}}$
respectively.

\section{$K$ groups in the presence of order two symmetry \label{sec:Kgroup}}
Using the hermitian map introduced in the previous sections, the unitary loops
with an order-two space-time symmetry/antisymmetry
are successfully mapped into static Hamiltonians
with an order-two crystalline symmatry/antisymmetry, 
whose classfication has already been worked out
in Ref.~\cite{Shiozaki2014}. 
Thus, the latter result can be directly applied to the classification
of unitary loops. 

We first summarize the $K$-theory-based method used for classifying static Hamiltonians, and 
then finish the classification of unitary loops.
Let us consider static Hamiltonians defined on a base space of momentum $\boldsymbol{k}\in T^d$ 
and real space coordinate $\boldsymbol{r}\in S^D$. For the classification of strong topological 
phases, one can instead simply use $S^{d+D}$ as the base space \cite{Kitaev2009,Teo2010}.
To classify these Hamiltonians, we will use notion of stable homotopy equivalence
as we defined for unitaries in Sec.~\ref{sec:floquet_basics},
by identifying Hamiltonians which are continuously deformable into each other
up to adding extra trivial bands, while preserving an energy gap at the chemical
potential. These equivalence classes can be formally added and they form an abelian group.

For a given AZ symmetry class $s$, the classification of static Hamiltonians
is given by the set of stable equivalence classes of maps $\mathcal{H}(\boldsymbol{k},\boldsymbol{r})$,
from the base space $(\boldsymbol{k},\boldsymbol{r})\in S^{d+D}$ to the classifying space, 
denoted as $\mathcal{C}_s$ or $\mathcal{R}_s$, for complex and real symmetry classes, as listed in Table~\ref{tab:AZ-symmetry-classes}.
The abelian group structure inherited from the equivalence classes 
leads to the group structure in this set of maps, which is called the $K$ group, or classification group.

For static topological insulators and superconductors of dimension $(d,D)$ in an AZ class $s$ without additional spatial symmetries, 
the $K$ groups are denoted as $K_{\mathbb{C}}(s;d,D)$  and $K_{\mathbb{R}}(s;d,D)$, 
for complex and real symmetry classes, respectively.
Note that for complex symmetry classes, we have $s=0,1 \mod 2$, whereas
for real symmetry classes,  $s=0,1,\dots,7 \mod 8$.  

These $K$ groups have the following properties
\begin{gather}
K_{\mathbb{C}}(s;d,D) = K_{\mathbb{C}}(s-d+D;0,0)= \pi_0(\mathcal{C}_{s-d+D})\\
K_{\mathbb{R}}(s;d,D) = K_{\mathbb{R}}(s-d+D;0,0)= \pi_0(\mathcal{R}_{s-d+D})
\end{gather}
known as the Bott periodicity, where $\pi_0$ denotes the zeroth homotopy group
which counts the number of path connected components in a given space. 
In the following, we will introduce the $K$ groups for Hamiltonians supporting an additional
order-two spatial symmetry/antisymmetry following Ref.~\cite{Shiozaki2014}.
Because of the hermitian map, these $K$ groups can also be associtated with the unitary loops, 
whose classification is then obtained.

\subsection{Complex symmetry classes with an additional order-two unitary symmetry/antisymmetry}
When a spatial or space-time symmetry/antisymmetry is considered, one needs to include the number of ``flipped''
coordinates for both $\boldsymbol{k}$ and $\boldsymbol{r}$, into the dimensions. 
For a static Hamiltonian of dimension $(d,d_{\parallel},D,D_{\parallel})$ in complex AZ classes 
with an additional order-two unitary symmetry/antisymmetry,
the $K$ group is denoted as $K_{\mathbb{C}}^{U}(s,t;d,d_{\parallel},D,D_{\parallel})$,
where the additional parameter $t=0,1 \mod 2$, specifies the coexisting order-two
unitary symmetry/antisymmetry.
These $K$ groups satisfy the following relation
\begin{align}
K_{\mathbb{C}}^{U}(s,t;d,d_{\parallel},D,D_{\parallel})&=K_{\mathbb{C}}^{U}(s-\delta,t-\delta_{\parallel}; 0,0,0,0)
\nonumber \\
&\equiv K_{\mathbb{C}}^{U}(s-\delta,t-\delta_{\parallel}),\label{eq:complex_unitary_Kgroup}
\end{align}
where $\delta=d-D$, $\delta_{\parallel}=d_{\parallel}-D_{\parallel}$.
Thus, for classification purpose,
one can use the pair $(\delta,\delta_{\parallel})$ instead of $(d,d_{\parallel},D,D_{\parallel})$
to denote the dimensions of the base space, on which the static Hamiltonian is defined.

\begin{table}
\caption{\label{tab:complex_unitary}Possible types ($t=0,1\mod2$) of order-two
additional unitary symmetr $\hat{\mathcal{U}}_{\eta_{S}}^{\epsilon_{U}}$/$\overline{\mathcal{U}}_{\overline{\eta}_{S}}^{\overline{\epsilon}_{U}}$
in complex AZ classes ($s=0,1\mod2$). The superscript and subscript
are defined as $\epsilon_{U}=\hat{\mathcal{U}}^{2}$, $\overline{\epsilon}_{U}=\overline{\mathcal{U}}^{2}$,
$\hat{\mathcal{U}}\hat{\mathcal{S}}=\eta_{S}\hat{\mathcal{S}}\hat{\mathcal{U}}$,
$\overline{\mathcal{U}}\hat{\mathcal{S}}=\overline{\eta}_{S}\hat{\mathcal{S}}\overline{\mathcal{U}}$. }
\begin{ruledtabular}
\centering
\begin{tabular}{cccc}
s & AZ class & $t=0$ & $t=1$\\
\hline
0 & A & $\hat{\mathcal{U}}_{0}^{+}, \hat{\mathcal{U}}_{T/2}^{+}$ & $\overline{\mathcal{U}}^{+}_{s}$\\
1 & AIII & $\hat{\mathcal{U}}_{0,+}^{+}$, $\hat{\mathcal{U}}_{T/2,-}^+$ & $\hat{\mathcal{U}}_{0,-}^{+}$, $\hat{\mathcal{U}}_{T/2,+}^{+}$\\
\end{tabular}
\end{ruledtabular}
\end{table}

To define $K$ groups for unitary loops, 
we use the fact that the $K$ group for certain unitary loops should be the same 
as the one for the corresponding static Hamiltonians under the hermitian map.
The $K$ groups for unitary loops are explicitly defined in Table~\ref{tab:complex_unitary}, 
where the two arguements $s,t$ label the AZ class and the coexisting order-two space-time symmetry/antisymmetry.

\subsection{Complex symmetry classes with an additional order-two antiunitary symmetry/antisymmetry}
We now consider static Hamiltonians of dimension $(d,d_{\parallel},D,D_{\parallel})$, 
in complex AZ classes, with an order-two antiunitary symmetry/antisymmetry, realized by
$\hat{\mathcal{A}}$ or $\overline{\mathcal{A}}$.
It turns out that complex AZ classes acquire real structures because of the antiunitary symmetry \cite{Shiozaki2014}.
Indeed, effective time-reversal or particle-hole
symmetry realized by $\hat{\mathcal{A}}$ or $\overline{\mathcal{A}}$
emerges, if we regard $(\boldsymbol{k}_{\perp},\boldsymbol{r}_{\parallel})$
as ``momenta'', and $(\boldsymbol{k}_{\parallel},\boldsymbol{r}_{\perp})$
as ``spatial coordinates''. Thus, a system in complex AZ classes
with an antiunitary symmetry can be mapped into a
real AZ class without additional spatial symmetries.

\begin{table}
\caption{\label{tab:complex_antiunitary}Possible types ($s=0,\dots,7\mod8$)
of order-two additional antiunitary symmetry $\hat{\mathcal{A}}_{\eta_{S}}^{\epsilon_{A}}$/$\overline{\mathcal{A}}_{\overline{\eta}_{S}}^{\overline{\epsilon}_{A}}$
in complex AZ classes. The superscript and subscript are defined as
$\epsilon_{A}=\hat{\mathcal{A}}^{2}$, $\overline{\epsilon}_{A}=\overline{\mathcal{A}}^{2}$,
$\hat{\mathcal{A}}\hat{\mathcal{S}}=\eta_{S}\hat{\mathcal{S}}\hat{\mathcal{A}}$,
$\overline{\mathcal{A}}\hat{\mathcal{S}}=\overline{\eta}_{S}\hat{\mathcal{S}}\overline{\mathcal{A}}$. }
\begin{ruledtabular}
\centering
\begin{tabular}{cccc}
$s$ & AZ class & Coexisiting symmetry & Mapped AZ class\\
\hline 
$0$ & A & $\hat{\mathcal{A}}_{s}^{+}$ & AI\\
$1$ & AIII & $\hat{\mathcal{A}}_{0,+}^{+}$, $\hat{\mathcal{A}}_{T/2,-}^{+}$ & BDI\\
$2$ & A & $\overline{\mathcal{A}}_{0}^{+}$, $\overline{\mathcal{A}}_{T/2}^{-}$ & D\\
$3$ & AIII & $\hat{\mathcal{A}}_{0,-}^{-}$, $\hat{\mathcal{A}}_{T/2,+}^{-}$ & DIII\\
$4$ & A & $\hat{\mathcal{A}}_{s}^{-}$ & AII\\
$5$ & AIII & $\hat{\mathcal{A}}_{0,+}^{-}$, $\hat{\mathcal{A}}_{T/2,-}^{-}$ & CII\\
$6$ & A & $\overline{\mathcal{A}}_{0}^{-}$, $\overline{\mathcal{A}}_{T/2}^{+}$ & C\\
$7$ & AIII & $\hat{\mathcal{A}}_{0,-}^{+}$, $\hat{\mathcal{A}}_{T/2,+}^{+}$ & CI\\
\end{tabular}
\end{ruledtabular}
\end{table}

The $K$ groups for these Hamiltonians are denoted as
$K_{\mathbb{C}}^{A}(s;d,d_{\parallel},D,D_{\parallel})$,
which satisfies
\begin{align}
K_{\mathbb{C}}^{A}(s;d,d_{\parallel},D,D_{\parallel})&=K_{\mathbb{C}}^{A}(s-\delta+2\delta_{\parallel};0,0,0,0) \nonumber
\\
&\equiv K_{\mathbb{C}}^{A}(s-\delta+2\delta_{\parallel}).\label{eq:complex_antiunitary_Kgroup}.
\end{align}

Similar to the previous case, the unitary loops 
with an antiunitary space-time symmetry/antisymmetry
can also be associated with these $K$ groups. 

If we group these antiunitary symmetries and antisymmetries
in terms of the index $s = 0,\dots,7 \mod 8$, according to 
Table~\ref{tab:complex_antiunitary}, then 
$K_{\mathbb{C}}^{A}(s)$ can further be reduced to $K_{\mathbb{R}}(s) \equiv K_{\mathbb{R}}(s;0,0)$.

\subsection{Real symmetry classes with an additional order-two symmetry}
In real symmetry classes, there are equivalence relations between
order-two unitary and antiunitary symmetries/antisymmetries, as discussed
previously. Thus, one can focus on unitary symmetries/antisymmetries
only. The existence of an additional order-two unitary symmetry
divide each class into four families ($t=0,\dots,3\mod4$), as summarized
in Table \ref{tab:real_unitary}, where we have used the 
equivalence of $K$ groups for static Hamiltonians and unitary loops in terms of the hermitian map.

\begin{table*}
\caption{\label{tab:real_unitary}Possible types ($t=0,\dots,3\mod4$) of order-two
additional symmetry $\hat{\mathcal{U}}_{\eta_{S}}^{\epsilon_{U}}$/$\overline{\mathcal{U}}_{\overline{\eta}_{S}}^{\overline{\epsilon}_{U}}$
in real AZ classes. The superscript and subscript are defined as $\epsilon_{U}=\hat{\mathcal{U}}^{2}$,
$\overline{\epsilon}_{U}=\overline{\mathcal{U}}^{2}$, $\hat{\mathcal{U}}\hat{\mathcal{S}}=\eta_{S}\hat{\mathcal{S}}\hat{\mathcal{U}}$,
$\overline{\mathcal{U}}\hat{\mathcal{S}}=\overline{\eta}_{S}\hat{\mathcal{S}}\overline{\mathcal{U}}$.
We fix $\epsilon_{U}=\overline{\epsilon}_{U}=1$. }
\begin{ruledtabular}
\centering
\begin{tabular}{cccccc}
s & AZ Class & $t=0$ & $t=1$ & $t=2$ & $t=3$\\
\hline
0 & AI & $\hat{\mathcal{U}}_{0,+}^{+}$, $\hat{\mathcal{U}}_{T/2,+}^{+}$ & $\overline{\mathcal{U}}_{0,-}^{+}$, $\overline{\mathcal{U}}_{T/2,+}^{+}$ 
& $\hat{\mathcal{U}}_{0,-}^{+}$, $\hat{\mathcal{U}}_{T/2,-}^{+}$ & $\overline{\mathcal{U}}_{0,+}^{+}$, $\overline{\mathcal{U}}_{T/2,-}^{+}$\\
1 & BDI & $\hat{\mathcal{U}}_{0,++}^{+}$, $\hat{\mathcal{U}}_{T/2,+-}^{+}$ & $\hat{\mathcal{U}}_{0,+-}^{+}$, $\hat{\mathcal{U}}_{T/2,++}^{+}$ 
& $\hat{\mathcal{U}}_{0,--}^{+}$,$\hat{\mathcal{U}}_{T/2,-+}^{+}$  & $\hat{\mathcal{U}}_{0,-+}^{+}$, $\hat{\mathcal{U}}_{T/2,--}^{+}$\\
2 & D & $\hat{\mathcal{U}}_{0,+}^{+}$, $\hat{\mathcal{U}}_{T/2,-}^{+}$ & $\overline{\mathcal{U}}_{s,+}^{+}$ 
& $\hat{\mathcal{U}}_{0,-}^{+}$, $\hat{\mathcal{U}}_{T/2,+}^{+}$ & $\overline{\mathcal{U}}_{s,-}^{+}$\\
3 & DIII & $\hat{\mathcal{U}}_{0,++}^{+}$, $\hat{\mathcal{U}}_{T/2,+-}^{+}$ 
& $\hat{\mathcal{U}}_{0,-+}^{+}$, $\hat{\mathcal{U}}_{T/2,--}^{+}$ & $\hat{\mathcal{U}}_{0,--}^{+}$, $\hat{\mathcal{U}}_{0,-+}^{+}$
& $\hat{\mathcal{U}}_{0,+-}^{+}$, $\hat{\mathcal{U}}_{T/2,++}^{+}$\\
4 & AII & $\hat{\mathcal{U}}_{0,+}^{+}$, $\hat{\mathcal{U}}_{T/2,+}^{+}$ & $\overline{\mathcal{U}}_{0,-}^{+}$, $\overline{\mathcal{U}}_{T/2,+}^{+}$
& $\hat{\mathcal{U}}_{0,-}^{+}$, $\hat{\mathcal{U}}_{T/2,-}^{+}$ & $\overline{\mathcal{U}}_{0,+}^{+}$, $\overline{\mathcal{U}}_{T/2,-}^{+}$\\
5 & CII & $\hat{\mathcal{U}}_{0,++}^{+}$, $\hat{\mathcal{U}}_{T/2,+-}^{+}$ & $\hat{\mathcal{U}}_{0,+-}^{+}$,
$\hat{\mathcal{U}}_{T/2,++}^{+}$ & $\hat{\mathcal{U}}_{0,--}^{+}$, $\hat{\mathcal{U}}_{T/2,-+}^{+}$ 
& $\hat{\mathcal{U}}_{0++,-+}^{+}$, $\hat{\mathcal{U}}_{T/2,--}^{+}$\\
6 & C & $\hat{\mathcal{U}}_{0,+}^{+}$, $\hat{\mathcal{U}}_{T/2,-}^{+}$ & $\overline{\mathcal{U}}_{s,+}^{+}$ 
& $\hat{\mathcal{U}}_{0,-}^{+}$, $\hat{\mathcal{U}}_{T/2,+}^{+}$ & $\overline{\mathcal{U}}_{s,-}^{+}$\\
7 & CI & $\hat{\mathcal{U}}_{0,++}^{+}$, $\hat{\mathcal{U}}_{T/2,+-}^{+}$ & $\hat{\mathcal{U}}_{0,-+}^{+}$,
$\hat{\mathcal{U}}_{T/2,--}^{+}$ & $\hat{\mathcal{U}}_{0,--}^{+}$, $\hat{\mathcal{U}}_{T/2,-+}^{+}$ 
& $\hat{\mathcal{U}}_{0,+-}^{+}$, $\hat{\mathcal{U}}_{T/2,++}^{+}$\\
\end{tabular}
\end{ruledtabular}
\end{table*}

We denote the $K$ group for unitary loops in real AZ classes ($s=0,\dots,7\mod8$) with an
additional order-two unitary symmetry/antisymmetry ($t=0,\dots,3\mod4$) as $K_{\mathbb{R}}^{U}(s,t;d,d_{\parallel},D,D_{\parallel})$,
which satisfies
\begin{align}
K_{\mathbb{R}}^{U}(s,t;d,d_{\parallel},D,D_{\parallel})&=K_{\mathbb{R}}^{U}(s-\delta,t-\delta_{\parallel};0,0,0,0)
\nonumber \\
&\equiv K_{\mathbb{R}}^{U}(s-\delta,t-\delta_{\parallel}).\label{eq:real_unitary_Kgroup}
\end{align}

\subsection{Nontrivial space-time vs static spatial symmetries/antisymmetries \label{sec:substitution}}
The classification of unitary loops with an order-two space-time symmetry/antisymmetry is given by the
$K$ groups, $K_{\mathbb{C}}^{U}(s,t)$, $K_{\mathbb{C}}^{A}(s)$ or $K_{\mathbb{R}}^{U}(s,t)$.
As can be seen in Tables~\ref{tab:complex_unitary}--\ref{tab:real_unitary}, 
for every order-two space-time (anti)unitary symmetry/antisymmetry that is
nontrivial, namely the half-period time translation is involved, 
there always exists a unique static spatial (anti)unitary symmetry/antisymmetry, 
such that both symmetries/antisymmetries give rise to the same $K$ group.
It is worth mentioning that when looking at the static symmetries/antisymmetries alone, 
the corresponding $K$ groups for unitary loops
are defined in the same way as the ones for Hamiltonians introduced
in Ref.~\cite{Shiozaki2014}, as expected.

The explicit relations between the two types of symmetries/antisymmetries (nontrivial space-time vs static) with the
same $K$ group can be summarized as follows. Recall that we use $\eta_{S}$ ($\overline{\eta}_{S}$), $\eta_{T}$ ($\overline{\eta}_{T}$) and $\eta_{C}$ ($\overline{\eta}_{C}$)
to characterize the commutation relations between the order-two symmetry (antisymmetry) operator and the nonspatial symmetry operators.
For two unitary order-two symmetries giving rise to the same $K$ group, the $\eta_{S}$s and $\eta_{C}$s for the two symmetries
take opposite signs, whereas $\eta_{T}$s are the same.
For two antiunitary order-two symmetries, we have $\eta_{S}$s take opposite signs.
For two unitary antisymmetries, the $\overline{\eta}_{T}$s have opposite signs.
Finally, for class A, the antiunitary space-time antisymmetry operator
$\overline{\mathcal{A}}_{T/2}^{\pm}$ have the same $K$ group
as the one for $\overline{\mathcal{A}}_{0}^{\mp}$. 
These relations are summarized in Table~\ref{tab:summary},
and can be better understood 
after we introduce the frequency-domain formulation of the 
Floquet problem in Sec.~\ref{sec:harmonically_driven_models}. 

\subsection{Periodic table \label{sec:periodic_table}}
From the $K$ groups introduced previously, we see that
in addition to the mod $2$ or mod $8$ Bott periodicity in
$\delta$, there also exists a periodic structure in flipped dimensions
$\delta_{\parallel}$, because of the twofold or fourfold periodicity
in $t$, which accounts for the additional order-two symmetry/antisymmetry. 
In particular, for complex symmetry classes with an order-two unitary symmetry/antisymmetry,
the classification has a twofold periodicity in $\delta_{\parallel}$,
whereas for complex symmetry classes with an order-two antiunitary symmetry/antisymmetry,
and for real symmetry classes with an order-two unitary/antisymmetry,
the periodicity in $\delta_{\parallel}$ is fourfold.
These periodic features are the same as the ones obtained 
in Ref.~\cite{Shiozaki2014} for static Hamiltonians with
an order-two crystalline symmetry/antisymmetry.
We summarize the periodic tables 
for the four $(\text{\ensuremath{\delta_{\parallel}=0,\dots,3\mod4}})$
different families below in the supplemental material \cite{suppl}.

Note that in obtaining the classification Tables, we made use 
of the $K$ groups in their zero dimensional forms defined
in Eqs.~(\ref{eq:complex_unitary_Kgroup}), (\ref{eq:complex_antiunitary_Kgroup}) and (\ref{eq:real_unitary_Kgroup}), 
as well as the following relations
\begin{gather}
K_{\mathbb{C}}^{U}(s,t=0)=\pi_{0}(\mathcal{C}_{s}\times\mathcal{C}_{s})=\pi_{0}(\mathcal{C}_{s})\oplus\pi_{0}(\mathcal{C}_{s})
\nonumber \\
K_{\mathbb{C}}^{U}(s,t=1)=\pi_{0}(\mathcal{C}_{s+1}), \nonumber \\
K_{\mathbb{C}}^{A}(s) = \pi_0(\mathcal{R}_s), \nonumber \\
K_{\mathbb{R}}^{U}(s,t=0)=\pi_{0}(\mathcal{R}_{s}\times\mathcal{R}_{s})=\pi_{0}(\mathcal{R}_{s})\oplus\pi_{0}(\mathcal{R}_{s}),
\nonumber\\
K_{\mathbb{R}}^{U}(s,t=1)=\pi_{0}(\mathcal{R}_{s+7})\nonumber \\
K_{\mathbb{R}}^{U}(s,t=2)=\pi_{0}(\mathcal{C}_{s}) \nonumber \\
K_{\mathbb{R}}^{U}(s,t=3)=\pi_{0}(\mathcal{R}_{s+1}).
\end{gather}
where $\mathcal{C}_{s}$ ($s=0,1\mod 2$) and $\mathcal{R}_{s}$ ($s=0,\dots, 7 \mod 8$) represent the classifying
space of complex and real AZ classes, see Table \ref{tab:AZ-symmetry-classes}.

\section{Floquet higher-order topological insulators and superconductors \label{sec:FHOTI_extension}}
In the previous sections, we obtained a complete classification of
the anomalous Floquet TI/SCs using $K$ theory,
where the $K$ groups for the unitary loops were defined
as the same ones for the static Hamiltonians, according to the hermitian map.

Noticeably, the classification obtained in this way is a bulk classification,
since the only the bulk unitary evolution operators were considered. 
These bulk $K$ groups include the information of topological classification 
at any order. 
For static tenfold-way TI/SCs,
in which the topological property is determined from the nonspatial symmetries, 
there is a bulk-boundary correspondence which essentially says that 
the nontrivial topological bulk indicates protected gapless boundary modes living
in one dimension lower. 
This boundary modes is irrespective of boundary orientation and lattice termination. 
The same is true for tenfold-way Floquet TI/SCs with only nonspatial symmetries. 
In this situation, since only first-order topological phases are allowed, this bulk $K$ group
is enough to understand the existence of gapless boundary modes. 

However, when an additional crystalline symmetry/antisymmetry is taking into account, 
the existence of gapless boundary modes due to nontrivial topological bulk is not
guaranteed unless the boundary is invariant under the nonlocal
transformation of the symmetry/antisymmetry \cite{Fu2011, Ando2015}. 

\begin{table*}
  \caption{\label{tab:complex_unitary_hoti_2}Subgroup series $K^{(d)}\subseteq \dots \subseteq K' \subseteq K$ for zero-
  ($d=0$), one- ($d=1$), and two-dimensional ($d=2$) anomalous Floquet HOTI/SCs with a unitary order-two space-time
symmetry/antisymmetry in complex classes. The number of flipped dimensions for the symmetry/antisymmetry is denoted as
$d_{\parallel}$.  }
\begin{ruledtabular}
\centering
\begin{tabular}{cccccccc}
 &  &$d=0$ &$d=1$ &$d=1$ &$d=2$ &$d=2$ &$d=2$ \\
   Symmetry  &Class &$d_{\parallel}=0$  &$d_{\parallel}=0$ &$d_{\parallel}=1$ &$d_{\parallel}=0$
    &$d_{\parallel}=1$ &$d_{\parallel}=2$ \\
\hline
$\hat{\mathcal{U}}_{0}^{+}$, $\hat{\mathcal{U}}_{T/2}^{+}$ &A &$\mathbb{Z}^2$ &$0\subseteq 0$
&$\mathbb{Z}\subseteq\mathbb{Z}$ &$0 \subseteq 0 \subseteq \mathbb{Z}^{2}$ &$0 \subseteq 0 \subseteq 0$
&$\mathbb{Z}\subseteq \mathbb{Z} \subseteq \mathbb{Z}^{2}$ \\
$\hat{\mathcal{U}}_{0,+}^{+}$, $\hat{\mathcal{U}}_{T/2,-}^{+}$ &AIII &$0$ &$0 \subseteq \mathbb{Z}^{2}$ &$0 \subseteq 0$
&$0 \subseteq 0 \subseteq 0$ &$0\subseteq \mathbb{Z} \subseteq \mathbb{Z}$ &$0 \subseteq 0 \subseteq 0$ \\ 
\hline
$\overline{\mathcal{U}}_{s}^{+}$ &A &$0$ &$0 \subseteq \mathbb{Z}$ &$0 \subseteq 0$ &$0 \subseteq 0 \subseteq 0$
&$0 \subseteq \mathbb{Z} \subseteq \mathbb{Z}^{2}$ &$0 \subseteq 0 \subseteq 0$ \\
$\hat{\mathcal{U}}_{0,-}^{+}$, $\hat{\mathcal{U}}_{T/2,+}^+$ &AIII &$\mathbb{Z}$ &$0 \subseteq 0$ &$\mathbb{Z}\subseteq
\mathbb{Z}^{2}$ &$0 \subseteq 0 \subseteq \mathbb{Z}$ &$0 \subseteq 0 \subseteq 0$ 
&$2\mathbb{Z} \subseteq \mathbb{Z} \subseteq \mathbb{Z}$
\end{tabular}
\end{ruledtabular}
\end{table*}

\begin{table*}
  \caption{\label{tab:complex_antiunitary_hoti_2} Same as Table~\ref{tab:complex_unitary_hoti_2} for antiunitary
  symmetries/antisymmetries.}
\begin{ruledtabular}
\centering
\begin{tabular}{cccccccc}
 &  &$d=0$ &$d=1$ &$d=1$ &$d=2$ &$d=2$ &$d=2$ \\
   Symmetry  &Class &$d_{\parallel}=0$  &$d_{\parallel}=0$ &$d_{\parallel}=1$ &$d_{\parallel}=0$
    &$d_{\parallel}=1$ &$d_{\parallel}=2$ \\
\hline
$\hat{\mathcal{A}}_{s}^{+}$ & A &$\mathbb{Z}$ &$0 \subseteq 0$ &$\mathbb{Z}_2\subseteq \mathbb{Z}_2$ &$0 \subseteq 0
\subseteq 0$ &$0 \subseteq 0 \subseteq \mathbb{Z}$ &$\mathbb{Z}_{2} \subseteq \mathbb{Z}_2 \subseteq
\mathbb{Z}_{2}$ \\
$\hat{\mathcal{A}}_{0,+}^{+},\hat{\mathcal{A}}_{T/2,-}^{+}$ & AIII &$\mathbb{Z}_2$ &$0 \subseteq \mathbb{Z}$
&$\mathbb{Z}_2 \subseteq \mathbb{Z}_{2}$ &$0 \subseteq 0 \subseteq 0$ &$0 \subseteq \mathbb{Z}_2 \subseteq
\mathbb{Z}_2$ &$0 \subseteq 0 \subseteq 0$ \\
$\overline{\mathcal{A}}_{0}^{+},\overline{\mathcal{A}}_{T/2}^{-}$ & A &$\mathbb{Z}_2$ &$0 \subseteq \mathbb{Z}_2$
&$0 \subseteq 0$ &$0 \subseteq 0 \subseteq \mathbb{Z}$ &$0 \subseteq \mathbb{Z}_2 \subseteq \mathbb{Z}_2$
&$0 \subseteq 0 \subseteq 2\mathbb{Z}$ \\
$\hat{\mathcal{A}}_{0,-}^{-},\hat{\mathcal{A}}_{T/2,+}^{-}$ & AIII  &$0$ &$0 \subseteq \mathbb{Z}_2$ &$0 \subseteq
2\mathbb{Z}$ &$0 \subseteq 0 \subseteq \mathbb{Z}_2$ &$0 \subseteq 0 \subseteq 0$ &$0 \subseteq 0 \subseteq 0$ \\
$\hat{\mathcal{A}}_{s}^{-}$ & A  &$2\mathbb{Z}$ &$0 \subseteq 0$ &$0 \subseteq 0$ &$0 \subseteq 0 \subseteq 0 \subseteq
\mathbb{Z}_{2}$ &$0 \subseteq 0 \subseteq 2\mathbb{Z}$ &$0 \subseteq 0 \subseteq 0$ \\
$\hat{\mathcal{A}}_{0,+}^{-},\hat{\mathcal{A}}_{T/2,-}^{-}$ & AIII  &$0$ &$0 \subseteq 2\mathbb{Z}$ &$0 \subseteq 0$
&$0 \subseteq 0 \subseteq 0$ &$0 \subseteq 0 \subseteq 0$ &$0 \subseteq 0 \subseteq 0$ \\
\end{tabular}
\end{ruledtabular}
\end{table*}

\begin{table*}
\caption{\label{tab:real_unitary_hoti_2}
Subgroup series $K^{(d)}\subseteq \dots \subseteq K' \subseteq K$ for zero-
  ($d=0$), one- ($d=1$), and two-dimensional ($d=2$) anomalous Floquet HOTI/SCs with a unitary order-two space-time
symmetry/antisymmetry in real classes. The number of flipped dimensions for the symmetry/antisymmetry is denoted as
$d_{\parallel}$.
}
\begin{ruledtabular}
\centering
\begin{tabular}{cccccccc}
 &  &$d=0$ &$d=1$ &$d=1$ &$d=2$ &$d=2$ &$d=2$ \\
   Symmetry  &Class &$d_{\parallel}=0$  &$d_{\parallel}=0$ &$d_{\parallel}=1$ &$d_{\parallel}=0$
    &$d_{\parallel}=1$ &$d_{\parallel}=2$ \\
\hline
$\text{\ensuremath{\hat{\mathcal{U}}_{0,+}^{+}}},\text{\ensuremath{\hat{\mathcal{U}}_{T/2,+}^{+}}}$ & AI
&$\mathbb{Z}^{2}$ &$0 \subseteq 0$ &$\mathbb{Z} \subseteq \mathbb{Z}$ &$0 \subseteq 0 \subseteq 0$
&$0 \subseteq 0 \subseteq 0$ &$\mathbb{Z} \subseteq \mathbb{Z} \subseteq \mathbb{Z}$ \\

$\hat{\mathcal{U}}_{0,++}^{+},\hat{\mathcal{U}}_{T/2,+-}^{+}$ & BDI &$\mathbb{Z}_{2}^{2}$ &$0 \subseteq
\mathbb{Z}^2$ &$\mathbb{Z}_{2} \subseteq \mathbb{Z}_2$ &$0 \subseteq 0 \subseteq 0$
&$0 \subseteq \mathbb{Z} \subseteq \mathbb{Z}$ &$0 \subseteq 0 \subseteq 0$ \\

$\ensuremath{\hat{\mathcal{U}}_{0,+}^{+}},\text{\ensuremath{\hat{\mathcal{U}}_{T/2,-}^{+}}}$ & D 
&$\mathbb{Z}_{2}^{2}$ &$0 \subseteq \mathbb{Z}_{2}^{2}$ &$\mathbb{Z}_{2} \subseteq \mathbb{Z}_{2}$
&$0 \subseteq 0 \subseteq \mathbb{Z}^{2}$ &$0 \subseteq \mathbb{Z}_{2} \subseteq \mathbb{Z}_{2}$
&$0 \subseteq 0 \subseteq \mathbb{Z}$ \\

$\hat{\mathcal{U}}_{0,++}^{+},\hat{\mathcal{U}}_{T/2,+-}^{+}$ & DIII &$0$
&$0 \subseteq \mathbb{Z}_{2}^2$ &$0 \subseteq 0$
&$0 \subseteq 0 \subseteq \mathbb{Z}_2^2$ &$0 \subseteq \mathbb{Z}_2 \subseteq \mathbb{Z}_2$
&$0 \subseteq 0 \subseteq 0$ \\

$\ensuremath{\hat{\mathcal{U}}_{0,+}^{+}},\ensuremath{\hat{\mathcal{U}}_{T/2,+}^{+}}$ & AII 
&$2\mathbb{Z}^2$ &$0 \subseteq 0$ &$2\mathbb{Z} \subseteq 2\mathbb{Z}$
&$0 \subseteq 0 \subseteq \mathbb{Z}_{2}^{2}$ &$0 \subseteq 0 \subseteq 0$
&$2\mathbb{Z} \subseteq 2\mathbb{Z} \subseteq \mathbb{Z}$ \\

$\hat{\mathcal{U}}_{0,++}^{+},\hat{\mathcal{U}}_{T/2,+-}^{+}$ & CII &$0$
&$0 \subseteq 2\mathbb{Z}^{2}$ &$0 \subseteq 0$ &$0 \subseteq 0 \subseteq 0$
&$0 \subseteq 2\mathbb{Z} \subseteq 2\mathbb{Z}$  &$0 \subseteq 0 \subseteq 0$ \\
\end{tabular}
\end{ruledtabular}
\end{table*}

\begin{table*}
\caption{\label{tab:complex_unitary_hoti_3} Subgroup series $K^{(d)}\subseteq \dots \subseteq K' \subseteq K$ for 
  three-dimensional ($d=3$) anomalous Floquet HOTI/SCs with a unitary order-two space-time
symmetry/antisymmetry in complex classes. The number of flipped dimensions for the symmetry/antisymmetry is denoted as
$d_{\parallel}$.}
\begin{ruledtabular}
\centering
\begin{tabular}{cccccc}
   Symmetry  &Class &$d_{\parallel}=0$  &$d_{\parallel}=1$ &$d_{\parallel}=2$ &$d_{\parallel}=3$ \\
\hline
$\hat{\mathcal{U}}_{0}^{+}$, $\hat{\mathcal{U}}_{T/2}^{+}$ &A  &$0 \subseteq 0 \subseteq 0 \subseteq 0$
&$0 \subseteq 0 \subseteq \mathbb{Z} \subseteq \mathbb{Z}$
&$0 \subseteq 0 \subseteq 0 \subseteq 0$
&$2\mathbb{Z} \subseteq 2\mathbb{Z} \subseteq \mathbb{Z} \subseteq \mathbb{Z} $
\\
$\hat{\mathcal{U}}_{0,+}^{+}$, $\hat{\mathcal{U}}_{T/2,-}^{+}$ &AIII
&$0 \subseteq 0 \subseteq 0 \subseteq \mathbb{Z}^{2}$
&$0 \subseteq 0 \subseteq 0 \subseteq 0$
&$0 \subseteq \mathbb{Z} \subseteq \mathbb{Z} \subseteq \mathbb{Z}^{2}$
&$0 \subseteq 0 \subseteq 0 \subseteq 0$
\\ 
\hline
$\overline{\mathcal{U}}_{s}^{+}$ &A 
&$0 \subseteq 0 \subseteq 0 \subseteq \mathbb{Z}$
&$0 \subseteq 0 \subseteq 0 \subseteq 0$
&$0 \subseteq 2\mathbb{Z} \subseteq \mathbb{Z} \subseteq \mathbb{Z}$
&$0 \subseteq 0 \subseteq 0 \subseteq 0$
\\
$\hat{\mathcal{U}}_{0,-}^{+}$, $\hat{\mathcal{U}}_{T/2,+}^+$ &AIII 
&$0 \subseteq 0 \subseteq 0 \subseteq 0$
&$0 \subseteq 0 \subseteq \mathbb{Z} \subseteq \mathbb{Z}^{2}$
&$0 \subseteq 0 \subseteq 0 \subseteq 0$
&$2\mathbb{Z} \subseteq \mathbb{Z} \subseteq \mathbb{Z} \subseteq \mathbb{Z}^{2}$
\end{tabular}
\end{ruledtabular}
\end{table*}

\begin{table*}
  \caption{\label{tab:complex_antiunitary_hoti_3}Same as Table~\ref{tab:complex_unitary_hoti_3} for antiunitary
  symmetries/antisymmetries.}
\begin{ruledtabular}
\centering
\begin{tabular}{cccccc}
   Symmetry  &Class &$d_{\parallel}=0$  &$d_{\parallel}=1$ &$d_{\parallel}=2$ &$d_{\parallel}=3$ \\
\hline
$\hat{\mathcal{A}}_{s}^{+}$ & A
&$0 \subseteq 0 \subseteq 0 \subseteq 0$
&$0 \subseteq 0 \subseteq 0 \subseteq 0$
&$0 \subseteq 0 \subseteq \mathbb{Z}_2 \subseteq \mathbb{Z}_2$
&$0 \subseteq 0 \subseteq 0 \subseteq 0$
\\
$\hat{\mathcal{A}}_{0,+}^{+},\hat{\mathcal{A}}_{T/2,-}^{+}$ & AIII 
&$0 \subseteq 0 \subseteq 0 \subseteq 0$
&$0 \subseteq 0 \subseteq 0 \subseteq \mathbb{Z}$
&$0 \subseteq \mathbb{Z}_2 \subseteq \mathbb{Z}_2 \subseteq \mathbb{Z}_2$
&$0 \subseteq 0 \subseteq 0 \subseteq 2\mathbb{Z}$
\\
$\overline{\mathcal{A}}_{0}^{+},\overline{\mathcal{A}}_{T/2}^{-}$ & A 
&$0 \subseteq 0 \subseteq 0 \subseteq 0$
&$0 \subseteq 0 \subseteq \mathbb{Z}_2 \subseteq \mathbb{Z}_2$
&$0 \subseteq 0 \subseteq 0 \subseteq 0$
&$0 \subseteq 0 \subseteq 0 \subseteq 0$
\\
$\hat{\mathcal{A}}_{0,-}^{-},\hat{\mathcal{A}}_{T/2,+}^{-}$ & AIII  
&$0 \subseteq 0 \subseteq 0 \subseteq \mathbb{Z}$
&$0 \subseteq 0 \subseteq \mathbb{Z}_2 \subseteq \mathbb{Z}_2$
&$0 \subseteq 0 \subseteq 0 \subseteq 2\mathbb{Z}$
&$0 \subseteq 0 \subseteq 0 \subseteq 0$
\\
$\hat{\mathcal{A}}_{s}^{-}$ & A 
&$0 \subseteq 0 \subseteq 0 \subseteq \mathbb{Z}_2$
&$0 \subseteq 0 \subseteq 0 \subseteq 0$
&$0 \subseteq 0 \subseteq 0 \subseteq 0$
&$0 \subseteq 0 \subseteq 0 \subseteq 0$
\\
$\hat{\mathcal{A}}_{0,+}^{-},\hat{\mathcal{A}}_{T/2,-}^{-}$ & AIII  
&$0 \subseteq 0 \subseteq 0 \subseteq \mathbb{Z}_2$
&$0 \subseteq 0 \subseteq 0 \subseteq 2\mathbb{Z}$
&$0 \subseteq 0 \subseteq 0 \subseteq 0$
&$0 \subseteq 0 \subseteq 0 \subseteq \mathbb{Z}$
\\
$\overline{\mathcal{A}}_{0}^{-},\overline{\mathcal{A}}_{T/2}^{+}$ & A
&$0 \subseteq 0 \subseteq 0 \subseteq 0$
&$0 \subseteq 0 \subseteq 0 \subseteq 0$
&$0 \subseteq 0 \subseteq 0 \subseteq 0$
&$0 \subseteq 0 \subseteq \mathbb{Z}_2 \subseteq \mathbb{Z}_2$
\\
\end{tabular}
\end{ruledtabular}
\end{table*}

\begin{table*}
\caption{\label{tab:real_unitary_hoti_3}
Subgroup series $K^{(d)}\subseteq \dots \subseteq K' \subseteq K$ for 
  three-dimensional ($d=3$) anomalous Floquet HOTI/SCs with a unitary order-two space-time
symmetry/antisymmetry in real classes. The number of flipped dimensions for the symmetry/antisymmetry is denoted as
$d_{\parallel}$.
}
\begin{ruledtabular}
\centering
\begin{tabular}{cccccc}
Symmetry  &Class &$d_{\parallel}=0$  &$d_{\parallel}=1$ &$d_{\parallel}=2$ &$d_{\parallel}=3$ \\
\hline
$\text{\ensuremath{\hat{\mathcal{U}}_{0,+}^{+}}},\text{\ensuremath{\hat{\mathcal{U}}_{T/2,+}^{+}}}$ & AI
&$0 \subseteq 0 \subseteq 0 \subseteq 0$
&$0 \subseteq 0 \subseteq 0 \subseteq 0$
&$0 \subseteq 0 \subseteq 0 \subseteq 0$
&$2\mathbb{Z} \subseteq 2\mathbb{Z} \subseteq 2\mathbb{Z} \subseteq 2\mathbb{Z}$
\\
$\hat{\mathcal{U}}_{0,++}^{+},\hat{\mathcal{U}}_{T/2,+-}^{+}$ & BDI
&$0 \subseteq 0 \subseteq 0 \subseteq 0$
&$0 \subseteq 0 \subseteq 0 \subseteq 0$
&$0 \subseteq \mathbb{Z} \subseteq \mathbb{Z} \subseteq \mathbb{Z}$
&$0 \subseteq 0 \subseteq 0 \subseteq 0$
\\
$\ensuremath{\hat{\mathcal{U}}_{0,+}^{+}},\text{\ensuremath{\hat{\mathcal{U}}_{T/2,-}^{+}}}$ & D 
&$0 \subseteq 0 \subseteq 0 \subseteq 0$
&$0 \subseteq 0 \subseteq \mathbb{Z} \subseteq \mathbb{Z}$
&$0 \subseteq 0 \subseteq 0 \subseteq 0$
&$0 \subseteq 0 \subseteq 0 \subseteq 0$
\\
$\hat{\mathcal{U}}_{0,++}^{+},\hat{\mathcal{U}}_{T/2,+-}^{+}$ & DIII 
&$0 \subseteq 0 \subseteq 0 \subseteq \mathbb{Z}^2$
&$0 \subseteq 0 \subseteq \mathbb{Z}_2 \subseteq \mathbb{Z}_2$
&$0 \subseteq 0 \subseteq 0 \subseteq \mathbb{Z}$
&$0 \subseteq 0 \subseteq 0 \subseteq 0$
\\
$\ensuremath{\hat{\mathcal{U}}_{0,+}^{+}},\ensuremath{\hat{\mathcal{U}}_{T/2,+}^{+}}$ & AII 
&$0 \subseteq 0 \subseteq 0 \subseteq \mathbb{Z}_2^2$
&$0 \subseteq 0 \subseteq \mathbb{Z}_2 \subseteq \mathbb{Z}_2$
&$0 \subseteq 0 \subseteq 0 \subseteq 0$
&$4\mathbb{Z} \subseteq 4\mathbb{Z} \subseteq 2\mathbb{Z} \subseteq \mathbb{Z}$
\\
$\hat{\mathcal{U}}_{0,++}^{+},\hat{\mathcal{U}}_{T/2,+-}^{+}$ & CII 
&$0 \subseteq 0 \subseteq 0 \subseteq \mathbb{Z}_2^2$
&$0 \subseteq 0 \subseteq 0 \subseteq 0$
&$0 \subseteq 2\mathbb{Z} \subseteq 2\mathbb{Z} \subseteq \mathbb{Z}$
&$0 \subseteq 0 \subseteq 0 \subseteq \mathbb{Z}_{2}$
\\
\end{tabular}
\end{ruledtabular}
\end{table*}

A more intriguing fact regarding crystalline symmetries/antisymmetries is that they can give rise to 
boundary modes with codimension higher than one, such as corners of 2D or 3D systems, 
as well as hinges of 3D systems \cite{Benalcazar2017, Peng2017, Langbehn2017,
Benalcazar2017s, Song2017, Schindler2018, Geier2018, Khalaf2018, Khalaf2018prx}. 
Such systems are known as HOTI/SCs, in which
the existence of the high codimension gapless boundary modes 
is guarenteed when the boundaries are compatible with the crystalline symmetry/antisymmetry,
i.e. a group of boundaries with different orientations 
are mapped onto each other under the nonlocal transformation of a particular
crystalline symmetry/antisymmetry. For example, to have a HOTI/SC protected by 
inversion, one needs to create boundaries in pairs related by inversion \cite{Khalaf2018, Khalaf2018prx}. 

An additional requirement for these corner or hinge modes is that they should be intrinsic,
namely their existence should not depend on 
lattice termination, otherwise such high codimension boundary modes can be thought as
a (codimension one) boundary modes in the low dimensional system, which is then glued 
to the original boundary. In other words, an $n$th order TI/SCs has codimension-$n$ boundary
modes which cannot be destroyed through modifications of lattice terminations at the boundaries
while preserving the bulk gap and the symmetries. 
According to this definition, the tenfold-way TI/SCs are indeed intrinsic first-order TI/SCs.

In Ref.~\cite{Trifunovic2019}, a complete classification of these intrinsic corner or hinge modes
was derived and a higher-order bulk-boundary correspondence between these high codimension boundary modes and 
the topological bulk was obtained.
These were accomplished by considering a $K$ subgroup series for a $d$-dimensional crystal, 
\begin{equation}
  K^{(d)} \subseteq \dots \subseteq K'' \subseteq K' \subseteq K,
\end{equation}
where $K\equiv K^{(0)}$ is the $K$ group which classifies the bulk band structure 
of Hamiltonians with coexisting order-two symmetry/antisymmetry, defined in the previous section. 
$K^{(n)}\subseteq K$ is a subgroup excluding topological phases of order $n$
or lower, for any crystalline-symmetry compatible boundaries.
For example, $K'$ classifies the ``purely crystalline phases'' \cite{Geier2018, Trifunovic2019}, 
which exclude the tenfold-way topological phases, which are first-order topological phases
protected by nonspatial symmetries alone and have gapless modes at any codimension-one boundaries.
This purely crystalline phases can have gapless modes only when the boundary preserves the crystalline symmetry,
and the gapless modes will be gapped when the crystalline symmetry is broken.

From a boundary perspective, one can define the boundary $K$ group $\mathcal{K}'$,
which classifies the tenfold-way topological phases with gapless codimension-one boundary modes
irrespective of boundary orientations, as long as the crystal shape and lattice termination are compatible
crystalline symmetries.
According to the above definitions, $\mathcal{K}'$ can be identified as the quotient group
\begin{equation}
  \mathcal{K}'= K/K'.
\end{equation}
Generalizing this idea, a series of boundary $K$ groups denoted as $\mathcal{K}^{(n)}$ can be defined,
which classify the intrinsic $n$-th order TI/SCs with intrinsic gapless codimension-$n$ boundary modes,
when the crystal has crystalline-symmetry-compatible shape and lattice termination.
In Ref.~\cite{Trifunovic2019}, 
the authors proved the following relation,
\begin{equation}
  \mathcal{K}^{(n+1)} = K^{(n)}/K^{(n+1)},
\end{equation}
known as the higher-order bulk-boundary correspondence: an intrinsic higher-order topological phase
is uniquely associated with a topologically nontrivial bulk.
Moreover, the above equation provides a systematic way of obtaining the complete classification of intrinsic HOTI/SCs 
from $K$ subgroup series, which were computed for crystals up to three dimensions with order-two crystalline
symmetries/antisymmetries.

We can generalize these results to anomalous Floquet HOTI/SCs,
by considering unitary loops $U(\boldsymbol{k},t)$ in $d$ dimension without topological defect.
To define a $K$ subgroup series for unitary loops with an order-two space-time symmetry/antisymmetry,
one can exploit the hermitian map and introduce the $K$ groups according to their corresponding Hamiltonians
with an order-two crystalline symmetry/antisymmetry. 
One obtains that the $K$ subgroup series for each nontrivial space-time symmetry/antisymmetry 
are the same as the ones for a corresponding static order-two crystalline symmetry/antisymmetry, 
according to the substitution rules summarized in Sec.~\ref{sec:substitution} and Table~\ref{tab:summary}.
On the other hand, the $K$ groups are the same for unitary loops and Hamiltonians when static order-two symmetries/antisymmetries
are considered. 
Using the results from Ref.~\cite{Trifunovic2019},
we present the $K$ subgroup series for unitary loops with an order-two space-time symmetry/antisymmetry
in Tables~\ref{tab:complex_unitary_hoti_2}--\ref{tab:real_unitary_hoti_3}, for systems up to three dimensions.
In these tables, we use the notation $G^2$ to denote  $G \oplus G$, with $G = \mathbb{Z}, 2\mathbb{Z}, \mathbb{Z}_2$.
One also notices that the largest $K$ group $K^{(0)}$ in the series is actually the ones shown in 
Tables of the supplemental material \cite{suppl}. 
The classification of intrinsic codimension-$n$ anomalous Floquet boundary modes is then given by the quotient
$\mathcal{K}^{(n)} = K^{(n-1)}/K^{(n)}$.

\section{Floquet HOTI/SCs in frequency domain \label{sec:FHOTI_frequency_domain}} 
In this section, we take an alternative route to connect a Floquet HOTI/SC
with a nontrivial space-time symmetry/antisymmetry, to a static HOTI/SC
with a corresponding crystalline symmetry/antisymmetry.
This connection is based on the frequency-domain formulation of the Floquet problem \cite{Rudner2013},
which provides a more intuitive perspective to the results obtained by $K$ theory.

\subsection{Frequency-domain formulation \label{sec:frequency_domain}}
In the frequency-domain formulation of the Floquet problem,
the quasienergies are obtained by diagonalizing the enlarged Hamiltonian
\begin{equation}
\mathscr{H}(\boldsymbol{k},\boldsymbol{r})=\left(\begin{array}{ccccc}
\ddots\\
 & h_{0}+\omega & h_{1} & h_{2}\\
 & h_{1}^{\dagger} & h_{0} & h_{1}\\
 & h_{2}^{\dagger} & h_{1}^{\dagger} & h_{0}-\omega\\
 &  &  &  & \ddots,
\end{array}\right)
\end{equation}
where the matrix blocks are given by
\begin{equation}
	h_n(\boldsymbol{k},\boldsymbol{r}) =  \frac{1}{T} \int_0^T dt\, H(\boldsymbol{k},t)  e^{-in\omega t}.
\end{equation}

Here, the appearance of the infinite dimensional matrix $\mathscr{H}$ can be subtle, and
should be defined more carefully.
Since later we would like to discuss the gap at $\epsilon_{\rm gap}=\omega/2$, 
we will assume that the infinite dimensional matrix $\mathscr{H}$ should be obtained as taking the limit
$n \to \infty$ of a finite dimensional matrix 
whose diagonal blocks are given from $h_0+n\omega$ to $h_{0}-(n-1)\omega$, with $n$ a positive
integer. With this definition, $\omega/2$ will be the particle-hole/chiral symmetric energy
whenever the system has particle-hole/chiral symmetries.

As a static Hamiltonian, 
$\mathscr{H}(\boldsymbol{k},\boldsymbol{r})$ has the same nonspatial symmetries as the original $H(\boldsymbol{k},\boldsymbol{r},t)$
does. Indeed, one can define the effective time-reversal $\mathscr{T}$, particle-hole $\mathscr{C}$ and chiral $\mathscr{S}$ symmetries
for the enlarged Hamiltonian $\mathscr{H}(\boldsymbol{k},\boldsymbol{r})$ as
\begin{align}
&\mathscr{T}=\left(\begin{array}{ccccc}
\ddots\\
& \hat{\mathcal{T}}\\
&  & \hat{\mathcal{T}}\\
&  &  & \hat{\mathcal{T}}\\
 &  &  &  & \ddots
\end{array}\right),
\end{align}
\begin{align}
&\mathscr{C}=\left(\begin{array}{ccccc}
 &  &  &  & \dots\\
 &  &  & \hat{\mathcal{C}}\\
 &  & \hat{\mathcal{C}}\\
 & \hat{\mathcal{C}}\\
\dots
\end{array}\right),
\end{align}
\begin{align}
&\mathscr{S}=\left(\begin{array}{ccccc}
 &  &  &  & \dots\\
 &  &  & \hat{\mathcal{S}}\\
 &  & \hat{\mathcal{S}}\\
 & \hat{\mathcal{S}}\\
\dots
\end{array}\right).
\end{align}

On the other hand, when the original $H(\boldsymbol{k},\boldsymbol{r},t)$ has a nontrivial space-time symmetry/antisymmetry, the enlarged Hamiltonian
$\mathscr{H}(\boldsymbol{k},\boldsymbol{r})$ will acquire the spatial (crystalline) symmetry/antisymmetry inherited from the spatial part of the space-time
symmetry/antisymmetry. 

Let us first consider $\hat{\mathcal{U}}_{T/2}$ defined in Eq.~(\ref{eq:unitary_symmetry}) for $s=T/2$, 
which is an unitary operation together with a half-period time translation. 
Since 
\begin{equation}
\hat{\mathcal{U}}_{T/2}h_{n}(\boldsymbol{k},\boldsymbol{r})\hat{\mathcal{U}}_{T/2}^{-1}
=(-1)^{n} h_{n}(-\boldsymbol{k}_{\parallel}, \boldsymbol{k}_{\perp}, -\boldsymbol{r}_{\parallel},\boldsymbol{r}_{\perp}),
\end{equation}
the enlarged Hamiltonian thus respects a unitary spatial symmetry defined by
\begin{equation}
	\mathscr{U}\mathscr{H}(\boldsymbol{k},\boldsymbol{r})\mathscr{U}^{-1}=
	\mathscr{H}(-\boldsymbol{k}_{\parallel}, \boldsymbol{k}_{\perp}, -\boldsymbol{r}_{\parallel},\boldsymbol{r}_{\perp}),
\end{equation}
where the unitary operator
\begin{equation}
\mathscr{U} = \left(\begin{array}{ccccc}
\ddots\\
& \hat{\mathcal{U}}_{T/2}\\
&  & -\hat{\mathcal{U}}_{T/2}\\
&  &  & \hat{\mathcal{U}}_{T/2}\\
 &  &  &  & \ddots
\end{array}\right)
\end{equation}
is inherited from $\hat{\mathcal{U}}_{T/2}$.

Next, we consider $\overline{\mathcal{A}}_{T/2}$. 
Since 
\begin{equation}
\overline{\mathcal{A}}_{T/2}h_{n}(\boldsymbol{k},\boldsymbol{r})\overline{\mathcal{A}}_{T/2}^{-1}
=-(-1)^{n}h_{-n}(\boldsymbol{k}_{\parallel},-\boldsymbol{k}_{\perp},-\boldsymbol{r}_{\parallel},\boldsymbol{r}_{\perp}),
\end{equation}
one can define
\begin{equation}
\overline{\mathscr{A}}=\left(\begin{array}{ccccc}
 &  &  &  & \iddots\\
  &  &  & -i\overline{\mathcal{A}}_{T/2}\\
   &  & i\overline{\mathcal{A}}_{T/2}\\
    & -i\overline{\mathcal{A}}_{T/2}\\
    \iddots
    \end{array}\right)
\end{equation}
such that
\begin{equation}
\overline{\mathscr{A}}\mathscr{H}(\boldsymbol{k},\boldsymbol{r})\overline{\mathscr{A}}^{-1} = 
-\mathscr{H}(\boldsymbol{k}_{\parallel},-\boldsymbol{k}_{\perp},-\boldsymbol{r}_{\parallel},\boldsymbol{r}_{\perp}).
\end{equation}

We now consider symmetry operators $\hat{\mathcal{A}}_{T/2}$ and $\overline{\mathcal{U}}_{T/2}$, 
for symmetry classes other than A, C, and D.
For $\hat{\mathcal{A}}_{T/2}$, we have
\begin{equation}
\hat{\mathcal{A}}_{T/2}h_{n}(\boldsymbol{k},\boldsymbol{r})\hat{\mathcal{A}}_{T/2}^{-1}
= (-1)^{n}h_{n}(\boldsymbol{k}_{\parallel},-\boldsymbol{k}_{\perp},-\boldsymbol{r}_{\parallel},\boldsymbol{r}_{\perp}).
\end{equation}
Thus, the enlarged Hamiltonian $\mathscr{H}(\boldsymbol{k},\boldsymbol{r})$ also has an antiunitary spatial symmetry
inherited from $\hat{\mathcal{A}}_{T/2}$, given by
\begin{equation}
\mathscr{A}\mathscr{H}(\boldsymbol{k},\boldsymbol{r})\mathscr{A}^{-1}
= \mathscr{H}(\boldsymbol{k}_{\parallel},-\boldsymbol{k}_{\perp},-\boldsymbol{r}_{\parallel},\boldsymbol{r}_{\perp}),
\end{equation}
where the antiunitary operator
\begin{equation}
\mathscr{A}=\left(\begin{array}{ccccc}
    \ddots\\
    & \hat{\mathcal{A}}_{T/2}\\
    &  & -\hat{\mathcal{A}}_{T/2}\\
    &  &  & \hat{\mathcal{A}}_{T/2}\\
    &  &  &  & \ddots
    \end{array}\right).
\end{equation}

Finally, for $\overline{\mathcal{U}}_{T/2}$, it satisfies
\begin{equation}
\overline{\mathcal{U}}_{T/2}h_{n}(\boldsymbol{k},\boldsymbol{r})\overline{\mathcal{U}}_{T/2}^{-1}
=-(-1)^{n}h_{-n}(-\boldsymbol{k}_{\parallel},\boldsymbol{k}_{\perp},-\boldsymbol{r}_{\parallel},\boldsymbol{r}_{\perp}).
\end{equation} 
Hence, if we define
\begin{equation}
\overline{\mathscr{U}}=\left(\begin{array}{ccccc}
    &  &  &  & \iddots\\
    &  &  & -i\overline{\mathcal{U}}_{T/2}\\
    &  & i\overline{\mathcal{U}}_{T/2}\\
    & -i\overline{\mathcal{U}}_{T/2}\\
    \iddots
    \end{array}\right),
\end{equation}
the enlarged Hamiltonian will satisfy 
\begin{equation}
\overline{\mathscr{U}}\mathscr{H}(\boldsymbol{k},\boldsymbol{r})\overline{\mathscr{U}}^{-1} 
= -\mathscr{H}(-\boldsymbol{k}_{\parallel},\boldsymbol{k}_{\perp},-\boldsymbol{r}_{\parallel},\boldsymbol{r}_{\perp}).
\end{equation}

\subsection{Harmonically driven systems \label{sec:harmonically_driven_models}} 
To simplify the discussion, it is helpful to restrict ourselves to a specific class 
of periodically driven systems, the harmonically driven ones, whose
Hamiltonians have the following form
\begin{equation}
H(\boldsymbol{k},t) = h_{0}(\boldsymbol{k}) + h_{1}(\boldsymbol{k})e^{i\omega t} +
h_{1}^\dagger(\boldsymbol{k})e^{-i\omega t}.
\label{eq:harmonic_hamiltonian}
\end{equation}

To discuss the band topology around at $\epsilon_{\rm gap}=\omega/2$, 
one can further truncate the enlarged Hamiltonian $\mathscr{H}$ to the $2 \times 2$ block, 
containing two Floquet zones with energy difference $\omega$, namely
\begin{equation}
\mathscr{H}(\boldsymbol{k}) = 
\left(\begin{array}{cc}
h_{0}(\boldsymbol{k})+\frac{\omega}{2} & h_{1}(\boldsymbol{k})\\
h_{1}^{\dagger}(\boldsymbol{k}) & h_{0}(\boldsymbol{k})-\frac{\omega}{2}
\end{array}\right) + \frac{\omega}{2}\rho_0,
\label{eq:effective_2by2}
\end{equation}
where $\rho_{0}$ is the identity in the two Floquet-zone basis. For later convenience,
we use $\rho_{x,y,z}$ to denote the Pauli matrices this basis.
Since the last term in Eq.~(\ref{eq:effective_2by2}) is a shift in energy by $\omega/2$, 
we have a Floquet HOTI/SC at $\epsilon_{\rm gap}=\omega/2$ if and only if
the first term in Eq.~(\ref{eq:effective_2by2}) is a static HOTI/SC. 

When restricted to the two Floquet-zone basis, the nonspatial symmetries can be conveniently writtens
\begin{equation}
\mathscr{T} = \rho_0\hat{\mathcal{T}},\  \mathscr{C} = \rho_x\hat{\mathcal{C}}, \  \mathscr{S} =
\rho_x\hat{\mathcal{S}}.
\label{eq:effective_AZ}
\end{equation}
The spatial symmetries/antisymmetries for $\mathscr{H}$, which are inherited from the space-time symmetries/antisymmetries, 
can also be written simply as
\begin{align}
&\mathscr{U} = \rho_z\hat{\mathcal{U}}_{T/2},\quad 
\overline{\mathscr{A}} = \rho_y\overline{\mathcal{A}}_{T/2}, \nonumber \\
&\mathscr{A} = \rho_z\hat{\mathcal{A}}_{T/2},\quad
\overline{\mathscr{U}} = \rho_y\overline{\mathcal{U}}_{T/2}.
\label{eq:effective_spatial}
\end{align}

From these relations, one arrives at the same results as the ones from $K$ theory in the previous sections. 
When a spatial symmetry $\mathscr{O}$, with $\mathscr{O}=\mathscr{U},\mathscr{A}$,
coexists with the particle-hole or/and chiral symmetry operators $\mathscr{C}$,
$\mathscr{S}$, $\mathscr{O}$ will commute or anticommute with $\mathscr{C}$ or/and $\mathscr{S}$.
Let us write 
\begin{align}
\mathscr{O}\mathscr{C} &= \chi_C \mathscr{C}\mathscr{O},\\
\mathscr{O}\mathscr{S} &= \chi_S \mathscr{S}\mathscr{O},
\end{align}
with $\chi_C, \chi_S=\pm 1$. Because of the additional Pauli matrices $\rho_{x,y,z}$ in Eqs.~(\ref{eq:effective_AZ}) and
(\ref{eq:effective_spatial}), we have $\eta_C = -\chi_C$ and $\eta_S= -\chi_S$.

For $\mathscr{O}$, the commutation relation with respect to the time-reversal symmetry does not vary,
whereas for a spatial antisymmetry $\overline{\mathscr{O}}$, 
with $\overline{\mathscr{O}}=\overline{\mathscr{U}},\overline{\mathscr{A}}$,
coexisting with the time-reversal symmetry, the commutation relation with respect to the latter does
get switched. Let us write
\begin{equation}
\overline{\mathscr{O}}\mathscr{T} = \chi_T \mathscr{T}\overline{\mathscr{O}},
\end{equation}
with $\chi_T = \pm 1$, then we would have
\begin{equation}
\eta_{T} = -\chi_{T},
\end{equation}
because $\rho_{y}$ is imaginary.
Because of this, we can also obtain $\overline{\mathscr{A}}^2 = -\overline{\mathcal{A}}_{T/2}^2$. 

\section{Model Hamiltonians for Floquet HOTI/SCs \label{sec:models}}
In this section, we introduce model Hamiltonians, which
are simple but still sufficiently general, 
for Floquet HOTI/SCs in all symmetry classes.
Particularly, we consider harmonically driven Floquet HOTI/SCs 
Hamiltonians with a given nontrivial space-time symmetry/antisymmetry,
realized by $\hat{\mathcal{U}}_{T/2}$, $\overline{\mathcal{A}}_{T/2}$,
$\hat{\mathcal{A}}_{T/2}$, or $\overline{\mathcal{U}}_{T/2}$.
One should notice that the latter two symmetries/antisymmetries are only available when the system is not in classes A,C or D,
because in these classes, the symmetries with $s=0$ and $T/2$ are the same up to redefining the origin of
time coordinate. 

\subsection{Hamiltonians}
The harmonically driven Floquet HOTI/SCs in $d$-dimension to be constructed have Bloch Hamiltonians of the following general form
\begin{equation}
  H(\boldsymbol{k},t,m) = d_{0}(\boldsymbol{k},m)\Gamma_0 + \sum_{j = 1}^{d} d_{j}(\boldsymbol{k})\Gamma_j \cos(\omega t),
\end{equation}
where
\begin{align}
  d_{0}(\boldsymbol{k},m) &= m+\sum_{j=1}^{d} (1-\cos k_j) + \dots\\
  d_{j}(\boldsymbol{k}) &= \sin k_j, \quad j=1,\dots d,
\end{align}
and $\{\Gamma_{i}, \Gamma_{j}\} = 2\delta_{ij}\mathbb{I}$, with $\mathbb{I}$ the identity matrix. 
Here ``$\dots$'' represents $\boldsymbol{k}$-independent symmetry allowed perturbations that will in general gap out unprotected gapless modes. 

One can further choose a representation of these $\Gamma_{j}$s such that 
\begin{equation}
  \Gamma_{0}=\left(\begin{array}{cc}
\mathbb{I} & 0\\
0 & -\mathbb{I}
\end{array}\right) = \tau_{z},
\quad 
\Gamma_{j}=\left(\begin{array}{cc}
0 & \gamma_j\\
\gamma_j^{\dagger}  &0 
\end{array}\right),
\label{eq:gamma_representation}
\end{equation}
for $j = 1,\dots, d$.
By the transformation properties of the symmetry/antisymmetry operators, 
we have, in this representation, $\hat{\mathcal{T}}$, $\hat{\mathcal{U}}_{T/2}$ and $\hat{\mathcal{A}}_{T/2}$
are block diagonal, namely they act independently on the two subspaces with $\tau_{z} = \pm 1$, 
whereas the operators $\hat{\mathcal{C}}$, $\hat{\mathcal{S}}$, $\overline{\mathcal{U}}_{T/2}$ and 
$\overline{\mathcal{A}}_{T/2}$ are block off-diagonal, which couple the two subspaces. 

In this representation, the enlarged Hamiltonian $\mathscr{H}(\boldsymbol{k})$ truncated to two Floquet zones,
up to the constant shift $\omega/2$, can be decoupled into two sectors with $\rho_z\tau_z = \pm 1$. 
Hence, one can write it as a direct sum
\begin{equation}
  \mathscr{H}(\boldsymbol{k})=h(\boldsymbol{k},m+\omega/2) \oplus h(\boldsymbol{k},m-\omega/2),
  \label{eq:direct_sum}
\end{equation}
with 
\begin{equation}
  h(\boldsymbol{k},m) = d_{0}(\boldsymbol{k},m)\tilde{\Gamma}_0 + \sum_{j=1}^{d}d_{j}(\boldsymbol{k})\tilde{\Gamma}_j.
  \label{eq:static_models}
\end{equation}
Here the matrices $\tilde{\Gamma}_{j}$s have a two-by-two block structure when restricting to the $\rho_z\tau_z = \pm 1$
sectors of $\mathscr{H}(\boldsymbol{k})$. If we abuse the notation by still using $\tau_{x,y,z}$ for 
this two-by-two degree of freedom, we can identify $\tilde{\Gamma}_{j}= \Gamma_j$, for $j=0,\dots,d$.

It is straightforward to verify that the static Hamiltonian $h(\boldsymbol{k},m)$ respects the same nonspatial
symmetries as the harmonically driven Hamiltonian $H(\boldsymbol{k},t,m)$ does, with the same symmetry operators.
Moreover, if $H(\boldsymbol{k},t,m)$ respects a nontrivial space-time symmetry, realized by $\hat{\mathcal{U}}_{T/2}$ or $\hat{\mathcal{A}}_{T/2}$, 
then $h(\boldsymbol{k},m)$ will respect a spatial symmetry, realized by $\Gamma_0\hat{\mathcal{U}}_{T/2}$ or
$\Gamma_0\hat{\mathcal{A}}_{T/2}$, respectively.
However, if $H(\boldsymbol{k},t,m)$ respects a nontrivial space-time antisymmetry, 
realized by $\overline{\mathcal{U}}_{T/2}$ or $\overline{\mathcal{A}}_{T/2}$, 
then $h(\boldsymbol{k},m)$ will respect a spatial antisymmetry, realized by $-i\Gamma_0\hat{\mathcal{U}}_{T/2}$ or
$-i\Gamma_0\hat{\mathcal{A}}_{T/2}$, respectively.
These relations can be worked out by using the block diagonal or off-diagonal properties of the 
operators of space-time symmetries/antisymmetries, as well as the relations in
Eq.~(\ref{eq:effective_spatial}).

Thus, we have established a mapping between harmonically driven Hamiltonians $H(\boldsymbol{k},t,m)$ and static Hamiltonians
$h(\boldsymbol{k},m)$, as well as their transformation properties under symmetry/antisymmetry operators. 
On the other hand, $h(\boldsymbol{k},m)$ given in Eq.~(\ref{eq:static_models}) are well studied
models for static HOTI/SCs \cite{Geier2018, Trifunovic2019}.
It is known that for $-2<m<0$,  
the Hamiltonian $h(\boldsymbol{k},m)$ is in the topological phases (if the classification is nontrivial), 
whereas for $m>0$ the Hamiltonian is in a trivial phase. 
A topological phases transition occurs at $m=0$ with the band gap closing at $\boldsymbol{k}=0$.

Since the enlarged Hamiltonian $\mathscr{H}(\boldsymbol{k})$, up to a constant $\omega/2$ shift, can be written as a direct sum of 
$h(\boldsymbol{k},m\pm\omega/2)$, the static Hamiltonian $\mathscr{H}(\boldsymbol{k})$ will be in the topological phase
(with chemical potential inside the gap at $\omega/2$) if $-2<m-\omega/2<0$ and $m+\omega/2>0$.
This is also the condition when $H(k,t,m)$ is in a Floquet topological 
phase at $\epsilon_{\rm gap} = \omega/2$.

\subsection{Symmetry/antisymmetry-breaking mass terms \label{sec:symmetry_mass_terms}}
Let us consider $-2<m-\omega/2<0$ and $m+\omega/2>0$.
In this parameter regime, $h(\boldsymbol{k},m+\omega/2)$
is always in a trivial insulating phase, whereas
$h(\boldsymbol{k},m-\omega/2)$ is in a nontrivial
topological phase, if there exists no mass term $M$
that respect the nonspatial symmetries, as well as
the spatial symmetry/antisymmetry inherited from the space-time symmetry/antisymmetry of $H(\boldsymbol{k},t,m)$.
Here, the mass term in addition satisfies $M^2=1$, $M=M^{\dagger}$ and $\{M,h(\boldsymbol{k},m)\}=0$.
Such a mass term will gap out any gapless states that may appear in a finite-size system whose bulk 
is given by $h(\boldsymbol{k},m-\omega/2)$.
When $M$ exist, one can define a term $M\cos(\omega t)$ respecting all nonspatial symmetries and
the space-time symmetry/antisymmetry of $H(\boldsymbol{k},m,t)$, and it will
gap out any gapless Floquet boundary modes at quasienergy $\epsilon_{\rm gap} = \omega/2$. 

If no mass term $M$, which satisfies only the nonspatial symmetries irrespective of the spatial symmetry/antisymmetry, 
exists, then $h(\boldsymbol{k},m-\omega/2)$ ($H(\boldsymbol{k},m,t)$) is in the static (Floquet) tenfold-way topologogical phases,
as it remains nontrivial even when the spatial (space-time) symmetry/antisymmetry is broken.  
Thus, the tenfold-way phases are always first-order topological phases. 
However, if such a $M$ exists, $h(\boldsymbol{k},m-\omega/2)$ ($H(\boldsymbol{k},m,t)$) 
describes a static (Floquet) ``purely crystalline'' topological phase,
which can be higher-order topological phases, and the topological protection relies
on the spatial (space-time) symmetry/antisymmetry.  

As pointed out in Ref.~\cite{Trifunovic2019}, 
several mutually anticommuting spatial-symmetry/antisymmetry-breaking mass terms $M_{l}$ 
can exist for $h(\boldsymbol{k},m-\omega/2)$, where $M_{l}$ also anticommutes with $h$.
Furthermore, if $h$ has the minimum possible dimension for a given  ``purely crystalline''
topological phase, then the mass terms $M_{l}$ all
anticommute (commute) with the spatial symmetry (antisymmetry) operator 
of $h(\boldsymbol{k},m-\omega/2)$.  
In this case, one can relate the number of these mass terms $M_{l}$
and the order of the topological phase \cite{Trifunovic2019}: 
When $n$ mass terms $M_{l}$ exist, with $l=1,\dots,n$, 
boundaries of codimension up to $\min(n,d_{\parallel})$ are gapped,
and one has a topological phase of order $\min (n+1,d_{\parallel}+1)$ 
if $\min(n+1,d_{\parallel}+1) \leq d$. 
However,  if $\min(n+1,d_{\parallel}+1) > d$, the system does not
support any protected boundary modes at any codimension. 
See Ref.~\cite{Trifunovic2019}, or Appendix~\ref{app:order_mass_terms}
for the proof of this statement. 

Hence,  the order of the Floquet topological phase described by $H(\boldsymbol{k},t,m)$ 
is reflected in the number of symmetry/antisymmetry-breaking mass terms $M_{l}$, 
due to the mapping between $H(\boldsymbol{k},t,m)$ and
$h(\boldsymbol{k},m-\omega/2)$. In the following, we explicitly
construct model Hamiltonians for Floquet HOTI/SCs with a 
given space-time symmetry/antisymmetry.

\subsection{First-order phase in $d_{\parallel}=0$ family}
When $d_{\parallel} = 0$, the symmetries/antisymmetries are onsite. From
Tables~\ref{tab:complex_unitary_hoti_2}--\ref{tab:real_unitary_hoti_3}, we see that
the onsite symmetries/antisymmetries only give rise to first-order TI/SCs, since
only the $K^{(0)}$ in the subgroup series can be nonzero.
This can also be understood from the fact that $\min(n+1,d_{\parallel}+1) = 1$ in this case. 
We will in the following provide two examples in which
we have anomalous Floquet boundary modes of codimension one which are protected by the unitary onsite space-time
symmetry.

\subsubsection{2D system in class AII with $\hat{\mathcal{U}}_{T/2, -}^{+}$}
The simplest static topological insulator protected by unitary onsite symmetry is the 
quantum spin Hall insulator with additional two-fold spin rotation symmetry around the $z$ axis \cite{Shiozaki2014}.
This system is in class AII with time-reversal symmetry $\hat{\mathcal{T}}^2 = -1$.  
It is known that either a static or a Floquet system of class AII in 2D will have a $\mathbb{Z}_{2}$ topological invariant \cite{Chiu2016, Roy2017}.
However, with a static unitary $d_{\parallel}=0$ symmetry (such as a two-fold spin rotation symmetry),
realized by the operator $\hat{\mathcal{U}}^{+}_{0,-}$ that squares to one and anticommutes with the
time-reversal symmetry operator, a $K^{(0)}= \mathbb{Z}$ topological invariant known as the spin Chern number can be defined. 
In fact, such a $\mathbb{Z}$ topological invariant (see Table~\ref{tab:real_unitary_hoti_2}) can also appear due to the
existence of space-time symmetry realized by $\hat{\mathcal{U}}_{T/2,-}^{+}$ at quasienergy gap $\epsilon_{\rm gap}
=\omega/2$.

A lattice model that realizes a spin Chern insulator can be defined using the following Bloch Hamiltonian
\begin{align}
  h(\boldsymbol{k},m) &= (m+2-\cos k_{x} - \cos k_{y})\tau_z \nonumber \\ 
	& + (\sin k_{x}\tau_x s_z + \sin k_{y}\tau_{y}),
  \label{eq:hamiltonian_spin_chern}
\end{align}
where $s_{x,y,z}$ and $\tau_{x,y,z}$ are two sets of Pauli matrices for spins and orbitals.
This Hamiltonian has time-reversal symmetry realized by $\hat{\mathcal{T}} = -i s_{y}\hat{\mathcal{K}}$ 
as well as the unitary symmetry realized by operator $\hat{\mathcal{U}}_{0,-}^{+} = s_{z}$.
When we choose an open boundary condition along $x$ while keep the $y$ direction with 
a periodic boundary condition, there will be gapless helical edge states inside the bulk gap propagating along the $x$
edge at $k_y=0$ for $-2<m<0$. 
    
The corresponding harmonically driven Hamiltonian can be written as
\begin{align}
  H(\boldsymbol{k},t,m) &= (m+2-\cos k_{x} - \cos k_{y})\tau_z \nonumber \\ 
  & + (\sin k_{x}\tau_x s_z + \sin k_{y}\tau_{y})\cos(\omega t),
\end{align}
where the time-reversal and the
half-period time translation onsite symmetry operators are defined as
$\hat{\mathcal{T}}=-is_{y}\hat{\mathcal{K}}$ and $\hat{\mathcal{U}}_{T/2,-}^{+}=s_{z}\tau_z$
respectively.

When $-2<m-\omega/2<0$ and $m+\omega/2>0$ are satisfied,
this model supports gapless helical edge states at $k_{y}=0$ inside the 
bulk quasienergy gap $\epsilon_{\rm gap}=\omega/2$ when the x direction
has an open boundary condition. 
Furthermore, such gapless Floquet edge modes persist as one introduces more
perturbations that preserve the time-reversal and the $\hat{\mathcal{U}}_{T/2,-}^{+}$ symmetry. 

\subsubsection{2D system in class D with $\hat{\mathcal{U}}_{T/2,-}^{+}$}
For 2D, either static or Floquet, superconductors in class D with no additional symmetries, 
the topological invariant is $\mathbb{Z}$ given by the Chern number of the Bogoliubov--de Gennes (BdG)
bands. When there exists a static unitary $d_{\parallel}=0$ symmetry, realized by
$\hat{\mathcal{U}}^{+}_{0,+}$ which commutes with the particle-hole symmetry operator, 
the topological invariant instead becomes to $K^{(0)} = \mathbb{Z}\oplus \mathbb{Z}$,
see Table~\ref{tab:real_unitary_hoti_2}.
The same topological invariant can also be obtained from a
space-time unitary symmetry realized by $\hat{\mathcal{U}}_{T/2,-}^{+}$, which anticommutes with the particle-hole symmetry operator.
In the following, we construct a model Hamiltonian for such a Floquet system. 

Let us start from the static 2D Hamiltonian in class D given by
\begin{align}
  h(\boldsymbol{k},m) &= (m+2-\cos k_x -\cos k_y + b s_z)\tau_z  \nonumber \\
&+ \sin k_x s_z \tau_x + \sin k_y \tau_y,
\label{eq:static_D_del0}
\end{align}
with particle-hole symmetry and the unitary onsite symmetries
realized by $\hat{\mathcal{C}} = \tau_{x}\hat{\mathcal{K}}$ and $\hat{\mathcal{U}}_{0,+}^+ = s_z$,
where $\tau_{x,y,z}$ are the Pauli matrices for the Nambu space. 
Here, the unitary symmetry can be thought as the mirror reflection with respect to the $xy$ plane,
and $bs_{z}$ is the Zeeman term which breaks the time-reversal symmetry.

The $\mathbb{Z}\oplus \mathbb{Z}$ structure is coming from the fact that
$\hat{\mathcal{U}}_{0,+}^{+}$, $\hat{\mathcal{C}}$ and $h(\boldsymbol{k},m)$ can be 
simultaneously block diagonalized, according to the $\pm 1$ eigenvalues of $\hat{\mathcal{U}}_{0,+}^{+}$. 
Each block is a class D system with no additional symmetries, and thus has a $\mathbb{Z}$ topological invariant. 
Since the two blocks are independent, we have the topological invariant of the system should 
be a direct sum of the topological invariant for each block, leading to $\mathbb{Z} \oplus \mathbb{Z}$. 

The harmonically driven Hamiltonian with a unitary space-time onsite symmetry realized by
$\hat{\mathcal{U}}_{T/2,-}^{+} = s_z\tau_z$ can be written as
\begin{align}
  H(\boldsymbol{k},t,m) &= (m+2-\cos k_x -\cos k_y + b s_z)\tau_z \nonumber  \\
  &+ (\sin k_x s_z\tau_x - \sin k_y \tau_y)\cos(\omega t).
\end{align}
The particle-hole symmetry operator for this Hamiltonian is $\hat{\mathcal{C}} = \tau_{x}\hat{\mathcal{K}}$. 

\subsection{Second-order phase in $d_\parallel=1$ family}
When a $d_{\parallel}=1$ space-time symmetry/antisymmetry is present, 
the system can be at most a second-order topological phase, since
the order is given by $\min(n+1,d_{\parallel}+1)\leq 2$.
Note that the unitary symmetry in this case is the so-called time-glide symmetry,
which has been already discussed thoroughly in Refs.~\cite{Morimoto2017, Peng2019}, we will
in the following construct models for second-order topological phases 
with antiunitary symmetries, as well as models with unitary antisymmetries. 

\subsubsection{2D system in class AIII with $\hat{\mathcal{A}}_{T/2,-}^{+}$}
For 2D systems in class AIII without any additional symmetries, the topological classification is trivial,
since the chiral symmetry will set the Chern number of the occupied bands to zero. 
However, in Table~\ref{tab:complex_antiunitary_hoti_2}, we see that 
when the 2D system has an antiunitary symmetry realized by either $\hat{\mathcal{A}}_{0,+}^{+}$ or $\hat{\mathcal{A}}_{T/2,-}^{+}$, 
the $K$ subgroup series is $0 \subseteq \mathbb{Z}_2 \subseteq \mathbb{Z}_2$.

Let us first understand the $K^{(0)} = \mathbb{Z}_{2}$ classification in the case of $\hat{\mathcal{A}}_{0,+}^{+}$
in a static system with Hamiltonian $h(k_{x},k_{y})$. Let us assume that $\hat{\mathcal{A}}_{0,+}^{+}$
corresponds to the antiunitary reflection about the $x$ axis, then we have
\begin{equation}
\hat{\mathcal{A}}_{0,+}^{+} h(k_x,k_y) (\hat{\mathcal{A}}_{0,+}^{+})^{-1} = h(k_x,-k_y).
\end{equation} 
On the other hand, the chiral symmetry imposes the following condition
\begin{equation}
\hat{\mathcal{S}}h(k_x,k_y) \hat{\mathcal{S}}^{-1}= -h(k_x,k_{y}).
\end{equation}
Thus, if we regard $k_{x}\in S^{1}$ as a cyclic parameter, then at every $k_{x}$, 
$h(k_{x},k_{y})$ as a function of the Bloch momentum $k_{y}$ is actually a 1D system in class BDI.
Thus, the topological classification in this case is the same as the one
for a topological pumping for a 1D system in class BDI described by a Hamiltonian $h'(k,t)$, with momentum $k$ and 
periodic time $t$. This gives rise to a $\mathbb{Z}_{2}$ topological invariant, corresponding to 
either the fermion parity has changed or not after an adiabatic cycle \cite{Teo2010}, when the 1D system has an open
boundary condition. Since the bulk is gapped at any $t$, such a fermion pairty switch is allowed only when
the boundary becomes gapless at some intermediate time $t$. 
Since our original Hamiltonian $h(k_x,k_y)$ is related to $h'(k,t)$ by replacing $k\leftrightarrow k_y$ and $t
\leftrightarrow k_x$, a nontrivial phase for $h(k_x,k_y)$ implies the existance of a counter propagating edge modes on the 
$x$ edge when we choose an open boundary condition along $y$.

Let us understand the pure crystalline classification $K' = \mathbb{Z}_2$. 
One can consider the edge Hamiltonian for a pair of counter propagating gapless mode on the 
edge parallel to $x$ as $H_{\rm edge} = k_{x}\sigma_z$, 
with $\hat{S} = \sigma_x$ and $\hat{\mathcal{A}}_{0,+}^{+} = \hat{\mathcal{K}}$. 
This pair of gapless mode cannot be gapped by any mass term. 
However, if there exist two pairs of gapless modes, whose Hamiltonian can be written as $H_{\rm edge} = k_x \tau_0\sigma_z$, 
one can then add a mass term $m \tau_y\sigma_y$ to $H_{\rm edge}$ to gap it out.
On the other hand, if the edge does not preserve the antiunitary symmetry given by $\hat{\mathcal{A}}_{0,+}^{+}$, then 
a mass term $m\sigma_y$ can be added to gap out a single pair of gapless mode, which implies that there is no
intrinsic codimension-one boundary modes. Thus, $K'= \mathbb{Z}_2$, and $\mathcal{K}' =0$. 

Instead of intrinsic codimension-one boundary modes, the system supports intrinsic codimension-two boundary modes, 
implying it as a second-order TI.  If one creates a corner that is invariant under the reflection $x \to -x$, 
this corner will support a codimension-two zero mode, with a $\mathcal{K}'' = K'/K'' = \mathbb{Z}_{2}$ classification.

\begin{figure}[t]
	\centering
	\includegraphics[width=0.45\textwidth]{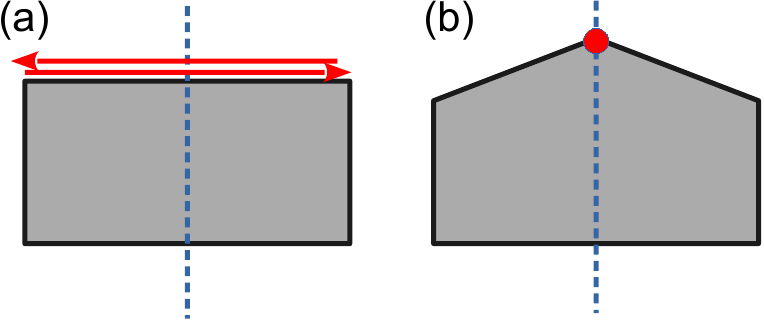}
	\caption{(a) Gapless modes at a reflection invariant edge. (b) Corner modes at a reflection invariant corner.
	The dashed line indicates the reflection (time-glide) plane.}
	\label{fig:2d_distort}
\end{figure}

An explicit Hamiltonian that realizes this phases can have the following form
\begin{align}
  h(\boldsymbol{k},m) &= (m+2-\cos k_x \cos k_y)\tau_z + \sin k_x\tau_x\sigma_x  \nonumber \\
&+ \sin k_y\tau_y + b\tau_z\sigma_z,
\end{align}
where $\tau_{x,y,z}$ and $\sigma_{x,y,z}$ are two sets of Pauli matrices, and the parameter $b$, which
gaps out the $y$ edge, is numerically small.  
One can show that this Hamiltonian has desired chiral and antiunitary reflection symmetries
given by $\hat{\mathcal{S}} = \tau_x\sigma_z$ and $\hat{\mathcal{A}}_{0,+}^{+} = \hat{\mathcal{K}}$, respectively.
When $-2<m<-0$, there are counter propagating edge modes on each $x$ edge at momentum $k_{x}=0$.
On the other hand, a corner, which is invariant under reflection $x \to -x$, will bound a zero mode. 
These two different boundary conditions are illustrated in Fig.~\ref{fig:2d_distort}.

The corresponding harmonically driven system has the following Hamiltonian 
\begin{align}
  H(\boldsymbol{k},t,m) &= (m+2 - \cos k_x - \cos k_y + b\sigma_z)\tau_z \\
  &+ (\sin k_x \tau_x\sigma_x - \sin k_y\tau_y)\cos(\omega t),
\end{align}
which has chiral and the antiunitary time-glide (antiunitary reflection together with half period time translation)
symmetries, realized by $\hat{\mathcal{S}} = \tau_x\sigma_z$ and $\hat{\mathcal{A}}_{T/2,-}^{+} =
\tau_z\hat{\mathcal{K}}$.
With appropriately chosen boundary conditions,  
one can either have counter propagating anomalous Floquet gapless mode at the reflection symmetric edge (Fig.~\ref{fig:2d_distort}(a)), 
or a corner mode at $\epsilon_{\rm gap} = \omega/2$ at the reflection symmetry corner (Fig.~\ref{fig:2d_distort}(b)).

\subsubsection{2D system in class AI with $\overline{\mathcal{U}}_{T/2,-}^{+}$}
For 2D systems in class AI, with only spinless time-reversal symmetry $\hat{\mathcal{T}}^{2}=1$, the topological
classification is trivial. However, with a unitary, either static or space-time,  $d_{\parallel}=1$ antisymmetry realized by 
$\overline{\mathcal{U}}_{0,+}^{+}$ or $\overline{\mathcal{U}}_{T/2,-}^{+}$, the $K$ group 
subseries is $0 \subseteq \mathbb{Z} \subseteq \mathbb{Z}$,
as given in Table~\ref{tab:real_unitary_hoti_2}.

Let us start by considering a Hamiltonian $h(k_x,k_y)$ with a static $d_{\parallel}=1$ antisymmetry, 
given by 
\begin{equation}
\overline{\mathcal{U}}_{0,+}^{+}h(k_x,k_y)(\overline{\mathcal{U}}_{0,+}^{+})^{-1} = -h(-k_x,k_y),
\end{equation}
in addition to the spinless time-reversal symmetry. 
At the reflection symmetric momenta $k_x = 0, \pi$, the Hamiltonian as a function of $k_y$
reduces to a 1D Hamiltonian in class BDI, which has a $\mathbb{Z}$ winding number topological 
invariant.

One can also undertand the topological classification from the edge perspective. 
At reflection invariant edge, the $x$ edge in this case, multiple pairs of counter propagating edge modes
can exist. One can write the edge Hamiltonian as $H_{\rm edge} = k_{x}\Gamma_x + m\Gamma_m$, 
with a possible mass term of magnitude $m$. Here
the matrices $\Gamma_x$ and $\Gamma_m$ anticommute with each other and squares to identity. 
Since the edge is reflection invariant, we have $[\Gamma_x, \overline{\mathcal{U}}_{0,+}^{+}]=0$, 
and $\{\Gamma_{m},\overline{\mathcal{U}}_{0,+}^{+}\} = 0$.
Hence we can simultaneously block diagonalize $\Gamma_x$ and $\overline{\mathcal{U}}_{0,+}^{+}$, 
and label the pair of gapless modes in terms of the eigenvalues $\pm 1$ of $\overline{\mathcal{U}}_{0,+}^{+}$.
If we denote the number of pairs of gapless modes with opposite $\overline{\mathcal{U}}_{0,+}^{+}$ parity by
$n_{\pm}$, then only $(n_{+} - n_{-}) \in \mathbb{Z}$ pairs of gapless modes are stable
because the mass $m\Gamma_m$ gaps out gapless modes with opposite eigenvalues of $\overline{\mathcal{U}}_{0,+}^{+}$. 

These gapless modes are purely protected by the $d_{\parallel}=1$ antisymmetry, and will be completely gapped
when the edge is not invariant under reflection, which implies $K' = K^{(0)} = \mathbb{Z}$. 
Indeed, we can assume there are $(n_{+} - n_{-})$ pairs of gapless modes which have positive parity
under $\overline{\mathcal{U}}_{0,+}^{+}$. The time-reversal operator can be chosen 
as $\hat{\mathcal{T}}=\hat{\mathcal{K}}$, because $[\hat{\mathcal{T}},\overline{\mathcal{U}}_{0,+}^{+}] = 0$.
We will write $\Gamma_{x} = \mathbb{I}_{(n_{+} - n_{-})}\otimes\sigma_y$, where $\mathbb{I}_{n}$ denotes the identity matrix of dimension $n$.
When the edge is deformed away symmetrically around a corner at $x=0$, mass terms 
$m_{1}(x)\sigma_x + m_{2}(x)\sigma_z$, with $m_{i}(x) = -m_{i}(-x)$, $i = 1,2$,  can be generated.
This gives rise to $(n_{+} - n_{-})$ zero energy corner modes, corresponding to $\mathcal{K}' = K'/K'' = \mathbb{Z}$.

An explicit Hamiltonian for $h(k_{x},k_{y})$ can have the following form
\begin{align}
  h(\boldsymbol{k},m) &= (m+2-\cos k_x - \cos k_y)\tau_z + \sin k_x \tau_x\sigma_y  \nonumber \\
&+ \sin k_y\tau_y + b\tau_z\sigma_z
\end{align}
with $\hat{\mathcal{T}}=\hat{\mathcal{K}}$ and $\overline{\mathcal{U}}_{0,+}^{+} = \tau_x$, 
and numerically small $b$. When $-2<m<0$, there exist counter propagating gapless
modes on the $x$ edges when the system has an open boundary condition in the $y$ direction. 

The corresponding harmonically driven Hamiltonian with a unitary space-time antisymmetry
has the following form
\begin{align}
  H(\boldsymbol{k},t,m) &= (m+2-\cos k_x - \cos_y + b\sigma_z)\tau_z \nonumber \\
  &+ (\sin k_x \tau_x\sigma_y - \sin k_y \tau_y)\cos(\omega t),
\end{align}
where the time-reversal symmetry and the unitary space-time antisymmetry are
realized by $\hat{\mathcal{T}}=\hat{\mathcal{K}}$ and $\overline{\mathcal{U}}_{T/2,-}^{+} = \tau_{y}$, respectively.
Gapless Floquet edge modes, or Floquet corner modes at $\epsilon_{\rm gap}=\omega/2$, can be created, 
with appropriately chosen boundary conditions,  when both $-2<(m-\omega/2)<0$ and $(m+\omega/2)>0$ are satisfied.

\subsection{Third-order phase in $d_{\parallel} = 2$ family}
When a Floquet system respects a $d_{\parallel}=2$ space-time
symmetry/antisymmetry, it can be at most a 
third-order topological phase, because $\min(n+1,d_{\parallel}+1)\leq 3$. 
In the following, we construct a model Hamiltonian for a third-order TI 
representing such systems.

\subsubsection*{3D system in class AIII with $\hat{\mathcal{A}}_{T/2,-}^{+}$}
It is known that for 3D system in class AIII without any additional spatial symmetries, the topological classification
is $\mathbb{Z}$ \cite{Chiu2016}, which counts the number of surface Dirac cones at the boundary of the 3D insulating
bulk. 
When there exists an antiunitary two-fold rotation symmetry, either $\hat{\mathcal{A}}_{0,+}^{+}$ or
$\hat{\mathcal{A}}_{T/2,-}^{+}$, the topological invariants are given by the
$K$ subgroup series $0 \subseteq \mathbb{Z}_2 \subseteq \mathbb{Z}_{2} \subseteq \mathbb{Z}_{2}$ in
Table~\ref{tab:complex_antiunitary_hoti_3}.

Indeed, because of the additional symmetry realized by $\hat{\mathcal{A}}_{0,+}^{+}$ or
$\hat{\mathcal{A}}_{T/2,-}^{+}$, the symmetry invariant boundary surface is able to support
gapless Dirac cone pairs. As will be shown in the following, it turns out that
the number of such pairs is maximum to be one, which gives rise to the 
$K^{(0)} = \mathbb{Z}_{2}$ topological invariant.

Let us first look at the static antiunitary two-fold rotation symmetry, realized by $\hat{\mathcal{A}}_{0,+}^{+}$, which
transforms a static Bloch Hamiltonian as
\begin{equation}
	\hat{\mathcal{A}}_{0,+}^{+} h(k_x,k_y,k_z) (\hat{\mathcal{A}}_{0,+}^{+})^{-1} = h(k_x,k_y,-k_z).
\end{equation}
With an appropriate basis, one can write $\hat{S} = \tau_z$, and $\hat{\mathcal{A}}_{0,+}^{+} = \hat{\mathcal{K}}$. 
At the symmetry invariant boundary surface perpendicular to $z$, while keeping the periodic boundary condition
in both $x$ and $y$ directions,  a single Dirac cone pair with a dispersion
$h_{\rm surf} = \tau_x(\sigma_{x}k_x + \sigma_zk_y)$ can exist. This Dirac cone pair cannot be gapped by an additional mass term
preserving the $\hat{\mathcal{A}}_{0,+}^{+}$ symmetry, which requires the mass term to be real.
However, when there are two pairs of Dirac cones, described by the surface Hamiltonian 
$h_{\rm surf} = \mu_0\tau_x(\sigma_x k_x + \sigma_z k_y)$, with $\mu_0$ a two-by-two identity matrix
for another spinor degree of freedom, for which we also introduce a new set of Pauli matrices $\mu_{x,y,z}$.
Noticeably, a mass term which couple the two pairs of Dirac cones and gap them out can be chosen
as $\mu_{y}\sigma_x\tau_y$, which preserves the antiunitary two-fold symmetry. Hence, we have a $K^{(0)} = \mathbb{Z}_{2}$.

When the surface is tilted away from the rotation invariant direction, two mutually anticommuting
rotation-symmetry-breaking mass terms exist, and can be written as $m_{1}\tau_y\sigma_0 + m_{2}\tau_{x}\sigma_y$, 
in which $m_{1,2}$ must change signs under two-fold rotation.
Hence, boundaries of codimension up to $\min(n,d_{\parallel}) = 2$ are gapped. This leads to 
$\mathcal{K}'' = \mathcal{K}' = 0$, which implies $K'' = K' = K^{(0)} = \mathbb{Z}_{2}$.
Moreover, at the symmetry invariant corner, this mass must vanish, and thus the system 
can host zero-energy corner mode. 

\begin{figure}[t]
	\centering
	\includegraphics[width=0.45\textwidth]{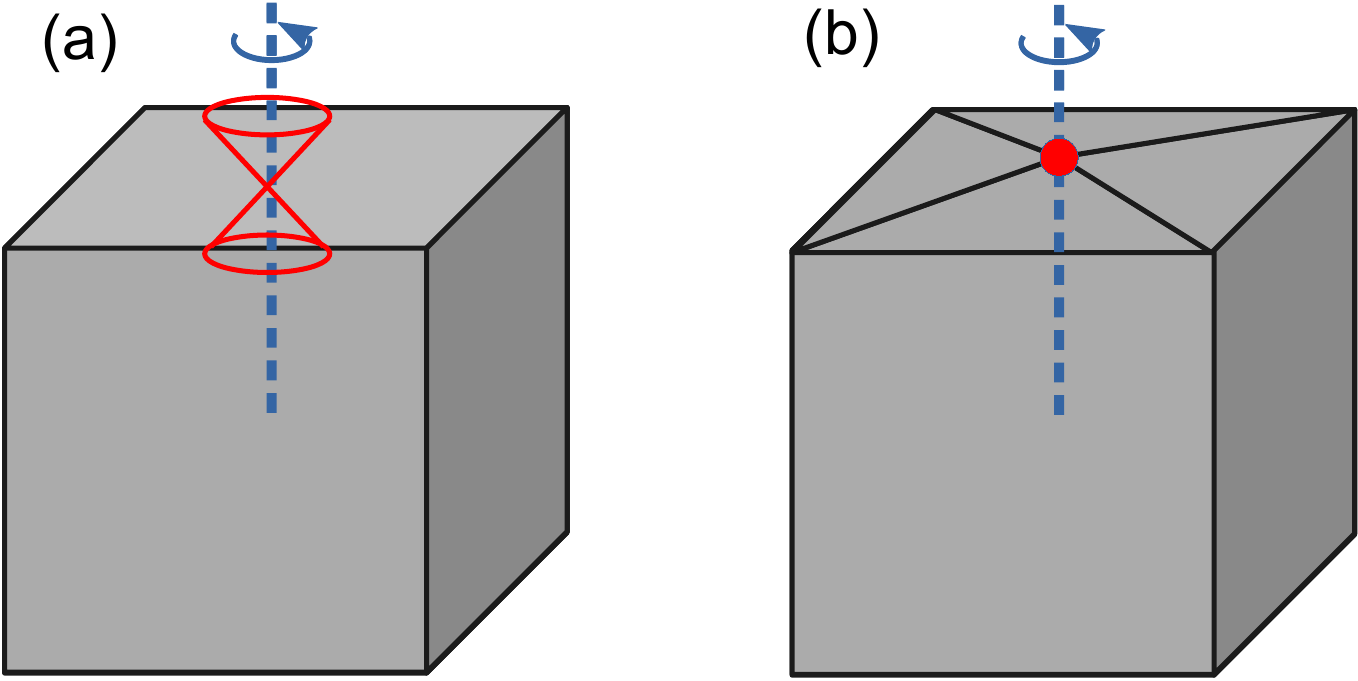}
	\caption{(a) Gapless surface mode (Dirac cone) on the rotation invariant surface. (b) Corner mode at rotation invariant corner.
		The dashed line indicates the two-fold rotation (time-screw) axis. 
	}
	\label{fig:3d_distort}
\end{figure}

One can write down the following concrete model Hamiltonian with eight bands,
\begin{align}
  &h(\boldsymbol{k},m) = (m+3 - \cos k_x -\cos k_y - \cos k_z)\tau_z  \nonumber \\
	&+ \sin k_x \tau_x\sigma_x  + \sin k_y \tau_x\sigma_z + \sin k_z \tau_y  \nonumber \\
	&+ b_1\mu_x\tau_z\sigma_z + b_2\mu_x\tau_z\sigma_x.
\end{align}
where the parameters $b_1$ and $b_2$ are numerically small.  Here, the chiral and the antiunitary two-fold rotation symmetries
are realized by $\hat{\mathcal{S}} = \mu_y\tau_x\sigma_y$, and $\hat{\mathcal{A}}_{0,+}^{+}=\hat{\mathcal{K}}$, respectively. 
This Hamiltonian supports a single pair of Dirac cones on the boundary surfaces perpendicular to the $z$ axis, 
at $k_{x}=k_{y}=0$ for $-2<m<0$, as illustrated in Fig.~\ref{fig:3d_distort}(a). 
When the surface perpendicular to the rotation axis gets deformed from Fig.~\ref{fig:3d_distort}(a) to (b), 
the rotation invariant corner then bounds a codimension-two boundary mode. 

The corresponding harmonically driven model has the following Hamiltonian
\begin{align}
  &H(\boldsymbol{k},t,m) = (m+3-\cos k_x - \cos k_y - \cos k_z  \nonumber \\
&+ b_1\mu_x\sigma_z + b_2\mu_x\sigma_x)\tau_z \nonumber \\
&+ \left[(\sin k_x \sigma_x + \sin k_y \sigma_y)\tau_x  - \sin k_z \tau_y\right]\cos(\omega t).
\end{align}
Here, the chiral symmetry is realized by
$\hat{\mathcal{S}}=\mu_y\tau_x\sigma_y$, while the the antiunitary two-fold time-screw symmetry
is realized by $\hat{\mathcal{A}}_{T/2,-}^{+}=\tau_z\hat{\mathcal{K}}$.
This Hamiltonian is able to support a pair of Dirac cones on the boundary surface perpendicular to
$z$ direction inside the bulk quasienergy gap around $\epsilon_{\rm gap} = \omega/2$ (Fig.~\ref{fig:3d_distort}(a)), 
as well as codimension-two mode with quasienergy $\omega/2$ localized at the rotation invariant corner of the system (Fig.~\ref{fig:3d_distort}(b)).

\subsection{Higher-order topological phases in $d_{\parallel} = 3$ family}
Unlike the symmetries discussed previously, the $d_{\parallel} = 3$
symmetry (antisymmetry) operator $\hat{\mathcal{P}}$ ($\overline{\mathcal{P}}$) 
does not leave any point invariant in our three dimensional world. 
In particular, since the surface of a 3D system naturally breaks the inversion symmetry, 
the topological classification of the gapless surface modes (if exist), should be 
the same as the 3D tenfold classification disregarding the crystalline symmetry, in the same symmetry class. 
Hence, we have the boundary $K$ group
\begin{equation}
  \mathcal{K}' = K^{(0)}/K' = 
  \begin{cases}
    K_{\rm TF} & K_{\rm TF} \subseteq K^{(0)}, \\
    0  &  \mathrm{otherwise}, 
  \end{cases}
\end{equation}
where $K_{\rm TF}$ is the corresponding $K$ group for the tenfold-way topological phase, with only nonspatial symmetries
considered. 

However, inversion related pairs of boundaries with codimension larger than one are able to host gapless modes, 
which can not be gapped out without breaking the symmetry (antisymmetry) realized by $\hat{\mathcal{P}}$
($\overline{\mathcal{P}}$). 
This can be understood by simply considering the surface Hamiltonian $h(\boldsymbol{p}_{\parallel},\hat{\boldsymbol{n}})$ 
with $\hat{\boldsymbol{n}}\in S^{2}$. Here $\boldsymbol{p}_{\parallel}$ is the momenta perpendicular to $\hat{\boldsymbol{n}}$.
Let us assume there are $n$ spatially dependent mass terms $m_{l}(\boldsymbol{\hat{\boldsymbol{n}}})M_{l}$, 
with $l=1,\dots, n$, 
that can gap out the surface Hamiltonian $h(\boldsymbol{p}_{\parallel},\hat{\boldsymbol{n}})$.
The inversion symmetry/antisymmetry restricts
$m_{l}(\hat{\boldsymbol{n}})=-m_{l}(-\hat{\boldsymbol{n}})$ (see Appendix~\ref{app:order_mass_terms} for details), 
which implies that there must exist a 1D inversion symmetric loop
$S^{1}\subseteq S^{2}$, such that $m_{l}(\hat{\boldsymbol{n}}) = 0$, for $\hat{\boldsymbol{n}} \in S^{1}$.
This 1D loop for different $l$s can be different,  but they all preserve the inversion symmetry, 
and cannot be removed. Hence, for $n=1$, we have a 1D massless great circle, whereas for $n=2$ we have a pair of
antipodal massless points. The 1D or 0D massless region are irremovable topological defects 
which are able to host gapless modes. 

Since inversion operation maps one point to another point, the stability of the gapless modes on the 
massless 1D or 0D region must be protected by the nonspatial symmetries alone \cite{Khalaf2018}. Hence, the codimension-$k$ gapless
modes are stable only when the $(4-k)$-dimensional system has a nontrivial tenfold classification, namely $K_{\rm
TF}\neq 0$. 

Moreover, the number of these gapless modes is at most one \cite{Khalaf2018}. 
Indeed, a system consisting of a pair of inversion symmetric systems with protected gapless modes can 
be deformed into a system with completely gapped boundaries without breaking the inversion symmetry. 
This statement can be understood
by considering a pair of inversion symmetric surface Hamiltonians 
\begin{equation}
  h'(\boldsymbol{p}_\parallel,\hat{\boldsymbol{n}})=\left(\begin{array}{cc}
    h(\boldsymbol{p}_\parallel,\hat{\boldsymbol{n}}) & 0\\
    0 & \pm h(\boldsymbol{p}_\parallel,-\hat{\boldsymbol{n}})
\end{array}\right),
\end{equation}
where the $+$ ($-$) sign is taken when we have a inversion symmetry (antisymmetry).
In this situation, the $h'(\boldsymbol{p}_\parallel,\hat{\boldsymbol{n}})$ has a inversion symmetry or antisymmetry realized by 
\begin{equation}
  \hat{\mathcal{P}}'=\left(\begin{array}{cc}
0 & \hat{\mathcal{P}}\\
\hat{\mathcal{P}} & 0
\end{array}\right)\  \mathrm{or}\ 
 \overline{\mathcal{P}}'=\left(\begin{array}{cc}
0 & \overline{\mathcal{P}}\\
\overline{\mathcal{P}} & 0
\end{array}\right).
\end{equation}
Now one can introduce mass terms 
\begin{equation}
  \left(\begin{array}{cc}
    m_{l}(\hat{\boldsymbol{n}})M_{l} & 0\\
      0 & -m_{l}(-\hat{\boldsymbol{n}})M_{l}
\end{array}\right),
\end{equation}
In this case $m_l(\hat{\boldsymbol{n}})$ can be nonzero for all $\hat{\boldsymbol{n}} \in S^{2}$, and therefore
$h'(\boldsymbol{p}_{\parallel},\hat{\boldsymbol{n}})$ can always be gapped.  

Hence, we obtain the boundary $K$ groups $\mathcal{K}^{(k)}$ which 
classifies boundary modes of codimension $k=2$ and $3$ as
\begin{equation}
  \mathcal{K}^{(k)} = K^{(k-1)}/K^{(k)}
  \begin{cases}
    \mathbb{Z}_{2}  & \mathbb{Z}_2 \subseteq K_{TF} \ \mathrm{in}\  (4-k)\mathrm{D} \\
    0 & \mathrm{otherwise}.
  \end{cases}
\end{equation}

Having understood the general structure of $K$ subgroup series, let us
in the following construct model Hamiltonians for Floquet HOTI/SCs 
in class DIII with a unitary space-time symmetry realized by $\hat{\mathcal{U}}_{T/2,++}^+$ ($d_{\parallel}=3$),
as an example. 

From Table~\ref{tab:real_unitary_hoti_3}, we see that the $K$ subgroup series is $4\mathbb{Z} \subseteq
2\mathbb{Z} \subseteq \mathbb{Z} \subseteq \mathbb{Z}^2$, which implies we can have first-order phase
classified by $\mathcal{K}' = \mathbb{Z}^2/\mathbb{Z} = \mathbb{Z}$, second-order phase
classified by $\mathcal{K}'' = \mathbb{Z}/2\mathbb{Z} = \mathbb{Z}_2 $, and third-order phase
classified by $\mathcal{K}^{(3)} = 2\mathbb{Z}/4\mathbb{Z} = \mathbb{Z}_2$.

\subsubsection{First-order topological phase}
Under the operator $\hat{\mathcal{U}}_{T/2,++}^+$, no points on the surface of a 3D bulk are left invariant. 
Hence, the existence of codimension-one boundary modes is due to the protection from the nonspatial symmetries alone. 
A tight-binding model realizing such a phase can be constructed from its static counter part, namely, a model
in class DIII with a static inversion symmetry realized by $\hat{\mathcal{U}}_{0,+-}^{+}$.

The static model can have the following Hamiltonian
\begin{align}
&h_{\pm}(\boldsymbol{k},m) = (m+3-\cos k_x - \cos k_y - \cos k_z)\tau_z  \nonumber \\
& \pm (\sin k_x \sigma_x + \sin k_y \sigma_y + \sin k_z \sigma_z)\tau_x,
\end{align}
where the time-reversal, particle-hole, chiral and the inversion symmetres are realized by
$\hat{\mathcal{T}} = -i\sigma_y \hat{\mathcal{K}}$, $\hat{\mathcal{C}} = \sigma_y \tau_y \hat{\mathcal{K}}$, 
$\hat{\mathcal{S}} = \tau_y$, and $\hat{\mathcal{U}}_{0,+-}^{+} = \tau_z$, respectively.
When $-2<m<0$, this model hosts hosts a gapless Dirac cone with chirality $\pm 1$ on any surfaces of the 3D bulk. 

Hence, the Hamiltonian for the corresponding Floquet first-order topological phase with a space-time symmetry
can be written as
\begin{align}
  &H_{\pm}(\boldsymbol{k},t,m) = (m+3-\cos k_x - \cos k_y - \cos k_z)\tau_z  \nonumber \\
  & \pm (\sin k_x \sigma_x + \sin k_y \sigma_y + \sin k_z \sigma_z)\tau_x\cos(\omega t),
\end{align}
where the space-time symmetry is realized by $\hat{\mathcal{U}}_{T/2,++}^{+} = \mathbb{I}$, 
and the nonspatial symmetry operators are the same as in the static model.
When $-2<(m-\omega/2)<0$ and $(m+\omega/2)>0$ are satisfied, $H_{\pm}(\boldsymbol{k},t,m)$
will host a gapless Dirac cone at quasienergy $\omega/2$ with chirality $\pm 1$.

\subsubsection{Second-order topological phase}
Similar to the construction of the first-order phase, let us start from the 
corresponding static model. A static second-order phase can be obtained
by couple $h_{+}(\boldsymbol{k},m_{1})$ and $h_{-}(\boldsymbol{k},m_{2})$. 
When both $m_1$ and $m_{2}$ are within the interval $(-2,0)$, the topological invariant for the codimension-one boundary modes
vanishes and their exists a mass term on the surface which gaps out all boundary modes of codimenion one.

Explicitly, one can define the following Hamiltonian
\begin{equation}
  h(\boldsymbol{k},m_{1},m_{2})=\left(\begin{array}{cc}
h_{+}(\boldsymbol{k},m_{1}) & 0\\
0 & h_{-}(\boldsymbol{k},m_{2})
\end{array}\right),
\end{equation}
and introduce a set of Pauli matrices $\mu_{x,y,z}$ for this newly introduced spinor degrees of freedom. 
There is only one mass term $M_{l} = \tau_x \mu_x$, which satisfies $\{M_1, h(\boldsymbol{k},m_{1},m_{2})\}=0$, 
$\{M_1,\hat{\mathcal{S}}\}=0$, $\{M_1,\hat{\mathcal{C}}\}=0$, and $[M_1,\hat{\mathcal{T}}]=0$.
According to the discussion on relation between mass terms and the codimension of boundary modes 
in Sec.~\ref{sec:symmetry_mass_terms}, 
as well as Appendix~\ref{app:order_mass_terms},
one can add a perturbation
\begin{equation}
  V = b_{1}^{(1)}\sigma_x\tau_z\mu_x+b_{2}^{(1)}\sigma_y\tau_z\mu_x + b_{3}^{(1)}\sigma_z\tau_z\mu_x
\end{equation}
that preserves all symmetries, to $h(\boldsymbol{k},m_{1},m_{2})$.
This perturbation gaps out all codimension-one surfaces and left a codimension-two inversion invariant loop
gapless, giving rise to a second-order topological phase. 

\begin{figure}[t]
  \centering
  \includegraphics[width=0.40\textwidth]{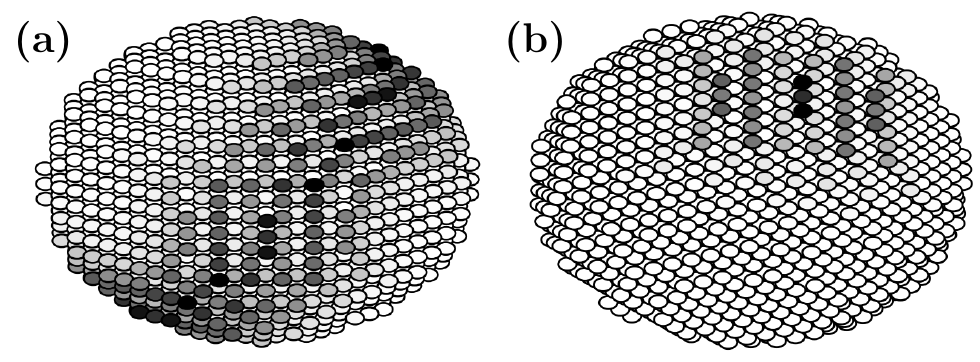}
  \caption{Spectral weight (darkness) of the Floquet boundary mode at $\omega/2$, cutted to an approximate sphere
    geometry of radius 10 lattice spacing. (a) Codimension-two boundary mode, computed with parameters $m_{1} = m_{2} = 0.5$, $\omega =3$, $b_{1}^{(1)} =
  b_{2}^{(1)} = b_{3}^{(1)} = 0.3$. (b) Codimension-three boundary mode, computed with  $m_{1} = m_{2} = m_{3} =
m_{4} = 0.5$,  $\omega =3$, $b_{1}^{(1)} = b_{2}^{(1)} = b_{3}^{(1)} = b_{3}^{(2)} = -b_{1}^{(2)} = -b_{2}^{(2)}=0.3$.}
  \label{fig:inversion_hoti}
\end{figure}

The Floquet second-order topological phase can therefore be constructed
by addition the perturbation $V$ to 
the following Hamiltonian
\begin{equation}
  H(\boldsymbol{k},t, m_{1},m_{2})=\left(\begin{array}{cc}
      H_{+}(\boldsymbol{k},t,m_{1}) & \\
      0 & H_{-}(\boldsymbol{k},t,m_{2}) 
  \end{array}\right).
\end{equation}

In Fig.~\ref{fig:inversion_hoti}(a), we show the spectral weight of the codimension-one Floquet boundary mode at
$\omega/2$, when the system is cutted to an approximate sphere geometry. This boundary mode is localized on an 
inversion invariant loop.

\subsubsection{Third-order topological phase}
To construct a model for the third-order topological phase, 
one needs to find two anticommuting masses $M_{1},M_{2}$, which satisfy 
the same conditions discussed previously. 
This can be realized by introducing another spinor degrees of freedom, 
as one couples two copies of $h(\boldsymbol{k},m_{1},m_{2})$. 
Explicitly, one can take the following Hamiltonian
\begin{align}
  &\tilde{h}(\boldsymbol{k},m_{1},m_{2},m_3,m_4) \nonumber \\
  &=\left(\begin{array}{cc}
h(\boldsymbol{k},m_{1},m_2) & 0\\
0 & h(\boldsymbol{k},m_{3},m_4)
\end{array}\right),
\end{align}
as well as the corresponding Pauli matrices $\tilde{\mu}_{x,y,z}$ for the spinor degrees of freedom. 

Thus, two anticommuting mass terms $M_{1} =\tau_x\mu_x$ and $M_{2} = \tau_x\mu_y\tilde{\mu}_{y}$ can be found. 
Therefore, one can introduce the symmetry preserving perturbation 
\begin{align}
  \tilde{V} &= (b_{1}^{(1)}\sigma_x+b_{2}^{(1)}\mu_x + b_{3}^{(1)}\sigma_z)\tau_z\mu_x \nonumber
  \\
  &+(b_{1}^{(2)}\sigma_x+b_{2}^{(2)}\mu_x + b_{3}^{(2)}\sigma_z)\tau_z\mu_y\tilde{\mu}_{y},
\end{align}
which in general gaps out all boundary modes except at two antipodal points, at
which codimension-three modes can exist. 

The Floquet version of such a third-order topological phase is constructed 
by adding the perturbation $\tilde{V}$ to the following
periodically driven Hamiltonian
\begin{align}
  &\tilde{H}(\boldsymbol{k},t,m_{1},m_{2},m_3,m_4) \nonumber \\
  &=\left(\begin{array}{cc}
      H(\boldsymbol{k},t,m_{1},m_2) & 0\\
      0 & H(\boldsymbol{k},t,m_{3},m_4)
  \end{array}\right).
\end{align}

In Fig.~\ref{fig:inversion_hoti}(b), the spectral weight of the zero-dimensional (codimension-three) Floquet 
modes at quasienergy $\omega/2$ is shown in a system with an approximate sphere geometry. The other zero-dimensional
mode is located at the antipodal point. 

\section{Conclusions \label{sec:conclusion}}
In this work, we have completed the classification of the Floquet HOTI/SCs with an order-two space-time symmetry/antisymmetry.
By introducing a hermitian map, we are able to map the unitary loops into hermitian matrices, and thus
define bulk $K$ groups as well as $K$ subgroup series for unitary loops. 
In particular, we show that 
for every order-two nontrivial space-time (anti)unitary symmetry/antisymmetry involving
a half-period time translation, there always exists a unique order-two
static spatial (anti)unitary symmetry/antisymmetry, such that the two symmetries/antisymmetries 
share the same $K$ group, as well as the subgroup series, and thus
have the same topological classification. 

Further, by exploiting the frequency-domain formulation, 
we introduce a general recipe of constructing
tight-binding model Hamiltonians for Floquet HOTI/SCs, 
which provides a more intuitive way of
understanding the topological classification table. 

It is also worth mentioning that although in this work we only
classify the Floquet HOTI/SCs with an order-two space-time symmetry/antisymmetry, 
the hermitian map introduced here can also be used to 
map the classification of unitary loops involving more complicated space-time symmetry,
to the classification of Hamiltonians with other point group symmetries. 
Similarly, the frequency-domain formulation and the recipe of constructing 
Floquet HOTI/SCs should also work with some modifications. 
In this sense, our approach can be more general than what we have shown in 
this work. 

Finally, we comment on one possible experimental realization of Floquet HOTI/SCs. 
As lattice vibrations naturally break some spatial symmetries instantaneously, while preserving the 
certain space-time symmetries, one way to engineer a Floquet HOTI/SC 
may involve exciting a particular phonon mode with a desired space-time symmetry,
which is investigated in Ref.~\cite{Swati2019}.

\acknowledgments
Y.P. acknowledges support from
the startup fund from California State University,
Northridge, 
as well as the IQIM, an NSF physics frontier center funded
in part by the Moore Foundation, and the support from
the Walter Burke Institute for Theoretical Physics at Caltech.
Y.P. also acknowledges partial support from  Y.P. is grateful for the helpful discussions with Gil Refeal at Caltech,
and with Luka Trifunovic.

\appendix
\section{Equivalent classification with symmetrized evolution operators\label{app:proof_symmetric_evolution}}
Let us prove the statement that the ordinary evolution operators $U_1(\boldsymbol{k},\boldsymbol{r},t)$ and $U_2(\boldsymbol{k},\boldsymbol{r},t)$
are homotopic if and only if the symmetric evolution operators $U_{\tau,1}(\boldsymbol{k},\boldsymbol{r},t)$
and $U_{\tau,2}(\boldsymbol{k},\boldsymbol{r},t)$ are homotopic.

When $U_1(\boldsymbol{k},\boldsymbol{r},t)$ and $U_2(\boldsymbol{k},\boldsymbol{r},t)$ are homotopic, there exists a continuous unitary-matrix-valued function 
$f(s,\boldsymbol{k},\boldsymbol{r},t)$ with $s\in[0,1]$, such that $f(0,\boldsymbol{k},\boldsymbol{r},t) =U_1(\boldsymbol{k},\boldsymbol{r},t)$, and $f(1,\boldsymbol{k},\boldsymbol{r},t)
=U_2(\boldsymbol{k},\boldsymbol{r},t)$. Hence, we can define a continuous unitary-matrix-valued function $g(s,\boldsymbol{k},\boldsymbol{r},t) =
f(s,\boldsymbol{k},\boldsymbol{r},\tau-t/2)f^\dagger(s,\boldsymbol{k},\boldsymbol{r},\tau+t/2)$, such that
$g(0,\boldsymbol{k},\boldsymbol{r},t) =U_{\tau,1}(\boldsymbol{k},\boldsymbol{r},t)$, and $g(1,\boldsymbol{k},\boldsymbol{r},t)
=U_{\tau,2}(\boldsymbol{k},\boldsymbol{r},t)$. We have that $U_{\tau,1}(\boldsymbol{k},\boldsymbol{r},t)$
and $U_{\tau,2}(\boldsymbol{k},\boldsymbol{r},t)$ are homotopic.

The other direction goes as follows. If $U_{\tau,1}(\boldsymbol{k},\boldsymbol{r},t)$
and $U_{\tau,2}(\boldsymbol{k},\boldsymbol{r},t)$ are homotopic,
then there exists a continuous unitary-matrix-valued function $g(s,\boldsymbol{k},\boldsymbol{r},t)$
such that $g(0,\boldsymbol{k},\boldsymbol{r},t)=U_{\tau,1}(\boldsymbol{k},\boldsymbol{r},t)$ and $g(1,\boldsymbol{k},\boldsymbol{r},t)=U_{\tau,2}(\boldsymbol{k},\boldsymbol{r},t)$.
Further, there exists another continuous unitary-matrix-valued function $f(s,\boldsymbol{k},\boldsymbol{r},t)$
such that $g(s,\boldsymbol{k},\boldsymbol{r},t) = f(s,\boldsymbol{k},\boldsymbol{r},t)f^\dagger(s,\boldsymbol{k},\boldsymbol{r},-t)$, 
because one requires the symmetry property is always satisfied during the deformation
when increasing $s$ from zero to $1$.
Hence, we have $f(0,\boldsymbol{k},\boldsymbol{r},t) = U_1(\boldsymbol{k},\boldsymbol{r},\frac{\tau+t}{2})$ and
$f(1,\boldsymbol{k},\boldsymbol{r},t) = U_2(\boldsymbol{k},\boldsymbol{r},\frac{\tau+t}{2})$.
This implies that the function $f(s,\boldsymbol{k},\boldsymbol{r},2t-\tau)$ would be the continuous
deformation between  $U_1(\boldsymbol{k},\boldsymbol{r},t)$ and $U_2(\boldsymbol{k},\boldsymbol{r},t)$.

\section{Decomposition of time evolution operators\label{app:two_theorems}}
In this section, we will follow Ref.~\cite{Roy2017} to show two theorems.
First, a generic time evolution can be decomposed as a unitary loop followed
by a constant Hamiltonian evolution, up to homotopy.  
Second, $L_{\tau,1}*C_{\tau,1} \approx L_{\tau,2} * C_{\tau,2}$ if and only if
$L_{\tau,1}\approx L_{\tau,2}$ and $C_{\tau,1}\approx C_{\tau,2}$, $L_{\tau,1}$, $L_{\tau,2}$ are unitary loops, 
and $C_{\tau,1}$, $C_{\tau,2}$ are constant Hamiltonian evolutions.

To prove the first theorem, let us assume $U_{\tau}$ is a symmetrized time evolution operator, and $H_F$ is its Floquet Hamlitonian. 
If $C_{\pm}(s)$ is the evolution with constant Hamiltonian $\pm sH_F$, then one can define
the continuous deformation 
\begin{equation}
f(s) = [U_{\tau} * C_{-}(s)] * C_{+}(s).
\end{equation}
We have $f(0) = U$, and $f(1) = L * C_{+}(1)$, which is a composition of a unitary loop followed by a constant
Hamiltonian evolution. 

Let us now prove the second theorem. If  $L_{\tau,1}*C_{\tau,1} \approx L_{\tau,2} * C_{\tau,2}$, 
then there exists a  continuous deformation $f(s)$ such that 
\begin{equation}
f(0) = L_{\tau,1}*C_{\tau,1}, \quad f(1) = L_{\tau,2} * C_{\tau,2}.
\end{equation}
If $H_F(s)$ is the corresponding Floquet Hamiltonian of the evolution $f(s)$, 
and $C_{+}(s)$ is the time evolution operator with constant Hamiltonian $H_{F}(s)$,
then $C_{+}(0)=C_{\tau,1}$ and $C_{+}(1)=C_{\tau,2}$, which implies $C_{\tau,1} \approx C_{\tau,2}$. 

Let $g(s) = f(s)*C_{-}(s)$, with $C_{-}(s)$ be the time evolution with constant Hamiltonian $-H_{F}(s)$,
then $g(s)$ is a unitary loop for all intermediat $s$. Moreover, we have $g(0) = L_{\tau,1}$ and $g(1)=L_{\tau,2}$. 
Thus, $L_{\tau,1} \approx L_{\tau,2}$.

The proof in the opposite direction is more straightforward. If $L_{\tau,1} \approx L_{\tau,2}$ and $C_{\tau,1} \approx
C_{\tau,2}$, then there exist two continuous deformations $f(s)$ and $g(s)$, which interpolate 
the two pairs. If we make the composition $h(s) = f(s) * g(s)$, then $h(s)$ continuously 
deforms $L_{\tau,1}*C_{\tau,1}$ into $L_{\tau,2}*C_{\tau,2}$.

\section{Order of HOTI/SCs and symmetry-breaking mass terms \label{app:order_mass_terms}}
Consider a static HOTI/SCs in $d$-dimension described by the Hamiltonian $h(\boldsymbol{k},m)$ given in
Eq.~(\ref{eq:static_models}). Let us denote the spatial symmetry (antisymmetry) operator as
$\hat{\mathcal{P}}$ ($\overline{\mathcal{P}}$), 
and assume there are $n$ mutually anticommuting $M_{l}$, with $l=1,\dots, n$,  $\{M_{l},h(\boldsymbol{k},m)\}=0$, and
$\{M_{l},\hat{\mathcal{P}}\}=0$ ($[M_{l},\overline{\mathcal{P}}]=0$). We further consider a slowly position-dependent 
parameter $m = m(\boldsymbol{r})$, which produces a position-dependent Hamiltonian
$h(\boldsymbol{k},m(\boldsymbol{r}))$. If there is a region with $m(\boldsymbol{r})<0$ 
and $m(\boldsymbol{r})>0$ outside this region, such that the boundary defined by $m(\boldsymbol{r})=0$ 
is topologically the same as $S^{d-1}$, then there may exist gapless modes localized at the boundary.  
One can try to gap out the possible gapless modes, while preserving the spatial symmetry of
$h(\boldsymbol{k},m(\boldsymbol{r}))$, by introducing a perturbation
\begin{equation}
  V = i \sum_{l=1}^{n}\sum_{j=1}^{d_{\parallel}} b_{j}^{(l)} M_{l} \Gamma_0 \Gamma_j.
\end{equation}

Let us focus on a point on the boundary defined by its normal unit vector $\hat{\boldsymbol{n}}$ (pointing toward the $m>0$ region), 
one can then define $p_{\perp} = \boldsymbol{k} \cdot \hat{\boldsymbol{n}}$, $\boldsymbol{p}_{\parallel}\cdot
\hat{\boldsymbol{n}} = 0$, and $x_{\perp}=\boldsymbol{r}\cdot \hat{\boldsymbol{n}}$.  
Thus, the low energy Hamiltonian near this point at the boundary can be written as
\begin{equation}
  h_{\rm boundary}(\boldsymbol{p}_{\parallel}) = m(x_{\perp})\Gamma_0 + \boldsymbol{p}_{\parallel}\cdot
  \boldsymbol{\Gamma} -
  i(\hat{\boldsymbol{n}}\cdot \boldsymbol{\Gamma}) \partial_{x_{\perp}}, 
\end{equation}
where $\boldsymbol{\Gamma} = (\Gamma_{1},\dots,\Gamma_d)$. 
The wave function for a bound state of $h_{\rm boundary}$ can be written as 
\begin{equation}
  \psi(x_{\perp}, \boldsymbol{p}_{\parallel}) = \exp(-\int_{0}^{x_{\perp}} dx' m(x'))
  \tilde{\psi}(\boldsymbol{p}_{\parallel}).
\end{equation}
The gapless mode corresponds to the solution $(\Gamma_0 + i\hat{\boldsymbol{n}} \cdot \boldsymbol{\Gamma})
\tilde{\psi}(\boldsymbol{p}_{\parallel}) = 0$. According to this, one can define the projector
into this gapless sector as
\begin{equation}
  P(\hat{\boldsymbol{n}}) = \frac{1}{2}(1+i(\hat{\boldsymbol{n}}\cdot \boldsymbol{\Gamma})\Gamma_0).
\end{equation}

Hence, we have the Hamiltonian with the additional perturbation $V$ projected into
the boundary low-energy sector
\begin{align}
  &P(\hat{\boldsymbol{n}})(h_{\rm boundary}(\boldsymbol{p}_{\parallel}) + V)P(\hat{\boldsymbol{n}})  \nonumber \\
  &=\boldsymbol{p}_{\parallel}\cdot P(\hat{\boldsymbol{n}})\boldsymbol{\Gamma} P(\hat{\boldsymbol{n}})-
  \frac{1}{2}\sum_{l=1}^{n}\sum_{j=1}^{d_{\parallel}}b_{l}^{(j)}M_{l}\hat{n}_{j},
\end{align}
where $\hat{n}_{j}$ is the $j$th component of $\hat{\boldsymbol{n}}$. Note that the second term gaps out the 
boundary, and we can have gapless boundary modes only at locations satisfying
\begin{equation}
  \sum_{j=1}^{d_{\parallel}}b_{l}^{(j)}\hat{n}_{j} = 0, \quad \forall l=1,\dots,n.
\end{equation}
This condition is equivalent to finding the intersection $\ker \boldsymbol{B} \cap S^{d-1}$, 
where $\ker \boldsymbol{B}$ denotes the kernal of matrix $\boldsymbol{B}$ whose
elements  are defined as $B_{ij} = b_{i}^{(j)}$.
Since $\ker \boldsymbol{B}$ is a linear subspace of $\mathbb{R}^{d}$ of dimension $d - \min(n,d_{\parallel})$, 
we find the gapless set is given by
\begin{align}
  &\ker\boldsymbol{B}\cap S^{d-1} \nonumber \\
  &=\begin{cases}
S^{d-\min(n+1,d_{\parallel}+1)} & \min(n+1,d_{\parallel}+1)\leq d\\
\emptyset & \min(n+1,d_{\parallel}+1)>d.
\end{cases}
\end{align}
This means one can have gapless boundary modes of codimension $\min(n+1,d_{\parallel}+1)$ if
$\min(n+1,d_{\parallel}+1)\leq d$, otherwise the boundary is completely gapped. 


\begin{thebibliography}{45}%
\makeatletter
\providecommand \@ifxundefined [1]{%
 \@ifx{#1\undefined}
}%
\providecommand \@ifnum [1]{%
 \ifnum #1\expandafter \@firstoftwo
 \else \expandafter \@secondoftwo
 \fi
}%
\providecommand \@ifx [1]{%
 \ifx #1\expandafter \@firstoftwo
 \else \expandafter \@secondoftwo
 \fi
}%
\providecommand \natexlab [1]{#1}%
\providecommand \enquote  [1]{``#1''}%
\providecommand \bibnamefont  [1]{#1}%
\providecommand \bibfnamefont [1]{#1}%
\providecommand \citenamefont [1]{#1}%
\providecommand \href@noop [0]{\@secondoftwo}%
\providecommand \href [0]{\begingroup \@sanitize@url \@href}%
\providecommand \@href[1]{\@@startlink{#1}\@@href}%
\providecommand \@@href[1]{\endgroup#1\@@endlink}%
\providecommand \@sanitize@url [0]{\catcode `\\12\catcode `\$12\catcode
  `\&12\catcode `\#12\catcode `\^12\catcode `\_12\catcode `\%12\relax}%
\providecommand \@@startlink[1]{}%
\providecommand \@@endlink[0]{}%
\providecommand \url  [0]{\begingroup\@sanitize@url \@url }%
\providecommand \@url [1]{\endgroup\@href {#1}{\urlprefix }}%
\providecommand \urlprefix  [0]{URL }%
\providecommand \Eprint [0]{\href }%
\providecommand \doibase [0]{http://dx.doi.org/}%
\providecommand \selectlanguage [0]{\@gobble}%
\providecommand \bibinfo  [0]{\@secondoftwo}%
\providecommand \bibfield  [0]{\@secondoftwo}%
\providecommand \translation [1]{[#1]}%
\providecommand \BibitemOpen [0]{}%
\providecommand \bibitemStop [0]{}%
\providecommand \bibitemNoStop [0]{.\EOS\space}%
\providecommand \EOS [0]{\spacefactor3000\relax}%
\providecommand \BibitemShut  [1]{\csname bibitem#1\endcsname}%
\let\auto@bib@innerbib\@empty
\bibitem [{\citenamefont {Hasan}\ and\ \citenamefont {Kane}(2010)}]{Hasan2010}%
  \BibitemOpen
  \bibfield  {author} {\bibinfo {author} {\bibfnamefont {M.~Z.}\ \bibnamefont
  {Hasan}}\ and\ \bibinfo {author} {\bibfnamefont {C.~L.}\ \bibnamefont
  {Kane}},\ }\bibfield  {title} {\enquote {\bibinfo {title} {Colloquium:
  Topological insulators},}\ }\href {\doibase 10.1103/RevModPhys.82.3045}
  {\bibfield  {journal} {\bibinfo  {journal} {Rev. Mod. Phys.}\ }\textbf
  {\bibinfo {volume} {82}},\ \bibinfo {pages} {3045--3067} (\bibinfo {year}
  {2010})}\BibitemShut {NoStop}%
\bibitem [{\citenamefont {Qi}\ and\ \citenamefont {Zhang}(2011)}]{Qi2011}%
  \BibitemOpen
  \bibfield  {author} {\bibinfo {author} {\bibfnamefont {Xiao-Liang}\
  \bibnamefont {Qi}}\ and\ \bibinfo {author} {\bibfnamefont {Shou-Cheng}\
  \bibnamefont {Zhang}},\ }\bibfield  {title} {\enquote {\bibinfo {title}
  {Topological insulators and superconductors},}\ }\href {\doibase
  10.1103/RevModPhys.83.1057} {\bibfield  {journal} {\bibinfo  {journal} {Rev.
  Mod. Phys.}\ }\textbf {\bibinfo {volume} {83}},\ \bibinfo {pages}
  {1057--1110} (\bibinfo {year} {2011})}\BibitemShut {NoStop}%
\bibitem [{\citenamefont {Bernevig}\ and\ \citenamefont
  {Hughes}(2013)}]{Bernevig2013book}%
  \BibitemOpen
  \bibfield  {author} {\bibinfo {author} {\bibfnamefont {B~Andrei}\
  \bibnamefont {Bernevig}}\ and\ \bibinfo {author} {\bibfnamefont {Taylor~L}\
  \bibnamefont {Hughes}},\ }\href@noop {} {\emph {\bibinfo {title} {Topological
  insulators and topological superconductors}}}\ (\bibinfo  {publisher}
  {Princeton University Press},\ \bibinfo {year} {2013})\BibitemShut {NoStop}%
\bibitem [{\citenamefont {Schnyder}\ \emph {et~al.}(2008)\citenamefont
  {Schnyder}, \citenamefont {Ryu}, \citenamefont {Furusaki},\ and\
  \citenamefont {Ludwig}}]{Schnyder2008}%
  \BibitemOpen
  \bibfield  {author} {\bibinfo {author} {\bibfnamefont {Andreas~P.}\
  \bibnamefont {Schnyder}}, \bibinfo {author} {\bibfnamefont {Shinsei}\
  \bibnamefont {Ryu}}, \bibinfo {author} {\bibfnamefont {Akira}\ \bibnamefont
  {Furusaki}}, \ and\ \bibinfo {author} {\bibfnamefont {Andreas W.~W.}\
  \bibnamefont {Ludwig}},\ }\bibfield  {title} {\enquote {\bibinfo {title}
  {Classification of topological insulators and superconductors in three
  spatial dimensions},}\ }\href {\doibase 10.1103/PhysRevB.78.195125}
  {\bibfield  {journal} {\bibinfo  {journal} {Phys. Rev. B}\ }\textbf {\bibinfo
  {volume} {78}},\ \bibinfo {pages} {195125} (\bibinfo {year}
  {2008})}\BibitemShut {NoStop}%
\bibitem [{\citenamefont {Kitaev}(2009)}]{Kitaev2009}%
  \BibitemOpen
  \bibfield  {author} {\bibinfo {author} {\bibfnamefont {Alexei}\ \bibnamefont
  {Kitaev}},\ }\bibfield  {title} {\enquote {\bibinfo {title} {Periodic table
  for topological insulators and superconductors},}\ }\href@noop {} {\bibfield
  {journal} {\bibinfo  {journal} {AIP Conf. Proc.}\ }\textbf {\bibinfo {volume}
  {1134}},\ \bibinfo {pages} {22} (\bibinfo {year} {2009})}\BibitemShut
  {NoStop}%
\bibitem [{\citenamefont {Ryu}\ \emph {et~al.}(2010)\citenamefont {Ryu},
  \citenamefont {Schnyder}, \citenamefont {Furusaki},\ and\ \citenamefont
  {Ludwig}}]{Ryu2010}%
  \BibitemOpen
  \bibfield  {author} {\bibinfo {author} {\bibfnamefont {Shinsei}\ \bibnamefont
  {Ryu}}, \bibinfo {author} {\bibfnamefont {Andreas~P}\ \bibnamefont
  {Schnyder}}, \bibinfo {author} {\bibfnamefont {Akira}\ \bibnamefont
  {Furusaki}}, \ and\ \bibinfo {author} {\bibfnamefont {Andreas~WW}\
  \bibnamefont {Ludwig}},\ }\bibfield  {title} {\enquote {\bibinfo {title}
  {Topological insulators and superconductors: tenfold way and dimensional
  hierarchy},}\ }\href@noop {} {\bibfield  {journal} {\bibinfo  {journal} {New
  Journal of Physics}\ }\textbf {\bibinfo {volume} {12}},\ \bibinfo {pages}
  {065010} (\bibinfo {year} {2010})}\BibitemShut {NoStop}%
\bibitem [{\citenamefont {Teo}\ and\ \citenamefont {Kane}(2010)}]{Teo2010}%
  \BibitemOpen
  \bibfield  {author} {\bibinfo {author} {\bibfnamefont {Jeffrey C.~Y.}\
  \bibnamefont {Teo}}\ and\ \bibinfo {author} {\bibfnamefont {C.~L.}\
  \bibnamefont {Kane}},\ }\bibfield  {title} {\enquote {\bibinfo {title}
  {Topological defects and gapless modes in insulators and superconductors},}\
  }\href {\doibase 10.1103/PhysRevB.82.115120} {\bibfield  {journal} {\bibinfo
  {journal} {Phys. Rev. B}\ }\textbf {\bibinfo {volume} {82}},\ \bibinfo
  {pages} {115120} (\bibinfo {year} {2010})}\BibitemShut {NoStop}%
\bibitem [{\citenamefont {Chiu}\ \emph {et~al.}(2016)\citenamefont {Chiu},
  \citenamefont {Teo}, \citenamefont {Schnyder},\ and\ \citenamefont
  {Ryu}}]{Chiu2016}%
  \BibitemOpen
  \bibfield  {author} {\bibinfo {author} {\bibfnamefont {Ching-Kai}\
  \bibnamefont {Chiu}}, \bibinfo {author} {\bibfnamefont {Jeffrey C.~Y.}\
  \bibnamefont {Teo}}, \bibinfo {author} {\bibfnamefont {Andreas~P.}\
  \bibnamefont {Schnyder}}, \ and\ \bibinfo {author} {\bibfnamefont {Shinsei}\
  \bibnamefont {Ryu}},\ }\bibfield  {title} {\enquote {\bibinfo {title}
  {Classification of topological quantum matter with symmetries},}\ }\href
  {\doibase 10.1103/RevModPhys.88.035005} {\bibfield  {journal} {\bibinfo
  {journal} {Rev. Mod. Phys.}\ }\textbf {\bibinfo {volume} {88}},\ \bibinfo
  {pages} {035005} (\bibinfo {year} {2016})}\BibitemShut {NoStop}%
\bibitem [{\citenamefont {Fu}(2011)}]{Fu2011}%
  \BibitemOpen
  \bibfield  {author} {\bibinfo {author} {\bibfnamefont {Liang}\ \bibnamefont
  {Fu}},\ }\bibfield  {title} {\enquote {\bibinfo {title} {Topological
  crystalline insulators},}\ }\href {\doibase 10.1103/PhysRevLett.106.106802}
  {\bibfield  {journal} {\bibinfo  {journal} {Phys. Rev. Lett.}\ }\textbf
  {\bibinfo {volume} {106}},\ \bibinfo {pages} {106802} (\bibinfo {year}
  {2011})}\BibitemShut {NoStop}%
\bibitem [{\citenamefont {Chiu}\ \emph {et~al.}(2013)\citenamefont {Chiu},
  \citenamefont {Yao},\ and\ \citenamefont {Ryu}}]{Chiu2013}%
  \BibitemOpen
  \bibfield  {author} {\bibinfo {author} {\bibfnamefont {Ching-Kai}\
  \bibnamefont {Chiu}}, \bibinfo {author} {\bibfnamefont {Hong}\ \bibnamefont
  {Yao}}, \ and\ \bibinfo {author} {\bibfnamefont {Shinsei}\ \bibnamefont
  {Ryu}},\ }\bibfield  {title} {\enquote {\bibinfo {title} {Classification of
  topological insulators and superconductors in the presence of reflection
  symmetry},}\ }\href {\doibase 10.1103/PhysRevB.88.075142} {\bibfield
  {journal} {\bibinfo  {journal} {Phys. Rev. B}\ }\textbf {\bibinfo {volume}
  {88}},\ \bibinfo {pages} {075142} (\bibinfo {year} {2013})}\BibitemShut
  {NoStop}%
\bibitem [{\citenamefont {Shiozaki}\ and\ \citenamefont
  {Sato}(2014)}]{Shiozaki2014}%
  \BibitemOpen
  \bibfield  {author} {\bibinfo {author} {\bibfnamefont {Ken}\ \bibnamefont
  {Shiozaki}}\ and\ \bibinfo {author} {\bibfnamefont {Masatoshi}\ \bibnamefont
  {Sato}},\ }\bibfield  {title} {\enquote {\bibinfo {title} {Topology of
  crystalline insulators and superconductors},}\ }\href {\doibase
  10.1103/PhysRevB.90.165114} {\bibfield  {journal} {\bibinfo  {journal} {Phys.
  Rev. B}\ }\textbf {\bibinfo {volume} {90}},\ \bibinfo {pages} {165114}
  (\bibinfo {year} {2014})}\BibitemShut {NoStop}%
\bibitem [{\citenamefont {Ando}\ and\ \citenamefont {Fu}(2015)}]{Ando2015}%
  \BibitemOpen
  \bibfield  {author} {\bibinfo {author} {\bibfnamefont {Yoichi}\ \bibnamefont
  {Ando}}\ and\ \bibinfo {author} {\bibfnamefont {Liang}\ \bibnamefont {Fu}},\
  }\bibfield  {title} {\enquote {\bibinfo {title} {Topological crystalline
  insulators and topological superconductors: from concepts to materials},}\
  }\href@noop {} {\bibfield  {journal} {\bibinfo  {journal} {Annu. Rev.
  Condens. Matter Phys.}\ }\textbf {\bibinfo {volume} {6}},\ \bibinfo {pages}
  {361--381} (\bibinfo {year} {2015})}\BibitemShut {NoStop}%
\bibitem [{\citenamefont {Kruthoff}\ \emph {et~al.}(2017)\citenamefont
  {Kruthoff}, \citenamefont {de~Boer}, \citenamefont {van Wezel}, \citenamefont
  {Kane},\ and\ \citenamefont {Slager}}]{Kruthoff2017}%
  \BibitemOpen
  \bibfield  {author} {\bibinfo {author} {\bibfnamefont {Jorrit}\ \bibnamefont
  {Kruthoff}}, \bibinfo {author} {\bibfnamefont {Jan}\ \bibnamefont {de~Boer}},
  \bibinfo {author} {\bibfnamefont {Jasper}\ \bibnamefont {van Wezel}},
  \bibinfo {author} {\bibfnamefont {Charles~L.}\ \bibnamefont {Kane}}, \ and\
  \bibinfo {author} {\bibfnamefont {Robert-Jan}\ \bibnamefont {Slager}},\
  }\bibfield  {title} {\enquote {\bibinfo {title} {Topological classification
  of crystalline insulators through band structure combinatorics},}\ }\href
  {\doibase 10.1103/PhysRevX.7.041069} {\bibfield  {journal} {\bibinfo
  {journal} {Phys. Rev. X}\ }\textbf {\bibinfo {volume} {7}},\ \bibinfo {pages}
  {041069} (\bibinfo {year} {2017})}\BibitemShut {NoStop}%
\bibitem [{\citenamefont {Turner}\ \emph {et~al.}(2010)\citenamefont {Turner},
  \citenamefont {Zhang},\ and\ \citenamefont {Vishwanath}}]{Turner2010}%
  \BibitemOpen
  \bibfield  {author} {\bibinfo {author} {\bibfnamefont {Ari~M.}\ \bibnamefont
  {Turner}}, \bibinfo {author} {\bibfnamefont {Yi}~\bibnamefont {Zhang}}, \
  and\ \bibinfo {author} {\bibfnamefont {Ashvin}\ \bibnamefont {Vishwanath}},\
  }\bibfield  {title} {\enquote {\bibinfo {title} {Entanglement and inversion
  symmetry in topological insulators},}\ }\href {\doibase
  10.1103/PhysRevB.82.241102} {\bibfield  {journal} {\bibinfo  {journal} {Phys.
  Rev. B}\ }\textbf {\bibinfo {volume} {82}},\ \bibinfo {pages} {241102}
  (\bibinfo {year} {2010})}\BibitemShut {NoStop}%
\bibitem [{\citenamefont {Hughes}\ \emph {et~al.}(2011)\citenamefont {Hughes},
  \citenamefont {Prodan},\ and\ \citenamefont {Bernevig}}]{Hughes2011}%
  \BibitemOpen
  \bibfield  {author} {\bibinfo {author} {\bibfnamefont {Taylor~L.}\
  \bibnamefont {Hughes}}, \bibinfo {author} {\bibfnamefont {Emil}\ \bibnamefont
  {Prodan}}, \ and\ \bibinfo {author} {\bibfnamefont {B.~Andrei}\ \bibnamefont
  {Bernevig}},\ }\bibfield  {title} {\enquote {\bibinfo {title}
  {Inversion-symmetric topological insulators},}\ }\href {\doibase
  10.1103/PhysRevB.83.245132} {\bibfield  {journal} {\bibinfo  {journal} {Phys.
  Rev. B}\ }\textbf {\bibinfo {volume} {83}},\ \bibinfo {pages} {245132}
  (\bibinfo {year} {2011})}\BibitemShut {NoStop}%
\bibitem [{\citenamefont {Slager}\ \emph {et~al.}(2015)\citenamefont {Slager},
  \citenamefont {Rademaker}, \citenamefont {Zaanen},\ and\ \citenamefont
  {Balents}}]{Slager2015}%
  \BibitemOpen
  \bibfield  {author} {\bibinfo {author} {\bibfnamefont {Robert-Jan}\
  \bibnamefont {Slager}}, \bibinfo {author} {\bibfnamefont {Louk}\ \bibnamefont
  {Rademaker}}, \bibinfo {author} {\bibfnamefont {Jan}\ \bibnamefont {Zaanen}},
  \ and\ \bibinfo {author} {\bibfnamefont {Leon}\ \bibnamefont {Balents}},\
  }\bibfield  {title} {\enquote {\bibinfo {title} {Impurity-bound states and
  green's function zeros as local signatures of topology},}\ }\href {\doibase
  10.1103/PhysRevB.92.085126} {\bibfield  {journal} {\bibinfo  {journal} {Phys.
  Rev. B}\ }\textbf {\bibinfo {volume} {92}},\ \bibinfo {pages} {085126}
  (\bibinfo {year} {2015})}\BibitemShut {NoStop}%
\bibitem [{\citenamefont {Benalcazar}\ \emph
  {et~al.}(2017{\natexlab{a}})\citenamefont {Benalcazar}, \citenamefont
  {Bernevig},\ and\ \citenamefont {Hughes}}]{Benalcazar2017}%
  \BibitemOpen
  \bibfield  {author} {\bibinfo {author} {\bibfnamefont {Wladimir~A.}\
  \bibnamefont {Benalcazar}}, \bibinfo {author} {\bibfnamefont {B.~Andrei}\
  \bibnamefont {Bernevig}}, \ and\ \bibinfo {author} {\bibfnamefont
  {Taylor~L.}\ \bibnamefont {Hughes}},\ }\bibfield  {title} {\enquote {\bibinfo
  {title} {Electric multipole moments, topological multipole moment pumping,
  and chiral hinge states in crystalline insulators},}\ }\href {\doibase
  10.1103/PhysRevB.96.245115} {\bibfield  {journal} {\bibinfo  {journal} {Phys.
  Rev. B}\ }\textbf {\bibinfo {volume} {96}},\ \bibinfo {pages} {245115}
  (\bibinfo {year} {2017}{\natexlab{a}})}\BibitemShut {NoStop}%
\bibitem [{\citenamefont {Peng}\ \emph {et~al.}(2017)\citenamefont {Peng},
  \citenamefont {Bao},\ and\ \citenamefont {von Oppen}}]{Peng2017}%
  \BibitemOpen
  \bibfield  {author} {\bibinfo {author} {\bibfnamefont {Yang}\ \bibnamefont
  {Peng}}, \bibinfo {author} {\bibfnamefont {Yimu}\ \bibnamefont {Bao}}, \ and\
  \bibinfo {author} {\bibfnamefont {Felix}\ \bibnamefont {von Oppen}},\
  }\bibfield  {title} {\enquote {\bibinfo {title} {Boundary green functions of
  topological insulators and superconductors},}\ }\href {\doibase
  10.1103/PhysRevB.95.235143} {\bibfield  {journal} {\bibinfo  {journal} {Phys.
  Rev. B}\ }\textbf {\bibinfo {volume} {95}},\ \bibinfo {pages} {235143}
  (\bibinfo {year} {2017})}\BibitemShut {NoStop}%
\bibitem [{\citenamefont {Langbehn}\ \emph {et~al.}(2017)\citenamefont
  {Langbehn}, \citenamefont {Peng}, \citenamefont {Trifunovic}, \citenamefont
  {von Oppen},\ and\ \citenamefont {Brouwer}}]{Langbehn2017}%
  \BibitemOpen
  \bibfield  {author} {\bibinfo {author} {\bibfnamefont {Josias}\ \bibnamefont
  {Langbehn}}, \bibinfo {author} {\bibfnamefont {Yang}\ \bibnamefont {Peng}},
  \bibinfo {author} {\bibfnamefont {Luka}\ \bibnamefont {Trifunovic}}, \bibinfo
  {author} {\bibfnamefont {Felix}\ \bibnamefont {von Oppen}}, \ and\ \bibinfo
  {author} {\bibfnamefont {Piet~W.}\ \bibnamefont {Brouwer}},\ }\bibfield
  {title} {\enquote {\bibinfo {title} {Reflection-symmetric second-order
  topological insulators and superconductors},}\ }\href {\doibase
  10.1103/PhysRevLett.119.246401} {\bibfield  {journal} {\bibinfo  {journal}
  {Phys. Rev. Lett.}\ }\textbf {\bibinfo {volume} {119}},\ \bibinfo {pages}
  {246401} (\bibinfo {year} {2017})}\BibitemShut {NoStop}%
\bibitem [{\citenamefont {Benalcazar}\ \emph
  {et~al.}(2017{\natexlab{b}})\citenamefont {Benalcazar}, \citenamefont
  {Bernevig},\ and\ \citenamefont {Hughes}}]{Benalcazar2017s}%
  \BibitemOpen
  \bibfield  {author} {\bibinfo {author} {\bibfnamefont {Wladimir~A}\
  \bibnamefont {Benalcazar}}, \bibinfo {author} {\bibfnamefont {B~Andrei}\
  \bibnamefont {Bernevig}}, \ and\ \bibinfo {author} {\bibfnamefont {Taylor~L}\
  \bibnamefont {Hughes}},\ }\bibfield  {title} {\enquote {\bibinfo {title}
  {Quantized electric multipole insulators},}\ }\href@noop {} {\bibfield
  {journal} {\bibinfo  {journal} {Science}\ }\textbf {\bibinfo {volume}
  {357}},\ \bibinfo {pages} {61--66} (\bibinfo {year}
  {2017}{\natexlab{b}})}\BibitemShut {NoStop}%
\bibitem [{\citenamefont {Song}\ \emph {et~al.}(2017)\citenamefont {Song},
  \citenamefont {Fang},\ and\ \citenamefont {Fang}}]{Song2017}%
  \BibitemOpen
  \bibfield  {author} {\bibinfo {author} {\bibfnamefont {Zhida}\ \bibnamefont
  {Song}}, \bibinfo {author} {\bibfnamefont {Zhong}\ \bibnamefont {Fang}}, \
  and\ \bibinfo {author} {\bibfnamefont {Chen}\ \bibnamefont {Fang}},\
  }\bibfield  {title} {\enquote {\bibinfo {title} {$d-2$-dimensional edge
  states of rotation symmetry protected topological states},}\ }\href {\doibase
  10.1103/PhysRevLett.119.246402} {\bibfield  {journal} {\bibinfo  {journal}
  {Phys. Rev. Lett.}\ }\textbf {\bibinfo {volume} {119}},\ \bibinfo {pages}
  {246402} (\bibinfo {year} {2017})}\BibitemShut {NoStop}%
\bibitem [{\citenamefont {Schindler}\ \emph {et~al.}(2018)\citenamefont
  {Schindler}, \citenamefont {Cook}, \citenamefont {Vergniory}, \citenamefont
  {Wang}, \citenamefont {Parkin}, \citenamefont {Bernevig},\ and\ \citenamefont
  {Neupert}}]{Schindler2018}%
  \BibitemOpen
  \bibfield  {author} {\bibinfo {author} {\bibfnamefont {Frank}\ \bibnamefont
  {Schindler}}, \bibinfo {author} {\bibfnamefont {Ashley~M}\ \bibnamefont
  {Cook}}, \bibinfo {author} {\bibfnamefont {Maia~G}\ \bibnamefont
  {Vergniory}}, \bibinfo {author} {\bibfnamefont {Zhijun}\ \bibnamefont
  {Wang}}, \bibinfo {author} {\bibfnamefont {Stuart~SP}\ \bibnamefont
  {Parkin}}, \bibinfo {author} {\bibfnamefont {B~Andrei}\ \bibnamefont
  {Bernevig}}, \ and\ \bibinfo {author} {\bibfnamefont {Titus}\ \bibnamefont
  {Neupert}},\ }\bibfield  {title} {\enquote {\bibinfo {title} {Higher-order
  topological insulators},}\ }\href@noop {} {\bibfield  {journal} {\bibinfo
  {journal} {Sci. Adv.}\ }\textbf {\bibinfo {volume} {4}},\ \bibinfo {pages}
  {eaat0346} (\bibinfo {year} {2018})}\BibitemShut {NoStop}%
\bibitem [{\citenamefont {Geier}\ \emph {et~al.}(2018)\citenamefont {Geier},
  \citenamefont {Trifunovic}, \citenamefont {Hoskam},\ and\ \citenamefont
  {Brouwer}}]{Geier2018}%
  \BibitemOpen
  \bibfield  {author} {\bibinfo {author} {\bibfnamefont {Max}\ \bibnamefont
  {Geier}}, \bibinfo {author} {\bibfnamefont {Luka}\ \bibnamefont
  {Trifunovic}}, \bibinfo {author} {\bibfnamefont {Max}\ \bibnamefont
  {Hoskam}}, \ and\ \bibinfo {author} {\bibfnamefont {Piet~W.}\ \bibnamefont
  {Brouwer}},\ }\bibfield  {title} {\enquote {\bibinfo {title} {Second-order
  topological insulators and superconductors with an order-two crystalline
  symmetry},}\ }\href {\doibase 10.1103/PhysRevB.97.205135} {\bibfield
  {journal} {\bibinfo  {journal} {Phys. Rev. B}\ }\textbf {\bibinfo {volume}
  {97}},\ \bibinfo {pages} {205135} (\bibinfo {year} {2018})}\BibitemShut
  {NoStop}%
\bibitem [{\citenamefont {Khalaf}(2018)}]{Khalaf2018}%
  \BibitemOpen
  \bibfield  {author} {\bibinfo {author} {\bibfnamefont {Eslam}\ \bibnamefont
  {Khalaf}},\ }\bibfield  {title} {\enquote {\bibinfo {title} {Higher-order
  topological insulators and superconductors protected by inversion
  symmetry},}\ }\href {\doibase 10.1103/PhysRevB.97.205136} {\bibfield
  {journal} {\bibinfo  {journal} {Phys. Rev. B}\ }\textbf {\bibinfo {volume}
  {97}},\ \bibinfo {pages} {205136} (\bibinfo {year} {2018})}\BibitemShut
  {NoStop}%
\bibitem [{\citenamefont {Khalaf}\ \emph {et~al.}(2018)\citenamefont {Khalaf},
  \citenamefont {Po}, \citenamefont {Vishwanath},\ and\ \citenamefont
  {Watanabe}}]{Khalaf2018prx}%
  \BibitemOpen
  \bibfield  {author} {\bibinfo {author} {\bibfnamefont {Eslam}\ \bibnamefont
  {Khalaf}}, \bibinfo {author} {\bibfnamefont {Hoi~Chun}\ \bibnamefont {Po}},
  \bibinfo {author} {\bibfnamefont {Ashvin}\ \bibnamefont {Vishwanath}}, \ and\
  \bibinfo {author} {\bibfnamefont {Haruki}\ \bibnamefont {Watanabe}},\
  }\bibfield  {title} {\enquote {\bibinfo {title} {Symmetry indicators and
  anomalous surface states of topological crystalline insulators},}\ }\href
  {\doibase 10.1103/PhysRevX.8.031070} {\bibfield  {journal} {\bibinfo
  {journal} {Phys. Rev. X}\ }\textbf {\bibinfo {volume} {8}},\ \bibinfo {pages}
  {031070} (\bibinfo {year} {2018})}\BibitemShut {NoStop}%
\bibitem [{\citenamefont {Trifunovic}\ and\ \citenamefont
  {Brouwer}(2019)}]{Trifunovic2019}%
  \BibitemOpen
  \bibfield  {author} {\bibinfo {author} {\bibfnamefont {Luka}\ \bibnamefont
  {Trifunovic}}\ and\ \bibinfo {author} {\bibfnamefont {Piet~W.}\ \bibnamefont
  {Brouwer}},\ }\bibfield  {title} {\enquote {\bibinfo {title} {Higher-order
  bulk-boundary correspondence for topological crystalline phases},}\ }\href
  {\doibase 10.1103/PhysRevX.9.011012} {\bibfield  {journal} {\bibinfo
  {journal} {Phys. Rev. X}\ }\textbf {\bibinfo {volume} {9}},\ \bibinfo {pages}
  {011012} (\bibinfo {year} {2019})}\BibitemShut {NoStop}%
\bibitem [{\citenamefont {Oka}\ and\ \citenamefont {Aoki}(2009)}]{Oka2009}%
  \BibitemOpen
  \bibfield  {author} {\bibinfo {author} {\bibfnamefont {Takashi}\ \bibnamefont
  {Oka}}\ and\ \bibinfo {author} {\bibfnamefont {Hideo}\ \bibnamefont {Aoki}},\
  }\bibfield  {title} {\enquote {\bibinfo {title} {Photovoltaic hall effect in
  graphene},}\ }\href {\doibase 10.1103/PhysRevB.79.081406} {\bibfield
  {journal} {\bibinfo  {journal} {Phys. Rev. B}\ }\textbf {\bibinfo {volume}
  {79}},\ \bibinfo {pages} {081406} (\bibinfo {year} {2009})}\BibitemShut
  {NoStop}%
\bibitem [{\citenamefont {Inoue}\ and\ \citenamefont
  {Tanaka}(2010)}]{Inoue2010}%
  \BibitemOpen
  \bibfield  {author} {\bibinfo {author} {\bibfnamefont {Jun-ichi}\
  \bibnamefont {Inoue}}\ and\ \bibinfo {author} {\bibfnamefont {Akihiro}\
  \bibnamefont {Tanaka}},\ }\bibfield  {title} {\enquote {\bibinfo {title}
  {Photoinduced transition between conventional and topological insulators in
  two-dimensional electronic systems},}\ }\href {\doibase
  10.1103/PhysRevLett.105.017401} {\bibfield  {journal} {\bibinfo  {journal}
  {Phys. Rev. Lett.}\ }\textbf {\bibinfo {volume} {105}},\ \bibinfo {pages}
  {017401} (\bibinfo {year} {2010})}\BibitemShut {NoStop}%
\bibitem [{\citenamefont {Kitagawa}\ \emph {et~al.}(2011)\citenamefont
  {Kitagawa}, \citenamefont {Oka}, \citenamefont {Brataas}, \citenamefont
  {Fu},\ and\ \citenamefont {Demler}}]{Kitagawa2011}%
  \BibitemOpen
  \bibfield  {author} {\bibinfo {author} {\bibfnamefont {Takuya}\ \bibnamefont
  {Kitagawa}}, \bibinfo {author} {\bibfnamefont {Takashi}\ \bibnamefont {Oka}},
  \bibinfo {author} {\bibfnamefont {Arne}\ \bibnamefont {Brataas}}, \bibinfo
  {author} {\bibfnamefont {Liang}\ \bibnamefont {Fu}}, \ and\ \bibinfo {author}
  {\bibfnamefont {Eugene}\ \bibnamefont {Demler}},\ }\bibfield  {title}
  {\enquote {\bibinfo {title} {Transport properties of nonequilibrium systems
  under the application of light: Photoinduced quantum hall insulators without
  landau levels},}\ }\href {\doibase 10.1103/PhysRevB.84.235108} {\bibfield
  {journal} {\bibinfo  {journal} {Phys. Rev. B}\ }\textbf {\bibinfo {volume}
  {84}},\ \bibinfo {pages} {235108} (\bibinfo {year} {2011})}\BibitemShut
  {NoStop}%
\bibitem [{\citenamefont {Lindner}\ \emph {et~al.}(2011)\citenamefont
  {Lindner}, \citenamefont {Refael},\ and\ \citenamefont
  {Galitski}}]{Lindner2011}%
  \BibitemOpen
  \bibfield  {author} {\bibinfo {author} {\bibfnamefont {Netanel~H}\
  \bibnamefont {Lindner}}, \bibinfo {author} {\bibfnamefont {Gil}\ \bibnamefont
  {Refael}}, \ and\ \bibinfo {author} {\bibfnamefont {Victor}\ \bibnamefont
  {Galitski}},\ }\bibfield  {title} {\enquote {\bibinfo {title} {Floquet
  topological insulator in semiconductor quantum wells},}\ }\href@noop {}
  {\bibfield  {journal} {\bibinfo  {journal} {Nature Physics}\ }\textbf
  {\bibinfo {volume} {7}},\ \bibinfo {pages} {490--495} (\bibinfo {year}
  {2011})}\BibitemShut {NoStop}%
\bibitem [{\citenamefont {Lindner}\ \emph {et~al.}(2013)\citenamefont
  {Lindner}, \citenamefont {Bergman}, \citenamefont {Refael},\ and\
  \citenamefont {Galitski}}]{Lindner2013}%
  \BibitemOpen
  \bibfield  {author} {\bibinfo {author} {\bibfnamefont {Netanel~H.}\
  \bibnamefont {Lindner}}, \bibinfo {author} {\bibfnamefont {Doron~L.}\
  \bibnamefont {Bergman}}, \bibinfo {author} {\bibfnamefont {Gil}\ \bibnamefont
  {Refael}}, \ and\ \bibinfo {author} {\bibfnamefont {Victor}\ \bibnamefont
  {Galitski}},\ }\bibfield  {title} {\enquote {\bibinfo {title} {Topological
  floquet spectrum in three dimensions via a two-photon resonance},}\ }\href
  {\doibase 10.1103/PhysRevB.87.235131} {\bibfield  {journal} {\bibinfo
  {journal} {Phys. Rev. B}\ }\textbf {\bibinfo {volume} {87}},\ \bibinfo
  {pages} {235131} (\bibinfo {year} {2013})}\BibitemShut {NoStop}%
\bibitem [{\citenamefont {Roy}\ and\ \citenamefont {Harper}(2017)}]{Roy2017}%
  \BibitemOpen
  \bibfield  {author} {\bibinfo {author} {\bibfnamefont {Rahul}\ \bibnamefont
  {Roy}}\ and\ \bibinfo {author} {\bibfnamefont {Fenner}\ \bibnamefont
  {Harper}},\ }\bibfield  {title} {\enquote {\bibinfo {title} {Periodic table
  for floquet topological insulators},}\ }\href {\doibase
  10.1103/PhysRevB.96.155118} {\bibfield  {journal} {\bibinfo  {journal} {Phys.
  Rev. B}\ }\textbf {\bibinfo {volume} {96}},\ \bibinfo {pages} {155118}
  (\bibinfo {year} {2017})}\BibitemShut {NoStop}%
\bibitem [{\citenamefont {Yao}\ \emph {et~al.}(2017)\citenamefont {Yao},
  \citenamefont {Yan},\ and\ \citenamefont {Wang}}]{Yao2017}%
  \BibitemOpen
  \bibfield  {author} {\bibinfo {author} {\bibfnamefont {Shunyu}\ \bibnamefont
  {Yao}}, \bibinfo {author} {\bibfnamefont {Zhongbo}\ \bibnamefont {Yan}}, \
  and\ \bibinfo {author} {\bibfnamefont {Zhong}\ \bibnamefont {Wang}},\
  }\bibfield  {title} {\enquote {\bibinfo {title} {Topological invariants of
  floquet systems: General formulation, special properties, and floquet
  topological defects},}\ }\href {\doibase 10.1103/PhysRevB.96.195303}
  {\bibfield  {journal} {\bibinfo  {journal} {Phys. Rev. B}\ }\textbf {\bibinfo
  {volume} {96}},\ \bibinfo {pages} {195303} (\bibinfo {year}
  {2017})}\BibitemShut {NoStop}%
\bibitem [{\citenamefont {Shirley}(1965)}]{Shirley1965}%
  \BibitemOpen
  \bibfield  {author} {\bibinfo {author} {\bibfnamefont {Jon~H.}\ \bibnamefont
  {Shirley}},\ }\bibfield  {title} {\enquote {\bibinfo {title} {Solution of the
  schr\"odinger equation with a hamiltonian periodic in time},}\ }\href
  {\doibase 10.1103/PhysRev.138.B979} {\bibfield  {journal} {\bibinfo
  {journal} {Phys. Rev.}\ }\textbf {\bibinfo {volume} {138}},\ \bibinfo {pages}
  {B979--B987} (\bibinfo {year} {1965})}\BibitemShut {NoStop}%
\bibitem [{\citenamefont {Huang}\ and\ \citenamefont {Liu}(2018)}]{Huang2018}%
  \BibitemOpen
  \bibfield  {author} {\bibinfo {author} {\bibfnamefont {Biao}\ \bibnamefont
  {Huang}}\ and\ \bibinfo {author} {\bibfnamefont {W~Vincent}\ \bibnamefont
  {Liu}},\ }\bibfield  {title} {\enquote {\bibinfo {title} {Higher-order
  floquet topological insulators with anomalous corner states},}\ }\href@noop
  {} {\bibfield  {journal} {\bibinfo  {journal} {arXiv preprint
  arXiv:1811.00555}\ } (\bibinfo {year} {2018})}\BibitemShut {NoStop}%
\bibitem [{\citenamefont {Bomantara}\ \emph {et~al.}(2019)\citenamefont
  {Bomantara}, \citenamefont {Zhou}, \citenamefont {Pan},\ and\ \citenamefont
  {Gong}}]{Bomantara2019}%
  \BibitemOpen
  \bibfield  {author} {\bibinfo {author} {\bibfnamefont {Raditya~Weda}\
  \bibnamefont {Bomantara}}, \bibinfo {author} {\bibfnamefont {Longwen}\
  \bibnamefont {Zhou}}, \bibinfo {author} {\bibfnamefont {Jiaxin}\ \bibnamefont
  {Pan}}, \ and\ \bibinfo {author} {\bibfnamefont {Jiangbin}\ \bibnamefont
  {Gong}},\ }\bibfield  {title} {\enquote {\bibinfo {title} {Coupled-wire
  construction of static and floquet second-order topological insulators},}\
  }\href {\doibase 10.1103/PhysRevB.99.045441} {\bibfield  {journal} {\bibinfo
  {journal} {Phys. Rev. B}\ }\textbf {\bibinfo {volume} {99}},\ \bibinfo
  {pages} {045441} (\bibinfo {year} {2019})}\BibitemShut {NoStop}%
\bibitem [{\citenamefont {Rodriguez-Vega}\ \emph {et~al.}(2019)\citenamefont
  {Rodriguez-Vega}, \citenamefont {Kumar},\ and\ \citenamefont
  {Seradjeh}}]{Rodriguez2019}%
  \BibitemOpen
  \bibfield  {author} {\bibinfo {author} {\bibfnamefont {Martin}\ \bibnamefont
  {Rodriguez-Vega}}, \bibinfo {author} {\bibfnamefont {Abhishek}\ \bibnamefont
  {Kumar}}, \ and\ \bibinfo {author} {\bibfnamefont {Babak}\ \bibnamefont
  {Seradjeh}},\ }\bibfield  {title} {\enquote {\bibinfo {title} {Higher-order
  floquet topological phases with corner and bulk bound states},}\ }\href
  {\doibase 10.1103/PhysRevB.100.085138} {\bibfield  {journal} {\bibinfo
  {journal} {Phys. Rev. B}\ }\textbf {\bibinfo {volume} {100}},\ \bibinfo
  {pages} {085138} (\bibinfo {year} {2019})}\BibitemShut {NoStop}%
\bibitem [{\citenamefont {Peng}\ and\ \citenamefont {Refael}(2019)}]{Peng2019}%
  \BibitemOpen
  \bibfield  {author} {\bibinfo {author} {\bibfnamefont {Yang}\ \bibnamefont
  {Peng}}\ and\ \bibinfo {author} {\bibfnamefont {Gil}\ \bibnamefont
  {Refael}},\ }\bibfield  {title} {\enquote {\bibinfo {title} {Floquet
  second-order topological insulators from nonsymmorphic space-time
  symmetries},}\ }\href {\doibase 10.1103/PhysRevLett.123.016806} {\bibfield
  {journal} {\bibinfo  {journal} {Phys. Rev. Lett.}\ }\textbf {\bibinfo
  {volume} {123}},\ \bibinfo {pages} {016806} (\bibinfo {year}
  {2019})}\BibitemShut {NoStop}%
\bibitem [{\citenamefont {Seshadri}\ \emph {et~al.}(2019)\citenamefont
  {Seshadri}, \citenamefont {Dutta},\ and\ \citenamefont {Sen}}]{Seshadri2019}%
  \BibitemOpen
  \bibfield  {author} {\bibinfo {author} {\bibfnamefont {Ranjani}\ \bibnamefont
  {Seshadri}}, \bibinfo {author} {\bibfnamefont {Anirban}\ \bibnamefont
  {Dutta}}, \ and\ \bibinfo {author} {\bibfnamefont {Diptiman}\ \bibnamefont
  {Sen}},\ }\bibfield  {title} {\enquote {\bibinfo {title} {Generating a
  second-order topological insulator with multiple corner states by periodic
  driving},}\ }\href {\doibase 10.1103/PhysRevB.100.115403} {\bibfield
  {journal} {\bibinfo  {journal} {Phys. Rev. B}\ }\textbf {\bibinfo {volume}
  {100}},\ \bibinfo {pages} {115403} (\bibinfo {year} {2019})}\BibitemShut
  {NoStop}%
\bibitem [{\citenamefont {Nag}\ \emph {et~al.}(2019)\citenamefont {Nag},
  \citenamefont {Juricic},\ and\ \citenamefont {Roy}}]{Nag2019}%
  \BibitemOpen
  \bibfield  {author} {\bibinfo {author} {\bibfnamefont {Tanay}\ \bibnamefont
  {Nag}}, \bibinfo {author} {\bibfnamefont {Vladimir}\ \bibnamefont {Juricic}},
  \ and\ \bibinfo {author} {\bibfnamefont {Bitan}\ \bibnamefont {Roy}},\
  }\bibfield  {title} {\enquote {\bibinfo {title} {Higher-order topological
  insulator out of equilibrium: Floquet engineering and quench dynamics},}\
  }\href@noop {} {\bibfield  {journal} {\bibinfo  {journal} {arXiv preprint
  arXiv:1904.07247}\ } (\bibinfo {year} {2019})}\BibitemShut {NoStop}%
\bibitem [{\citenamefont {Morimoto}\ \emph {et~al.}(2017)\citenamefont
  {Morimoto}, \citenamefont {Po},\ and\ \citenamefont
  {Vishwanath}}]{Morimoto2017}%
  \BibitemOpen
  \bibfield  {author} {\bibinfo {author} {\bibfnamefont {Takahiro}\
  \bibnamefont {Morimoto}}, \bibinfo {author} {\bibfnamefont {Hoi~Chun}\
  \bibnamefont {Po}}, \ and\ \bibinfo {author} {\bibfnamefont {Ashvin}\
  \bibnamefont {Vishwanath}},\ }\bibfield  {title} {\enquote {\bibinfo {title}
  {Floquet topological phases protected by time glide symmetry},}\ }\href
  {\doibase 10.1103/PhysRevB.95.195155} {\bibfield  {journal} {\bibinfo
  {journal} {Phys. Rev. B}\ }\textbf {\bibinfo {volume} {95}},\ \bibinfo
  {pages} {195155} (\bibinfo {year} {2017})}\BibitemShut {NoStop}%
\bibitem [{\citenamefont {Rudner}\ \emph {et~al.}(2013)\citenamefont {Rudner},
  \citenamefont {Lindner}, \citenamefont {Berg},\ and\ \citenamefont
  {Levin}}]{Rudner2013}%
  \BibitemOpen
  \bibfield  {author} {\bibinfo {author} {\bibfnamefont {Mark~S.}\ \bibnamefont
  {Rudner}}, \bibinfo {author} {\bibfnamefont {Netanel~H.}\ \bibnamefont
  {Lindner}}, \bibinfo {author} {\bibfnamefont {Erez}\ \bibnamefont {Berg}}, \
  and\ \bibinfo {author} {\bibfnamefont {Michael}\ \bibnamefont {Levin}},\
  }\bibfield  {title} {\enquote {\bibinfo {title} {Anomalous edge states and
  the bulk-edge correspondence for periodically driven two-dimensional
  systems},}\ }\href {\doibase 10.1103/PhysRevX.3.031005} {\bibfield  {journal}
  {\bibinfo  {journal} {Phys. Rev. X}\ }\textbf {\bibinfo {volume} {3}},\
  \bibinfo {pages} {031005} (\bibinfo {year} {2013})}\BibitemShut {NoStop}%
\bibitem [{\citenamefont {Fulga}\ \emph {et~al.}(2012)\citenamefont {Fulga},
  \citenamefont {Hassler},\ and\ \citenamefont {Akhmerov}}]{Fulga2012}%
  \BibitemOpen
  \bibfield  {author} {\bibinfo {author} {\bibfnamefont {I.~C.}\ \bibnamefont
  {Fulga}}, \bibinfo {author} {\bibfnamefont {F.}~\bibnamefont {Hassler}}, \
  and\ \bibinfo {author} {\bibfnamefont {A.~R.}\ \bibnamefont {Akhmerov}},\
  }\bibfield  {title} {\enquote {\bibinfo {title} {Scattering theory of
  topological insulators and superconductors},}\ }\href {\doibase
  10.1103/PhysRevB.85.165409} {\bibfield  {journal} {\bibinfo  {journal} {Phys.
  Rev. B}\ }\textbf {\bibinfo {volume} {85}},\ \bibinfo {pages} {165409}
  (\bibinfo {year} {2012})}\BibitemShut {NoStop}%
\bibitem [{sup()}]{suppl}%
  \BibitemOpen
  \href@noop {} {\bibinfo  {journal} {Supplemental Material}\ }\BibitemShut
  {NoStop}%
\bibitem [{\citenamefont {Chaudhary}\ \emph {et~al.}(2019)\citenamefont
  {Chaudhary}, \citenamefont {Haim}, \citenamefont {Peng},\ and\ \citenamefont
  {Refael}}]{Swati2019}%
  \BibitemOpen
\bibfield  {journal} {  }\bibfield  {author} {\bibinfo {author} {\bibfnamefont
  {Swati}\ \bibnamefont {Chaudhary}}, \bibinfo {author} {\bibfnamefont {Arbel}\
  \bibnamefont {Haim}}, \bibinfo {author} {\bibfnamefont {Yang}\ \bibnamefont
  {Peng}}, \ and\ \bibinfo {author} {\bibfnamefont {Gil}\ \bibnamefont
  {Refael}},\ }\bibfield  {title} {\enquote {\bibinfo {title} {Phonon-induced
  floquet second-order topological phases protected by space-time
  symmetries},}\ }\href@noop {} {\bibfield  {journal} {\bibinfo  {journal}
  {arXiv preprint arXiv:1911.07892}\ } (\bibinfo {year} {2019})}\BibitemShut
  {NoStop}%
\end{thebibliography}

%

\newpage

\begin{widetext}
\section*{Supplemental Material}

\section*{$K$ groups for unitary loops with an order-two space-time symmetry/antisymmetry}
In this supplement, the explicit form of the $K$ groups for unitary loops with an order-two space-time
symmetry/antisymmetry, in different dimensions are listed. 

\subsubsection{$\delta_{\parallel}=0$ family}
In this family, the additional symmetry includes nonspatial symmetry,
such as spin rotations with and without a simultaneous half-period
time translation. We summarize the classification table for $\delta_{\parallel}=0\mod2$
in complex symmetry classes with an order-two unitary symmetry
in Table~\ref{tab:complex_unitary_classification_0}.
In Table~\ref{tab:complex_antiunitary_classification_0}
and \ref{tab:real_unitary_classification_0}, we give the classification
for $\delta_{\parallel}=0\mod4$ in complex symmetry classes with an
order-two antiunitary symmetry, and in real symmetry classes
with an order-two unitary symmetry, respectively.

\subsubsection{$\delta_{\parallel}=1$ family}
This family includes Floquet topological phases protected by reflection
symmetry and time-glide symmetry, where only one direction
of the momenta is flipped. We summarize the classification table for
$\delta_{\parallel}=1\mod2$ in complex symmetry classes with an order-two
unitary symmetry in Table \ref{tab:complex_unitary_classification_1}.
In Table \ref{tab:complex_antiunitary_classification_1} and \ref{tab:real_unitary_classification_1},
we give the classification for $\delta_{\parallel}=0\mod4$ in
complex symmetry classes with an order-two antiunitary symmetry,
and in real symmetry classes with an order-two unitary symmetry,
respectively.

\subsubsection{$\delta_{\parallel}=2$ family}
The additional symmetry includes twofold spatial rotation and twofold
time-screw rotation, in which the momenta along two directions are
flipped. For $\delta_{\parallel}=2$, whose classification is the
same as $\delta_{\parallel}=0\mod2$ in complex symmetry classes with an
order-two unitary symmetry, as shown in Table \ref{tab:complex_unitary_classification_0}.
We summarize the classification for $\delta_{\parallel}=2\mod4$ in
complex symmetry classes with an order-two antiunitary symmetry,
and in real symmetry classes with an order-two unitary symmetry
in Table~\ref{tab:complex_antiunitary_classification_2} and \ref{tab:real_unitary_classification_2},
respectively.

\subsubsection{$\delta_{\parallel}=3$ family}
The order-two symmetry in this family includes inversion with and
without a simultaneous half-period time translation. For $\delta_{\parallel}=3$,
whose classification is the same as $\delta_{\parallel}=1\mod2$ in
complex symmetry classes with an order-two unitary symmetry,
as shown in Table \ref{tab:complex_unitary_classification_1}. We
summarize the classification for $\delta_{\parallel}=3\mod4$ in complex
symmetry classes with an order-two antiunitary symmetry, and
in real symmetry classes with an order-two unitary symmetry
in Table \ref{tab:complex_antiunitary_classification_3} and \ref{tab:real_unitary_classification_3},
respectively.

\begin{table*}
\caption{\label{tab:complex_unitary_classification_0}Classification table
for Floquet topological phases in complex symmetry classes supporting an additional order-two space-time unitary
symmetry with flipped parameters $\delta_{\parallel}=d_{\parallel}-D_{\parallel}=0\mod2$.
Here, $\delta=d-D$.}
\begin{ruledtabular}
\centering
\begin{tabular}{ccccccccccc}
Symmetry & Class & Classifying space & $\delta=0$ & $\delta=1$ & $\delta=2$ & $\delta=3$ & $\delta=4$ & $\delta=5$ & $\delta=6$ & $\delta=7$\tabularnewline
\hline
$\hat{\mathcal{U}}_{0}^{+},\hat{\mathcal{U}}_{T/2}^{+}$ & A & $\mathcal{C}_{-\delta}\times\mathcal{C}_{-\delta}$ & $\mathbb{Z}\oplus\mathbb{Z}$ & 0 & $\mathbb{Z}\oplus\mathbb{Z}$ & 0 & $\mathbb{Z}\oplus\mathbb{Z}$ & 0 & $\mathbb{Z}\oplus\mathbb{Z}$ & 0\tabularnewline
$\hat{\mathcal{U}}_{0,+}^{+},\hat{\mathcal{U}}_{T/2,-}^{+}$ & AIII & $\mathcal{C}_{1-\delta}\times\mathcal{C}_{1-\delta}$ & 0 & $\mathbb{Z}\oplus\mathbb{Z}$ & 0 & $\mathbb{Z}\oplus\mathbb{Z}$ & 0 & $\mathbb{Z}\oplus\mathbb{Z}$ & 0 & $\mathbb{Z}\oplus\mathbb{Z}$\tabularnewline
$\overline{\mathcal{U}}_{s}^{+}$ & A & $\mathcal{C}_{1-\delta}$ & 0 & $\mathbb{Z}$ & 0 & $\mathbb{Z}$ & 0 & $\mathbb{Z}$ & 0 & $\mathbb{Z}$\tabularnewline
$\hat{\mathcal{U}}_{0,-}^{+},\hat{\mathcal{U}}_{T/2,+}^{+}$ & AIII & $\mathcal{C}_{-\delta}$ & $\mathbb{Z}$ & 0 & $\mathbb{Z}$ & 0 & $\mathbb{Z}$ & 0 & $\mathbb{Z}$ & 0\tabularnewline
\end{tabular}
\end{ruledtabular}
\end{table*}

\begin{table*}
\caption{\label{tab:complex_antiunitary_classification_0}
Classification table
for Floquet topological phases in complex symmetry classes supporting an additional order-two space-time antiunitary
symmetry with flipped parameters $\delta_{\parallel}=d_{\parallel}-D_{\parallel}=0\mod4$.
Here, $\delta=d-D$.}
\begin{ruledtabular}
\begin{tabular}{ccccccccccc}
Symmetry & Class & Classifying space & $\delta=0$ & $\delta=1$ & $\delta=2$ & $\delta=3$ & $\delta=4$ & $\delta=5$ & $\delta=6$ & $\delta=7$\tabularnewline
\hline 
$\hat{\mathcal{A}}_{s}^{+}$ & A & $\mathcal{R}_{-\delta}$ & $\mathbb{Z}$ & 0 & 0 & 0 & 2$\mathbb{Z}$ & 0 & $\mathbb{\mathbb{Z}}_{2}$ & $\mathbb{\mathbb{Z}}_{2}$\tabularnewline
$\hat{\mathcal{A}}_{0,+}^{+},\hat{\mathcal{A}}_{T/2,-}^{+}$ & AIII & $\mathcal{R}_{1-\delta}$ & $\mathbb{\mathbb{Z}}_{2}$ & $\mathbb{Z}$ & 0 & 0 & 0 & 2$\mathbb{Z}$ & 0 & $\mathbb{\mathbb{Z}}_{2}$\tabularnewline
$\overline{\mathcal{A}}_{0}^{+},\overline{\mathcal{A}}_{T/2}^{-}$ & A & $\mathcal{R}_{2-\delta}$ & $\mathbb{\mathbb{Z}}_{2}$ & $\mathbb{\mathbb{Z}}_{2}$ & $\mathbb{Z}$ & 0 & 0 & 0 & 2$\mathbb{Z}$ & 0\tabularnewline
$\hat{\mathcal{A}}_{0,-}^{-},\hat{\mathcal{A}}_{T/2,+}^{-}$ & AIII & $\mathcal{R}_{3-\delta}$ & 0 & $\mathbb{\mathbb{Z}}_{2}$ & $\mathbb{\mathbb{Z}}_{2}$ & $\mathbb{Z}$ & 0 & 0 & 0 & 2$\mathbb{Z}$\tabularnewline
$\hat{\mathcal{A}}_{s}^{-}$ & A & $\mathcal{R}_{4-\delta}$ & 2$\mathbb{Z}$ & 0 & $\mathbb{\mathbb{Z}}_{2}$ & $\mathbb{\mathbb{Z}}_{2}$ & $\mathbb{Z}$ & 0 & 0 & 0\tabularnewline
$\hat{\mathcal{A}}_{0,+}^{-},\hat{\mathcal{A}}_{T/2,-}^{-}$ & AIII & $\mathcal{R}_{5-\delta}$ & 0 & 2$\mathbb{Z}$ & 0 & $\mathbb{\mathbb{Z}}_{2}$ & $\mathbb{\mathbb{Z}}_{2}$ & $\mathbb{Z}$ & 0 & 0\tabularnewline
$\overline{\mathcal{A}}_{0}^{-},\overline{\mathcal{A}}_{T/2}^{+}$ & A & $\mathcal{R}_{6-\delta}$ & 0 & 0 & 2$\mathbb{Z}$ & 0 & $\mathbb{\mathbb{Z}}_{2}$ & $\mathbb{\mathbb{Z}}_{2}$ & $\mathbb{Z}$ & 0\tabularnewline
$\hat{\mathcal{A}}_{0,-}^{+},\hat{\mathcal{A}}_{T/2,+}^{+}$ & AIII & $\mathcal{R}_{7-\delta}$ & 0 & 0 & 0 & 2$\mathbb{Z}$ & 0 & $\mathbb{\mathbb{Z}}_{2}$ & $\mathbb{\mathbb{Z}}_{2}$ & $\mathbb{Z}$\tabularnewline
\end{tabular}
\end{ruledtabular}
\end{table*}

\begin{table*}
\caption{\label{tab:real_unitary_classification_0}
Classification table
for Floquet topological phases in real symmetry classes supporting an additional order-two space-time unitary 
symmetry with flipped parameters $\delta_{\parallel}=d_{\parallel}-D_{\parallel}=0\mod4$.
Here, $\delta=d-D$.}
\begin{ruledtabular}
\begin{tabular}{ccccccccccc}
Symmetry or antisymmetry & Class & Classifying space & $\delta=0$ & $\delta=1$ & $\delta=2$ & $\delta=3$ & $\delta=4$ & $\delta=5$ & $\delta=6$ & $\delta=7$\tabularnewline
\hline 
$\text{\ensuremath{\hat{\mathcal{U}}_{0,+}^{+}}},\text{\ensuremath{\hat{\mathcal{U}}_{T/2,+}^{+}}}$ & AI & $\mathcal{R}_{-\delta}\times\mathcal{R}_{-\delta}$ & $\mathbb{Z}\oplus\mathbb{Z}$ & 0 & 0 & 0 & $2\mathbb{Z}$$\oplus$$2\mathbb{Z}$ & 0 & $\mathbb{\mathbb{Z}}_{2}\oplus\mathbb{\mathbb{Z}}_{2}$ & $\mathbb{\mathbb{Z}}_{2}\oplus\mathbb{\mathbb{Z}}_{2}$\tabularnewline
$\hat{\mathcal{U}}_{0,++}^{+},\hat{\mathcal{U}}_{T/2,+-}^{+}$ & BDI & $\mathcal{R}_{1-\delta}\times\mathcal{R}_{1-\delta}$ & $\mathbb{\mathbb{Z}}_{2}\oplus\mathbb{\mathbb{Z}}_{2}$ & $\mathbb{Z}\oplus\mathbb{Z}$ & 0 & 0 & 0 & $2\mathbb{Z}$$\oplus$$2\mathbb{Z}$ & 0 & $\mathbb{\mathbb{Z}}_{2}\oplus\mathbb{\mathbb{Z}}_{2}$\tabularnewline
$\ensuremath{\hat{\mathcal{U}}_{0,+}^{+}},\text{\ensuremath{\hat{\mathcal{U}}_{T/2,-}^{+}}}$ & D & $\mathcal{R}_{2-\delta}\times\mathcal{R}_{2-\delta}$ & $\mathbb{\mathbb{Z}}_{2}\oplus\mathbb{\mathbb{Z}}_{2}$ & $\mathbb{\mathbb{Z}}_{2}\oplus\mathbb{\mathbb{Z}}_{2}$ & $\mathbb{Z}\oplus\mathbb{Z}$ & 0 & 0 & 0 & $2\mathbb{Z}$$\oplus$$2\mathbb{Z}$ & 0\tabularnewline
$\hat{\mathcal{U}}_{0,++}^{+},\hat{\mathcal{U}}_{T/2,+-}^{+}$ & DIII & $\mathcal{R}_{3-\delta}\times\mathcal{R}_{3-\delta}$ & 0 & $\mathbb{\mathbb{Z}}_{2}\oplus\mathbb{\mathbb{Z}}_{2}$ & $\mathbb{\mathbb{Z}}_{2}\oplus\mathbb{\mathbb{Z}}_{2}$ & $\mathbb{Z}\oplus\mathbb{Z}$ & 0 & 0 & 0 & $2\mathbb{Z}$$\oplus$$2\mathbb{Z}$\tabularnewline
$\ensuremath{\hat{\mathcal{U}}_{0,+}^{+}},\ensuremath{\hat{\mathcal{U}}_{T/2,+}^{+}}$ & AII & $\mathcal{R}_{4-\delta}\times\mathcal{R}_{4-\delta}$ & $2\mathbb{Z}$$\oplus$$2\mathbb{Z}$ & 0 & $\mathbb{\mathbb{Z}}_{2}\oplus\mathbb{\mathbb{Z}}_{2}$ & $\mathbb{\mathbb{Z}}_{2}\oplus\mathbb{\mathbb{Z}}_{2}$ & $\mathbb{Z}\oplus\mathbb{Z}$ & 0 & 0 & 0\tabularnewline
\end{tabular}
\end{ruledtabular}
\end{table*}

\begin{table*}
\caption{\label{tab:complex_unitary_classification_1}
Classification table
for Floquet topological phases in complex symmetry classes supporting an additional order-two space-time unitary
symmetry with flipped parameters $\delta_{\parallel}=d_{\parallel}-D_{\parallel}=1\mod2$.
Here, $\delta=d-D$.}
\begin{ruledtabular}
\begin{tabular}{ccccccccccc}
Symmetry or antisymmetry & Class & Classifying space & $\delta=0$ & $\delta=1$ & $\delta=2$ & $\delta=3$ & $\delta=4$ & $\delta=5$ & $\delta=6$ & $\delta=7$\tabularnewline
\hline 
$\hat{\mathcal{U}}_{0}^{+},\hat{\mathcal{U}}_{T/2}^{+}$ & A & $\mathcal{C}_{1-\delta}$ & 0 & $\mathbb{Z}$ & 0 & $\mathbb{Z}$ & 0 & $\mathbb{Z}$ & 0 & $\mathbb{Z}$\tabularnewline
$\hat{\mathcal{U}}_{0,+}^{+},\hat{\mathcal{U}}_{T/2,-}^{+}$ & AIII & $\mathcal{C}_{-\delta}$ & $\mathbb{Z}$ & 0 & $\mathbb{Z}$ & 0 & $\mathbb{Z}$ & 0 & $\mathbb{Z}$ & 0\tabularnewline
$\overline{\mathcal{U}}_{s}^{+}$ & A & $\mathcal{C}_{-\delta}\times\mathcal{C}_{-\delta}$ & $\mathbb{Z}\oplus\mathbb{Z}$ & 0 & $\mathbb{Z}\oplus\mathbb{Z}$ & 0 & $\mathbb{Z}\oplus\mathbb{Z}$ & 0 & $\mathbb{Z}\oplus\mathbb{Z}$ & 0\tabularnewline
$\hat{\mathcal{U}}_{0,-}^{+},\hat{\mathcal{U}}_{T/2,+}^{+}$ & AIII & $\mathcal{C}_{1-\delta}\times\mathcal{C}_{1-\delta}$ & 0 & $\mathbb{Z}\oplus\mathbb{Z}$ & 0 & $\mathbb{Z}\oplus\mathbb{Z}$ & 0 & $\mathbb{Z}\oplus\mathbb{Z}$ & 0 & $\mathbb{Z}\oplus\mathbb{Z}$\tabularnewline
\hline 
\end{tabular}
\end{ruledtabular}
\end{table*}

\begin{table*}
\caption{\label{tab:complex_antiunitary_classification_1}
Classification table
for Floquet topological phases in complex symmetry classes supporting an additional order-two space-time antiunitary
symmetry with flipped parameters $\delta_{\parallel}=d_{\parallel}-D_{\parallel}=1\mod4$.
Here, $\delta=d-D$.}
\begin{ruledtabular}
\begin{tabular}{ccccccccccc}
Symmetry or antisymmetry & Class & Classifying space & $\delta=0$ & $\delta=1$ & $\delta=2$ & $\delta=3$ & $\delta=4$ & $\delta=5$ & $\delta=6$ & $\delta=7$\tabularnewline
\hline 
$\hat{\mathcal{A}}_{s}^{+}$ & A & $\mathcal{R}_{2-\delta}$ & $\mathbb{\mathbb{Z}}_{2}$ & $\mathbb{\mathbb{Z}}_{2}$ & $\mathbb{Z}$ & 0 & 0 & 0 & 2$\mathbb{Z}$ & 0\tabularnewline
$\hat{\mathcal{A}}_{0,+}^{+},\hat{\mathcal{A}}_{T/2,-}^{+}$ & AIII & $\mathcal{R}_{3-\delta}$ & 0 & $\mathbb{\mathbb{Z}}_{2}$ & $\mathbb{\mathbb{Z}}_{2}$ & $\mathbb{Z}$ & 0 & 0 & 0 & 2$\mathbb{Z}$\tabularnewline
$\overline{\mathcal{A}}_{0}^{+},\overline{\mathcal{A}}_{T/2}^{-}$ & A & $\mathcal{R}_{4-\delta}$ & 2$\mathbb{Z}$ & 0 & $\mathbb{\mathbb{Z}}_{2}$ & $\mathbb{\mathbb{Z}}_{2}$ & $\mathbb{Z}$ & 0 & 0 & 0\tabularnewline
$\hat{\mathcal{A}}_{0,-}^{-},\hat{\mathcal{A}}_{T/2,+}^{-}$ & AIII & $\mathcal{R}_{5-\delta}$ & 0 & 2$\mathbb{Z}$ & 0 & $\mathbb{\mathbb{Z}}_{2}$ & $\mathbb{\mathbb{Z}}_{2}$ & $\mathbb{Z}$ & 0 & 0\tabularnewline
$\hat{\mathcal{A}}_{s}^{-}$ & A & $\mathcal{R}_{6-\delta}$ & 0 & 0 & 2$\mathbb{Z}$ & 0 & $\mathbb{\mathbb{Z}}_{2}$ & $\mathbb{\mathbb{Z}}_{2}$ & $\mathbb{Z}$ & 0\tabularnewline
$\hat{\mathcal{A}}_{0,+}^{-},\hat{\mathcal{A}}_{T/2,-}^{-}$ & AIII & $\mathcal{R}_{7-\delta}$ & 0 & 0 & 0 & 2$\mathbb{Z}$ & 0 & $\mathbb{\mathbb{Z}}_{2}$ & $\mathbb{\mathbb{Z}}_{2}$ & $\mathbb{Z}$\tabularnewline
$\overline{\mathcal{A}}_{0}^{-},\overline{\mathcal{A}}_{T/2}^{+}$ & A & $\mathcal{R}_{-\delta}$ & $\mathbb{Z}$ & 0 & 0 & 0 & 2$\mathbb{Z}$ & 0 & $\mathbb{\mathbb{Z}}_{2}$ & $\mathbb{\mathbb{Z}}_{2}$\tabularnewline
$\hat{\mathcal{A}}_{0,-}^{+},\hat{\mathcal{A}}_{T/2,+}^{+}$ & AIII & $\mathcal{R}_{1-\delta}$ & $\mathbb{\mathbb{Z}}_{2}$ & $\mathbb{Z}$ & 0 & 0 & 0 & 2$\mathbb{Z}$ & 0 & $\mathbb{\mathbb{Z}}_{2}$\tabularnewline
\end{tabular}
\end{ruledtabular}
\end{table*}

\begin{table*}
\caption{\label{tab:real_unitary_classification_1}
Classification table
for Floquet topological phases in real symmetry classes supporting an additional order-two space-time unitary 
symmetry with flipped parameters $\delta_{\parallel}=d_{\parallel}-D_{\parallel}=1\mod4$.
Here, $\delta=d-D$.}
\begin{ruledtabular}
\begin{tabular}{ccccccccccc}
Symmetry or antisymmetry & Class & Classifying space & $\delta=0$ & $\delta=1$ & $\delta=2$ & $\delta=3$ & $\delta=4$ & $\delta=5$ & $\delta=6$ & $\delta=7$\tabularnewline
\hline 
$\text{\ensuremath{\hat{\mathcal{U}}_{0,+}^{+}}},\text{\ensuremath{\hat{\mathcal{U}}_{T/2,+}^{+}}}$ & AI & $\mathcal{R}_{1-\delta}$ & $\mathbb{\mathbb{Z}}_{2}$ & $\mathbb{Z}$ & 0 & 0 & 0 & $2\mathbb{Z}$ & 0 & $\mathbb{\mathbb{Z}}_{2}$\tabularnewline
$\hat{\mathcal{U}}_{0,++}^{+},\hat{\mathcal{U}}_{T/2,+-}^{+}$ & BDI & $\mathcal{R}_{2-\delta}$ & $\mathbb{\mathbb{Z}}_{2}$ & $\mathbb{\mathbb{Z}}_{2}$ & $\mathbb{Z}$ & 0 & 0 & 0 & $2\mathbb{Z}$ & 0\tabularnewline
$\ensuremath{\hat{\mathcal{U}}_{0,+}^{+}},\text{\ensuremath{\hat{\mathcal{U}}_{T/2,-}^{+}}}$ & D & $\mathcal{R}_{3-\delta}$ & 0 & $\mathbb{\mathbb{Z}}_{2}$ & $\mathbb{\mathbb{Z}}_{2}$ & $\mathbb{Z}$ & 0 & 0 & 0 & $2\mathbb{Z}$\tabularnewline
$\hat{\mathcal{U}}_{0,++}^{+},\hat{\mathcal{U}}_{T/2,+-}^{+}$ & DIII & $\mathcal{R}_{4-\delta}$ & $2\mathbb{Z}$ & 0 & $\mathbb{\mathbb{Z}}_{2}$ & $\mathbb{\mathbb{Z}}_{2}$ & $\mathbb{Z}$ & 0 & 0 & 0\tabularnewline
$\ensuremath{\hat{\mathcal{U}}_{0,+}^{+}},\ensuremath{\hat{\mathcal{U}}_{T/2,+}^{+}}$ & AII & $\mathcal{R}_{5-\delta}$ & 0 & $2\mathbb{Z}$ & 0 & $\mathbb{\mathbb{Z}}_{2}$ & $\mathbb{\mathbb{Z}}_{2}$ & $\mathbb{Z}$ & 0 & 0\tabularnewline
$\hat{\mathcal{U}}_{0,++}^{+},\hat{\mathcal{U}}_{T/2,+-}^{+}$ & CII & $\mathcal{R}_{6-\delta}$ & 0 & 0 & $2\mathbb{Z}$ & 0 & $\mathbb{\mathbb{Z}}_{2}$ & $\mathbb{\mathbb{Z}}_{2}$ & $\mathbb{Z}$ & 0\tabularnewline
$\ensuremath{\hat{\mathcal{U}}_{0,+}^{+}},\ensuremath{\hat{\mathcal{U}}_{T/2,-}^{+}}$ & C & $\mathcal{R}_{7-\delta}$ & 0 & 0 & 0 & $2\mathbb{Z}$ & 0 & $\mathbb{\mathbb{Z}}_{2}$ & $\mathbb{\mathbb{Z}}_{2}$ & $\mathbb{Z}$\tabularnewline
\end{tabular}
\end{ruledtabular}
\end{table*}

\begin{table*}
\caption{\label{tab:complex_antiunitary_classification_2}
Classification table
for Floquet topological phases in complex symmetry classes supporting an additional order-two space-time antiunitary
symmetry with flipped parameters $\delta_{\parallel}=d_{\parallel}-D_{\parallel}=2\mod4$.
Here, $\delta=d-D$.}
\begin{ruledtabular}
\begin{tabular}{ccccccccccc}
Symmetry or antisymmetry & Class & Classifying space & $\delta=0$ & $\delta=1$ & $\delta=2$ & $\delta=3$ & $\delta=4$ & $\delta=5$ & $\delta=6$ & $\delta=7$\tabularnewline
\hline 
$\hat{\mathcal{A}}_{s}^{+}$ & A & $\mathcal{R}_{4-\delta}$ & 2$\mathbb{Z}$ & 0 & $\mathbb{\mathbb{Z}}_{2}$ & $\mathbb{\mathbb{Z}}_{2}$ & $\mathbb{Z}$ & 0 & 0 & 0\tabularnewline
$\hat{\mathcal{A}}_{0,+}^{+},\hat{\mathcal{A}}_{T/2,-}^{+}$ & AIII & $\mathcal{R}_{5-\delta}$ & 0 & 2$\mathbb{Z}$ & 0 & $\mathbb{\mathbb{Z}}_{2}$ & $\mathbb{\mathbb{Z}}_{2}$ & $\mathbb{Z}$ & 0 & 0\tabularnewline
$\overline{\mathcal{A}}_{0}^{+},\overline{\mathcal{A}}_{T/2}^{-}$ & A & $\mathcal{R}_{6-\delta}$ & 0 & 0 & 2$\mathbb{Z}$ & 0 & $\mathbb{\mathbb{Z}}_{2}$ & $\mathbb{\mathbb{Z}}_{2}$ & $\mathbb{Z}$ & 0\tabularnewline
$\hat{\mathcal{A}}_{0,-}^{-},\hat{\mathcal{A}}_{T/2,+}^{-}$ & AIII & $\mathcal{R}_{7-\delta}$ & 0 & 0 & 0 & 2$\mathbb{Z}$ & 0 & $\mathbb{\mathbb{Z}}_{2}$ & $\mathbb{\mathbb{Z}}_{2}$ & $\mathbb{Z}$\tabularnewline
$\hat{\mathcal{A}}_{s}^{-}$ & A & $\mathcal{R}_{-\delta}$ & $\mathbb{Z}$ & 0 & 0 & 0 & 2$\mathbb{Z}$ & 0 & $\mathbb{\mathbb{Z}}_{2}$ & $\mathbb{\mathbb{Z}}_{2}$\tabularnewline
$\hat{\mathcal{A}}_{0,+}^{-},\hat{\mathcal{A}}_{T/2,-}^{-}$ & AIII & $\mathcal{R}_{1-\delta}$ & $\mathbb{\mathbb{Z}}_{2}$ & $\mathbb{Z}$ & 0 & 0 & 0 & 2$\mathbb{Z}$ & 0 & $\mathbb{\mathbb{Z}}_{2}$\tabularnewline
$\overline{\mathcal{A}}_{0}^{-},\overline{\mathcal{A}}_{T/2}^{+}$ & A & $\mathcal{R}_{2-\delta}$ & $\mathbb{\mathbb{Z}}_{2}$ & $\mathbb{\mathbb{Z}}_{2}$ & $\mathbb{Z}$ & 0 & 0 & 0 & 2$\mathbb{Z}$ & 0\tabularnewline
$\hat{\mathcal{A}}_{0,-}^{+},\hat{\mathcal{A}}_{T/2,+}^{+}$ & AIII & $\mathcal{R}_{3-\delta}$ & 0 & $\mathbb{\mathbb{Z}}_{2}$ & $\mathbb{\mathbb{Z}}_{2}$ & $\mathbb{Z}$ & 0 & 0 & 0 & 2$\mathbb{Z}$\tabularnewline
\end{tabular}
\end{ruledtabular}
\end{table*}

\begin{table*}
\caption{\label{tab:real_unitary_classification_2}
Classification table
for Floquet topological phases in real symmetry classes supporting an additional order-two space-time unitary 
symmetry with flipped parameters $\delta_{\parallel}=d_{\parallel}-D_{\parallel}=2\mod4$.
Here, $\delta=d-D$.}
\begin{ruledtabular}
\begin{tabular}{ccccccccccc}
Symmetry or antisymmetry & Class & Classifying space & $\delta=0$ & $\delta=1$ & $\delta=2$ & $\delta=3$ & $\delta=4$ & $\delta=5$ & $\delta=6$ & $\delta=7$\tabularnewline
\hline 
$\text{\ensuremath{\hat{\mathcal{U}}_{0,+}^{+}}},\text{\ensuremath{\hat{\mathcal{U}}_{T/2,+}^{+}}}$ & AI & \emph{$\mathcal{C}_{-\delta}$} & $\mathbb{Z}$ & 0 & $\mathbb{Z}$ & 0 & $\mathbb{Z}$ & 0 & $\mathbb{Z}$ & 0\tabularnewline
$\hat{\mathcal{U}}_{0,++}^{+},\hat{\mathcal{U}}_{T/2,+-}^{+}$ & BDI & $\mathcal{C}_{1-\delta}$ & 0 & $\mathbb{Z}$ & 0 & $\mathbb{Z}$ & 0 & $\mathbb{Z}$ & 0 & $\mathbb{Z}$\tabularnewline
$\ensuremath{\hat{\mathcal{U}}_{0,+}^{+}},\text{\ensuremath{\hat{\mathcal{U}}_{T/2,-}^{+}}}$ & D & \emph{$\mathcal{C}_{-\delta}$} & $\mathbb{Z}$ & 0 & $\mathbb{Z}$ & 0 & $\mathbb{Z}$ & 0 & $\mathbb{Z}$ & 0\tabularnewline
$\hat{\mathcal{U}}_{0,++}^{+},\hat{\mathcal{U}}_{T/2,+-}^{+}$ & DIII & $\mathcal{C}_{1-\delta}$ & 0 & $\mathbb{Z}$ & 0 & $\mathbb{Z}$ & 0 & $\mathbb{Z}$ & 0 & $\mathbb{Z}$\tabularnewline
$\ensuremath{\hat{\mathcal{U}}_{0,+}^{+}},\ensuremath{\hat{\mathcal{U}}_{T/2,+}^{+}}$ & AII & \emph{$\mathcal{C}_{-\delta}$} & $\mathbb{Z}$ & 0 & $\mathbb{Z}$ & 0 & $\mathbb{Z}$ & 0 & $\mathbb{Z}$ & 0\tabularnewline
$\hat{\mathcal{U}}_{0,++}^{+},\hat{\mathcal{U}}_{T/2,+-}^{+}$ & CII & $\mathcal{C}_{1-\delta}$ & 0 & $\mathbb{Z}$ & 0 & $\mathbb{Z}$ & 0 & $\mathbb{Z}$ & 0 & $\mathbb{Z}$\tabularnewline
$\ensuremath{\hat{\mathcal{U}}_{0,+}^{+}},\ensuremath{\hat{\mathcal{U}}_{T/2,-}^{+}}$ & C & \emph{$\mathcal{C}_{-\delta}$} & $\mathbb{Z}$ & 0 & $\mathbb{Z}$ & 0 & $\mathbb{Z}$ & 0 & $\mathbb{Z}$ & 0\tabularnewline
\end{tabular}
\end{ruledtabular}
\end{table*}

\begin{table*}
\caption{\label{tab:complex_antiunitary_classification_3}
Classification table
for Floquet topological phases in complex symmetry classes supporting an additional order-two space-time antiunitary
symmetry with flipped parameters $\delta_{\parallel}=d_{\parallel}-D_{\parallel}=3\mod4$.
Here, $\delta=d-D$.}
\begin{ruledtabular}
\begin{tabular}{ccccccccccc}
Symmetry or antisymmetry & Class & Classifying space & $\delta=0$ & $\delta=1$ & $\delta=2$ & $\delta=3$ & $\delta=4$ & $\delta=5$ & $\delta=6$ & $\delta=7$\tabularnewline
\hline 
$\hat{\mathcal{A}}_{s}^{+}$ & A & $\mathcal{R}_{6-\delta}$ & 0 & 0 & 2$\mathbb{Z}$ & 0 & $\mathbb{\mathbb{Z}}_{2}$ & $\mathbb{\mathbb{Z}}_{2}$ & $\mathbb{Z}$ & 0\tabularnewline
$\hat{\mathcal{A}}_{0,+}^{+},\hat{\mathcal{A}}_{T/2,-}^{+}$ & AIII & $\mathcal{R}_{7-\delta}$ & 0 & 0 & 0 & 2$\mathbb{Z}$ & 0 & $\mathbb{\mathbb{Z}}_{2}$ & $\mathbb{\mathbb{Z}}_{2}$ & $\mathbb{Z}$\tabularnewline
$\overline{\mathcal{A}}_{0}^{+},\overline{\mathcal{A}}_{T/2}^{-}$ & A & $\mathcal{R}_{-\delta}$ & $\mathbb{Z}$ & 0 & 0 & 0 & 2$\mathbb{Z}$ & 0 & $\mathbb{\mathbb{Z}}_{2}$ & $\mathbb{\mathbb{Z}}_{2}$\tabularnewline
$\hat{\mathcal{A}}_{0,-}^{-},\hat{\mathcal{A}}_{T/2,+}^{-}$ & AIII & $\mathcal{R}_{1-\delta}$ & $\mathbb{\mathbb{Z}}_{2}$ & $\mathbb{Z}$ & 0 & 0 & 0 & 2$\mathbb{Z}$ & 0 & $\mathbb{\mathbb{Z}}_{2}$\tabularnewline
$\hat{\mathcal{A}}_{s}^{-}$ & A & $\mathcal{R}_{2-\delta}$ & $\mathbb{\mathbb{Z}}_{2}$ & $\mathbb{\mathbb{Z}}_{2}$ & $\mathbb{Z}$ & 0 & 0 & 0 & 2$\mathbb{Z}$ & 0\tabularnewline
$\hat{\mathcal{A}}_{0,+}^{-},\hat{\mathcal{A}}_{T/2,-}^{-}$ & AIII & $\mathcal{R}_{3-\delta}$ & 0 & $\mathbb{\mathbb{Z}}_{2}$ & $\mathbb{\mathbb{Z}}_{2}$ & $\mathbb{Z}$ & 0 & 0 & 0 & 2$\mathbb{Z}$\tabularnewline
$\overline{\mathcal{A}}_{0}^{-},\overline{\mathcal{A}}_{T/2}^{+}$ & A & $\mathcal{R}_{4-\delta}$ & 2$\mathbb{Z}$ & 0 & $\mathbb{\mathbb{Z}}_{2}$ & $\mathbb{\mathbb{Z}}_{2}$ & $\mathbb{Z}$ & 0 & 0 & 0\tabularnewline
$\hat{\mathcal{A}}_{0,-}^{+},\hat{\mathcal{A}}_{T/2,+}^{+}$ & AIII & $\mathcal{R}_{5-\delta}$ & 0 & 2$\mathbb{Z}$ & 0 & $\mathbb{\mathbb{Z}}_{2}$ & $\mathbb{\mathbb{Z}}_{2}$ & $\mathbb{Z}$ & 0 & 0\tabularnewline
\end{tabular}
\end{ruledtabular}
\end{table*}

\begin{table*}
\caption{\label{tab:real_unitary_classification_3}
Classification table
for Floquet topological phases in real symmetry classes supporting an additional order-two space-time unitary 
symmetry with flipped parameters $\delta_{\parallel}=d_{\parallel}-D_{\parallel}=3\mod4$.
Here, $\delta=d-D$.}
\begin{ruledtabular}
\begin{tabular}{ccccccccccc}
Symmetry or antisymmetry & Class & Classifying space & $\delta=0$ & $\delta=1$ & $\delta=2$ & $\delta=3$ & $\delta=4$ & $\delta=5$ & $\delta=6$ & $\delta=7$\tabularnewline
\hline 
$\text{\ensuremath{\hat{\mathcal{U}}_{0,+}^{+}}},\text{\ensuremath{\hat{\mathcal{U}}_{T/2,+}^{+}}}$ & AI & $\mathcal{R}_{7-\delta}$ & 0 & 0 & 0 & $2\mathbb{Z}$ & 0 & $\mathbb{\mathbb{Z}}_{2}$ & $\mathbb{\mathbb{Z}}_{2}$ & $\mathbb{Z}$\tabularnewline
$\hat{\mathcal{U}}_{0,++}^{+},\hat{\mathcal{U}}_{T/2,+-}^{+}$ & BDI & $\mathcal{R}_{-\delta}$ & $\mathbb{Z}$ & 0 & 0 & 0 & $2\mathbb{Z}$ & 0 & $\mathbb{Z}_{2}$ & $\mathbb{\mathbb{Z}}_{2}$\tabularnewline
$\ensuremath{\hat{\mathcal{U}}_{0,+}^{+}},\text{\ensuremath{\hat{\mathcal{U}}_{T/2,-}^{+}}}$ & D & $\mathcal{R}_{1-\delta}$ & $\mathbb{\mathbb{Z}}_{2}$ & $\mathbb{Z}$ & 0 & 0 & 0 & $2\mathbb{Z}$ & 0 & $\mathbb{Z}_{2}$\tabularnewline
$\hat{\mathcal{U}}_{0,++}^{+},\hat{\mathcal{U}}_{T/2,+-}^{+}$ & DIII & $\mathcal{R}_{2-\delta}$ & $\mathbb{\mathbb{Z}}_{2}$ & $\mathbb{\mathbb{Z}}_{2}$ & $\mathbb{Z}$ & 0 & 0 & 0 & $2\mathbb{Z}$ & 0\tabularnewline
$\ensuremath{\hat{\mathcal{U}}_{0,+}^{+}},\ensuremath{\hat{\mathcal{U}}_{T/2,+}^{+}}$ & AII & $\mathcal{R}_{3-\delta}$ & 0 & $\mathbb{\mathbb{Z}}_{2}$ & $\mathbb{\mathbb{Z}}_{2}$ & $\mathbb{Z}$ & 0 & 0 & 0 & $2\mathbb{Z}$\tabularnewline
$\hat{\mathcal{U}}_{0,++}^{+},\hat{\mathcal{U}}_{T/2,+-}^{+}$ & CII & $\mathcal{R}_{4-\delta}$ & $2\mathbb{Z}$ & 0 & $\mathbb{\mathbb{Z}}_{2}$ & $\mathbb{\mathbb{Z}}_{2}$ & $\mathbb{Z}$ & 0 & 0 & 0\tabularnewline
$\ensuremath{\hat{\mathcal{U}}_{0,+}^{+}},\ensuremath{\hat{\mathcal{U}}_{T/2,-}^{+}}$ & C & $\mathcal{R}_{5-\delta}$ & 0 & $2\mathbb{Z}$ & 0 & $\mathbb{\mathbb{Z}}_{2}$ & $\mathbb{\mathbb{Z}}_{2}$ & $\mathbb{Z}$ & 0 & 0\tabularnewline
\end{tabular}
\end{ruledtabular}
\end{table*}

\end{widetext}
\end{document}